\documentclass[a4paper,11pt]{article}
\usepackage{jheppub}
\usepackage[utf8]{inputenc}
\usepackage[compat=1.1.0]{tikz-feynman}
\usepackage{tikz}
\usetikzlibrary{decorations.markings}
\usepackage{cancel}
\usepackage{bbold}
\usepackage{bbm}
\usepackage{bm}
\usepackage{caption}
\usepackage{subcaption}
\usepackage{hyperref}
\usepackage{mathtools,slashed}
\usepackage{mathtools}
\usepackage{xcolor}
\usepackage{appendix}
\usepackage{amsmath,amssymb}

\usepackage{graphicx}
\usepackage[capitalise]{cleveref}
\usepackage{todonotes}
\usepackage{enumitem}
\usepackage{multirow}
\usepackage[T1]{fontenc}
\usepackage{colortbl}
\usepackage{xspace}

\AddToHook{cmd/appendix/before}{\crefalias{section}{appendix}}

\DeclareMathOperator*{\SumInt}{
\mathchoice
  {\ooalign{$\displaystyle\sum$\cr\hidewidth$\displaystyle\int$\hidewidth\cr}}
  {\ooalign{\raisebox{.14\height}{\scalebox{.7}{$\textstyle\sum$}}\cr\hidewidth$\textstyle\int$\hidewidth\cr}}
  {\ooalign{\raisebox{.2\height}{\scalebox{.6}{$\scriptstyle\sum$}}\cr$\scriptstyle\int$\cr}}
  {\ooalign{\raisebox{.2\height}{\scalebox{.6}{$\scriptstyle\sum$}}\cr$\scriptstyle\int$\cr}}
}

\usepackage{bm}

\usepackage{booktabs}

\newcommand{\eg}{\textit{e.g.\ }}
\newcommand{\ie}{\textit{i.e.\ }}
\newcommand{\viz}{\textit{viz.\ }}
\newcommand{\dd}{\mathrm{d}}
\newcommand{\Veff}{V}
\newcommand{\msbar}{\ensuremath{\overline{\text{MS}}}\xspace}
\newcommand{\os}{\ensuremath{\widetilde{\text{OS}}}\xspace}
\newcommand{\mh}{m_h}
\newcommand{\ms}{m_s}
\newcommand{\rxi}{\ensuremath{R_\xi}\xspace}

\graphicspath{{./plots/}}

\hypersetup{
    colorlinks=true,
    linkcolor=blue,
    citecolor=cyan,
    urlcolor=blue,
    }

\title{Perturbative aspects of the electroweak phase
transition with a complex singlet and implications for
gravitational wave predictions}

\author[a]{Thomas Biekötter,}
\author[b,c]{Andrii Dashko,}
\author[b]{Maximilian Löschner,}
\author[b,d]{Georg Weiglein}

\affiliation[a]{
        Instituto de F\'isica Te\'orica UAM/CSIC,\\
        Calle Nicolás Cabrera 13-15,
        Cantoblanco, 28049, Madrid, Spain
}

\affiliation[b]{
	Deutsches Elektronen-Synchrotron DESY,\\
	Notkestr.~85, 22607 Hamburg, Germany
}

\affiliation[c]{
    Departamento de F\'isica Te\'orica y del Cosmos,\\ Universidad de Granada, E--18071 Granada, Spain}

\affiliation[d]{
	II. Institut für Theoretische Physik,\\
    Universität Hamburg, Luruper Chaussee 149, 22761 Hamburg,
    Germany
}

\emailAdd{thomas.biekoetter@desy.de}
\emailAdd{andrii.dashko@desy.de}
\emailAdd{maximilian.loeschner@desy.de}
\emailAdd{georg.weiglein@desy.de}

\abstract{
We present a detailed analysis of strong first-order electroweak phase transitions within the extension of the Standard Model 
by a complex scalar singlet (cxSM). Focusing on the impact of 
renormalization scale and gauge 
dependence, we systematically compare commonly used perturbative frameworks for 
predicting
thermodynamic observables that characterize the phase transition and the associated gravitational-wave (GW) spectrum. These include both the four-dimensional ($4D$) formalism and the dimensionally reduced three-dimensional effective field theory ($3D$ EFT) approach in different renormalization schemes. Within the $3D$ EFT, we compute the effective potential up to two-loop order in a general $R_\xi$ gauge, and demonstrate that applying the $\hbar$-expansion yields gauge-independent results in excellent agreement with those obtained from a direct minimization of the loop-corrected potential. In contrast, large discrepancies between the two methods persist in the $4D$ approaches. We find that, across most of the parameter space, the $3D$ EFT approach provides 
predictions for phase-transition parameters and GW spectra with theoretical uncertainties in the GW peak amplitude that are reduced by more than an order of magnitude compared to the $4D$ calculations. We point out, however, that the $3D$ EFT approach is subject to an additional theory uncertainty from truncating the EFT at finite operator dimension and show that 
higher-dimensional operators within the $3D$ EFT approach 
can substantially modify the predicted transition strength and GW signals. This indicates a potential breakdown of the high-temperature expansion precisely in the region with the lowest transition temperatures, where the strongest GW signals are expected and the detection prospects with LISA are most promising.
}

\preprint{DESY-25-131, IFT-UAM/CSIC-25-104}
\begin{document}

\maketitle

\section{Introduction}
\label{sec:intro}

The possibility of a First Order Electroweak Phase Transition (FOEWPT) in the early universe has intrigued physicists for decades, as it might 
play a fundamental role in solving pressing
open issues, such as 
providing an explanation for the
origin of the
observed baryon asymmetry
of the universe~(BAU)~\cite{Kuzmin:1985mm}.
Recently, with the first direct detection
of gravitational waves~(GW) by the LIGO and VIRGO
collaborations in 2015~\cite{LIGOScientific:2016aoc},
the interest in a cosmological FOEWPT has further increased
since it produces a primordial GW background that might be
detectable at current and future (space-based) GW
experiments~\cite{ET:2019dnz,
LISACosmologyWorkingGroup:2022jok,Affleck:1980ac}.

Moreover, a cosmological phase transition can serve,
for instance,
as a
mechanism to generate cosmological topological defects
(e.g.~domain walls)~\cite{Zeldovich:1974uw,Kibble:1976sj},
primordial black holes~\cite{Hawking:1982ga,Kodama:1982sf}
that potentially
contribute to the observed dark-matter relic
abundance (see Refs.~\cite{Lewicki:2019gmv,Liu:2021svg,Baldes:2023rqv}
for recent studies),
cosmic magnetic fields~\cite{Vachaspati:1991nm,Sigl:1996dm,
Olea-Romacho:2023rhh}
and for
the production of heavy particles~\cite{Azatov:2021ifm,Baldes:2023cih}.

In the Standard Model~(SM), with the mass of the detected 
Higgs boson
measured at about
125\,GeV~\cite{ATLAS:2012yve,CMS:2012qbp},
electroweak~(EW) symmetry breaking is predicted to
proceed via a smooth crossover instead
of a 
strong FOEWPT (SFOEWPT)~\cite{Kajantie:1996mn}.
As a consequence, the presence of a 
SFOEWPT
would be a clear indication of
physics beyond the Standard Model~(BSM).
A SFOEWPT can occur
even in simple extensions of the SM.
For instance, models with additional scalar
singlets~\cite{Profumo:2007wc,Espinosa:2011ax,Cline:2012hg},
SU(2) doublets~\cite{Dorsch:2013wja,
Basler:2016obg,Goncalves:2021egx,Biekotter:2021ysx,
Biekotter:2022kgf,Biekotter:2023eil}
and/or triplets~\cite{Patel:2012pi,Niemi:2018asa},
effective theories incorporating
higher dimensional operators~\cite{Grojean:2004xa,
Camargo-Molina:2021zgz,Camargo-Molina:2024sde},
classically conformal models~\cite{Nardini:2007me,Konstandin:2011dr,Kierkla:2022odc,Kierkla:2023von},
or models with extra dimensions~\cite{Creminelli:2001th}
can all exhibit 
a SFOEWPT.
As mentioned above, a sufficiently strong
FOEWPT can be probed by 
future space-based GW observatories.
The 
observation of a GW signal would be very helpful in this context to
pinpoint the underlying scalar sector
responsible for the FOEWPT. 

A reliable and precise way to study
phase transitions is via non-perturbative methods
and lattice field theory computations.
However, due to their high computational cost, it is not feasible to apply such methods
in scans over large parameter spaces of BSM theories
with several free parameters that can have an
impact on the transition.
This
necessitates the development of reliable perturbative 
methods for 
detailed studies of phase transition
dynamics and their phenomenological consequences 
within the allowed regions of the parameter space of the considered model.
A variety of perturbative methods for this purpose have been
developed and applied
in the literature.
They can be divided into two main approaches:

The traditional four-dimensional~($4D$) approach
relies on the inclusion of temperature
and quantum corrections to the effective potential in the UV theory,
typically incorporating thermal resummation techniques such
as daisy resummation to improve the infrared
behavior (see e.g.~Ref.~\cite{Quiros:1999jp} for a review).
The simplest and
most-often applied resummation prescriptions
are the Arnold--Espinosa resummation
approach~\cite{Arnold:1992rz},
where only the Matsubara zero modes are
resummed, and the Parwani resummation
approach~\cite{Parwani:1991gq},
where masses of all bosonic modes receive thermal
corrections.
An alternative thermal resummation 
prescription called ``partial dressing''
or ``tadpole resummation''
has been developed, which systematically accounts
for daisy and superdaisy diagrams
by computing thermal masses from the gap equation
and inserting them into the first derivative of
the effective potential with respect to the 
fields~\cite{Boyd:1993tz,Curtin:2016urg,
Bahl:2024ykv,Bittar:2025lcr}. 
Moreover, a method for constructing a resummed effective potential without resorting to the high-temperature expansion has recently been proposed \cite{Navarrete:2025yxy}, based on the separation of contributions from different scales. 
There are also recent efforts to reduce the renormalization scale dependence by reorganizing the perturbation theory, employing the so-called Optimized Perturbation Theory \cite{Camara:2025zmb}.

Irrespective of the specific thermal resummation
prescription,
the main advantage of the $4D$ approach is that 
the form of the effective potential at one-loop level is
relatively simple to implement, 
while capturing the leading
thermal corrections and providing a
qualitatively
correct picture 
of the phase transitions in many
extensions of the SM. Recent studies using the $4D$
approach 
have, for instance, been carried out
for
the singlet-extended
SM~\cite{Jiang:2015cwa,Chiang:2017nmu,Chiang:2018gsn,Carena:2019une,Cho:2021itv,
Fernandez-Martinez:2022stj,Ellis:2022lft,Zhang:2023mnu,
Braathen:2025svl},
the Two Higgs doublet model~\cite{Goncalves:2021egx,
Biekotter:2022kgf,Biekotter:2021ysx}
and the triplet-extended SM~\cite{Chala:2018opy,Zhou:2018zli}, taking into account current experimental constraints.
In some studies also two-loop contributions have
been considered in the $4D$ approach~\cite{Arnold:1992rz,
Espinosa:1996qw,Laine:2000kv,
Funakubo:2012qc,Bahl:2024ykv}.
A key drawback of the $4D$ approach is that
it is often affected by the poor behavior of the perturbative
expansion at finite temperature, leading to sizable
renormalization scale and
residual gauge dependencies in predictions for physical
observables characterizing
the phase transitions.
In addition, it suffers from infrared divergences
and an unsystematic treatment of thermal fluctuations in the evaluation of transition rates~\cite{Ekstedt:2020abj,
Athron:2022jyi}.
This leads to large theoretical uncertainties in the
predictions 
for these parameters and quantities that are derived from them, such as GW signals produced during the
transitions~\cite{Croon:2020cgk,Lewicki:2024xan,Gould:2024jjt,Zhu:2025pht}.

A more refined alternative to the traditional $4D$ approaches
is the dimensionally reduced three-dimensional~(3D)
approach~\cite{Ginsparg:1980ef,Appelquist:1981vg},
which systematically integrates out heavy modes to obtain
a low-energy effective field theory~(EFT)
in three spatial dimensions that captures
the relevant infrared dynamics of the phase transition.
This method is well-defined in the high-temperature limit
and provides more accurate predictions compared to the $4D$ approach
for the case
in which the temperature of the phase transition
is of the order or larger than the masses of the
particles contained in the dimensionally reduced EFT.
Since the EFT only contains bosonic degrees of freedom,
it also facilitates a more direct comparison with
lattice simulations~\cite{Niemi:2020hto,Niemi:2024axp,
Ramsey-Musolf:2024ykk} 
than the $4D$ approach.
In perturbation theory,
the $3D$ approach, which exploits the separation of thermal scales, allows the inclusion of
thermal next-to-leading
order~(NLO) two-loop contributions in a
simpler and more systematic way~\cite{Farakos:1994kx}.
A drawback of the $3D$ approach compared to
the traditional $4D$ approach is that one has to perform
a non-trivial matching procedure
between the underlying
UV theory and the EFT describing the infrared dynamics.
In particular, this matching has to be 
derived not
only in a model-dependent way, but it also depends
on the specific realization of the
particle spectrum within a model, and thus in general
has to be 
carried out differently
for different choices
of the free parameters of the model.
This makes the application of the $3D$ approach more
involved in the context of BSM theories, whereas
the $4D$ approach can be applied 
with less additional effort in comparison to the case of the SM.
Despite these 
difficulties, there has been significant
development in recent years regarding tools that
can be used to construct $3D$ EFTs in an automized
way for various classes of BSM theories.
This development has led to the public computer
program \texttt{DRalgo}~\cite{Ekstedt:2022bff}
that can be used to perform the matching
between the UV theory and the high-$T$ EFT in order to construct
a dimensionally reduced,
high-temperature EFT. This can be applied to generic models 
in which an appropriate hierarchy of scales can be exploited. 
\texttt{DRalgo} provides
two-loop thermal corrections to scalar and Debye
masses as well as one-loop thermal corrections to
couplings.
Dimensional reduction has nowadays become a widely used
tool 
for the study of the EW phase transition in
scalar extensions of the SM where the high-temperature expansion is applicable, see for instance
Refs.~\cite{Niemi:2018asa,Kainulainen:2019kyp,
Niemi:2024vzw}
for recent analyses.

Another problem occurring
both in the
$4D$ and the $3D$ approaches is the presence of residual
gauge dependence in predictions for physical observables
related to the phase transition.
A proposed solution to this issue is the so-called
$\hbar$-expansion~\cite{Patel:2011th},
which systematically expands physical quantities in
powers of a loop-counting parameter~$\hbar$.
This ensures that one obtains gauge-independent
results for physical observables order by order
in perturbation theory.
At the one-loop level in the $4D$ approach, applying
the $\hbar$-expansion removes gauge dependence,
but leads to results that, in many models, are not
sufficiently precise, deviating substantially
from more accurate predictions~\cite{Patel:2011th}.
Additionally, the $\hbar$-expansion must be
implemented consistently with the thermal
resummation procedure to obtain a gauge-independent formulation
at finite temperature~\cite{Ekstedt:2018ftj,
Athron:2022jyi,Qin:2024dfp}.

In this work, we point out that
in the $4D$ approach, the $\hbar$-expansion 
is difficult to pursue up to a sufficiently high order that yields the desired level of accuracy. 
In the $3D$ approach, we show that it is possible to achieve a
gauge-independent formulation with higher numerical precision for the predicted quantities than in the $4D$ approach.
By systematically including
thermal effects via dimensional reduction, we obtain results for
physical observables and predictions for GW signals
that are both gauge-independent and numerically accurate.
We compare gauge-independent results obtained via $\hbar$-expansion for the thermodynamic observables with the gauge-dependent ones, obtained by the direct minimization of the effective potential. For the latter, we evaluate the effective potential at the two-loop order in $\rxi$-gauge, cross-checking the two-loop potential in the Landau gauge, which is implemented in \texttt{DRalgo}.
Moreover, we compare the residual gauge dependence in \rxi- and Fermi gauges at the one-loop level.

We furthermore compare the uncertainty 
in the predictions of observables
describing
the FOEWPT arising from the gauge dependence 
with the uncertainties
related to the variation of renormalization and
matching scales in both $4D$ and $3D$ approaches.
For instance, we show how various matching orders of dimensional reduction affect the matching scale uncertainty.

Specifically,
we investigate the EW phase transition
in the SM extended by a complex gauge singlet
scalar field~(cxSM).
The cxSM provides 
an interesting
framework in view of the fact that no clear sign of new physics has been detected at the LHC up to now,
as the cxSM can accommodate
a SFOEWPT
without predicting detectable traces at current and future runs of the LHC ~\cite{Ashoorioon:2009nf}.
Consequently, 
the possibility of probing a
SFOEWPT in the cxSM via GWs becomes particularly important.
It is therefore crucial to make precise
and accurate predictions for the parameters
characterizing the transition
and to ensure theoretical
control over GW predictions.
The model has been studied previously using
the $4D$ approach in Refs.~\cite{Chiang:2017nmu,
Freitas:2021yng,Cho:2021itv,Zhang:2023mnu,Ghosh:2025rbt,Jiang:2015cwa} and
the $3D$ approach in Ref.~\cite{Schicho:2022wty}.
However, a comprehensive analysis of the parameter
space facilitating 
a SFOEWPT
and
of the prospects for 
future GW detection 
comparing different perturbative methods
has not yet been carried out.
In this paper, we study thermodynamic observables of FOEWPTs and determine the parameter
space regions that can potentially be probed
with the LISA GW observatory\footnote{The question whether
a predicted GW signal is detectable or not at a particular experiment can ultimately only be
answered if additional
theoretical uncertainties, e.g.~from the hydrodynamic modeling of GW production
during the FOEWPT~\cite{Caprini:2024hue} and ambiguities in choosing an appropriate transition temperature for the evaluation~\cite{Athron:2023rfq},
as well as
experimental uncertainties,
e.g.~from unknown (astrophysical or cosmological)
GW foregrounds are 
taken into account. These
considerations lie beyond
the scope of this work, where we focus on the
theoretical uncertainties in the perturbative
description of the EW phase transition.},
focusing especially
on the theoretical uncertainties in the predictions
and the impact
of residual gauge dependence
and renormalization- and matching-scale 
dependence.
We compare and contrast the $4D$ and the $3D$ approaches,
assessing their respective limitations in capturing
the dynamics of the transition. 
Moreover, we test the validity of the $3D$ approach --- which relies on the high-temperature expansion --- by including higher-dimensional operators in the EFT Lagrangian.

The outline of the paper is as follows:
in \cref{sec:cxsm} we introduce the cxSM and
specify our notation. In \cref{sec:effpot} we discuss the construction of the finite-temperature
effective potential in $4D$ and 3D,
focusing in particular on the gauge dependence
and scale dependence. In \cref{sec:thermo} we
discuss the related theoretical uncertainties
on thermodynamic parameters.
\Cref{sec:gravwave} is devoted to the predictions
for the stochastic GW backgrounds produced during
a SFOEWPT, 
and
we determine the parameter space that can potentially be
probed with future GW experiments.
In \cref{sec:highdim}, we analyze the range of validity for the $3D$ approach
by studying the impact
of higher-dimensional operators of the EFT.
We summarize our results and conclude in
\cref{sec:conclus}.
In addition, the relations between physical input and Lagrange parameters in
different schemes are discussed in \cref{app:renormalization}; 
mass eigenstates of the cxSM in \rxi and Fermi gauges are presented in \cref{app:masseigenstates}.

\section{The model: cxSM}
\label{sec:cxsm}

The model used for studying
the phase transition dynamics
in this work is
the Standard Model with an additional complex
scalar singlet~(cxSM).
The cxSM is one of the simplest
SM extensions
that can realize a (strong) FOEWPT. 
This
leads to a variety of phenomenologically interesting
signatures, such as GW signals produced in the early
universe which would potentially be accessible at
near-future space-based GW observatories.
Moreover, the cxSM can include a stable neutral
particle that could serve as a viable
dark matter candidate.
However, we will not discuss this feature
in our analysis below.

The scalar sector of the cxSM
consists of the doublet $\Phi$ with hypercharge $Y = 1$
and a complex gauge singlet $S$,
\begin{equation}
  \Phi = \left( \begin{array}{c}
    G^+ \\
    \frac{1}{\sqrt{2}}\left(\phi_h + h + i G^0\right)
  \end{array} \right) \, , \quad
  S =\frac{1}{\sqrt{2}}(\phi_s + s + i A) \, ,
\end{equation}
where we expanded the real part of the neutral component of $\Phi$
and the real part of the singlet field $S$ 
around the background field configurations $\phi_h$ and $\phi_s$, respectively. 
The Lagrangian density of the model in Minkowski space 
can be written as
\begin{equation}
	\mathcal{L}_\text{cxSM} = \mathcal{L}_\text{YM} + \mathcal{L}_\text{F} + \mathcal{L}_{\Phi S},
\end{equation}
where  $\mathcal{L}_\text{YM}$, $\mathcal{L}_\text{F}$ and $\mathcal{L}_{\Phi S}$ are the Yang-Mills, fermionic and scalar parts of the Lagrangian, respectively.
Since only the doublet $\Phi$ is charged under
the $\text{SU(2)}_L \times \text{U}(1)_Y$ gauge groups,
the gauge and fermionic sectors of the cxSM are identical
to the SM.
The electroweak part of the Yang-Mills Lagrangian is given by
\begin{equation}
    \mathcal{L}_\textrm{YM} \supset -\frac{1}{4}\left(\partial_{\mu}W^a_{\nu}-\partial_{\nu}W^a_{\mu}+g\varepsilon^{abc}W^b_{\mu}W^c_{\nu}\right)^2-\frac{1}{4}\left(\partial_{\mu}B_{\nu}-\partial_{\nu}B_{\mu}\right)^2 \, ,
\end{equation}
where $W^a_\mu$ and $B_\mu$ are the
corresponding gauge fields,
$\tau^a$ the generators of the $\text{SU(2)}_L$
gauge group, and $g$ is the SU(2)$_L$ gauge coupling.
The fermionic part is identical to the SM (for instance, see \cite{Denner:1991kt});
however, since only the top quark is relevant for
the phase transition dynamics due to its large mass, we only take into account its Yukawa coupling to the Higgs doublet,
\begin{equation}
    \mathcal{L}_\textrm{F} \supset - y_t \, \overline{Q}_L^3 \, (i \sigma_2\Phi^*) \, t_R + \text{h.c.} \, ,
\end{equation}
where $y_t$ denotes the Yukawa coupling of the top quark,
$Q^3_{L} = (t_L, b_L)^T$ is the third generation quark doublet, and 
$t_R$ is the right-handed top quark 
singlet state.
In this paper, we focus on a specific subclass of
the cxSM in which a global U(1)  symmetry is imposed
on the gauge singlet~$S$~\cite{Barger:2008jx}.
Then, the scalar part
of the Lagrangian is given by
\begin{equation}
	\mathcal{L}_{\Phi S} = D_\mu \Phi^\dagger D^\mu \Phi - \mu_h^2 \Phi^\dagger \Phi - \lambda_h (\Phi^\dagger \Phi)^2 
	+ |\partial^\mu S|^2 - \frac{1}{2} \mu_s^2 |S|^2 - \frac{1}{4}\lambda_s |S|^4 
	- \frac{1}{2} \lambda_{hs} |S|^2 \Phi^\dagger \Phi \, ,
\end{equation}
with the covariant derivative
\begin{equation}
	D_\mu = \partial_\mu - \frac{i}{2} g \tau^a W_\mu^a + i\frac{Y}{2} g' B_\mu \, ,
\end{equation}
and $g'$ and $g$ are the couplings of the
$\text{U}(1)_Y$ and $\text{SU(2)}_L$ gauge groups,
respectively.
If $S$ does not have a vev,
the global U(1)  symmetry is conserved.

\begin{figure}[ht]
    \centering
	\begin{tikzpicture}[scale=1.8,very thick,decoration={markings, mark=at position 0.5 with {\arrow{Stealth}}}] 
		\draw[->,thin] (0,0) -- (2,0) node[right]{$\phi_h$};
		\draw[->,thin] (0,0) -- (0,2) node[above]{$\phi_s$};
		\node (c1) at (0,1.41) [draw, circle, fill, label=left:{$ (0,v_s(T))$}, inner sep=2pt] {};
		\node (c2) at (1.41,0) [draw, circle, fill, label=below:{$(v_h(T),0)$}, inner sep=2pt] {};
		\draw[postaction={decorate}] (c1) -- (c2);
	\end{tikzpicture}
    \caption{The pattern of a possible strong FOEWPT in the cxSM model explored in this work.
    	The arrow points in the direction of decreasing temperature.}
    	    \label{fig:phasestructure}
\end{figure}
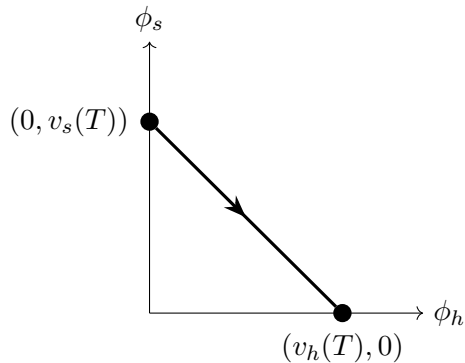

The tree-level scalar potential
as a function of the background fields $\phi_h$ and $\phi_s$ reads
\begin{equation}
\label{eq:vtree}
    V_0(\phi_h,\phi_s) = \frac{\mu_h^2}{2}\phi_h^2+\frac{\mu_s^2}{4}\phi_s^2+\frac{\lambda_h}{4}\phi_h^4 +\frac{\lambda_{hs} }{8} \phi_h^2\phi_s^2 + \frac{\lambda_s}{16}\phi_s^4.
\end{equation}
This also serves as the 
lowest-order
version of the effective potential at zero temperature.
In \cref{sec:effpot}, we will describe how the effective potential is modified by zero-temperature and thermal loop corrections, leading to a 
temperature dependence of the vacuum structure.
It is known that, unlike in the SM, 
as a consequence of the additional singlet
the phase evolution of the cxSM can exhibit a
SFOEWPT in certain regions of parameter space. 
In our study, we ensure that the tree-level potential is bounded from below by choosing $\lambda_h>0$, $\lambda_s>0$ and $\lambda_{hs}>-2\sqrt{\lambda_h\lambda_s}$.
In order to accommodate a SFOEWPT, 
we concentrate on positive values for the portal coupling,  $\lambda_{hs}>0$~\cite{Chiang:2017nmu}.
Moreover, we demand that only the
Higgs field $h$ 
has a non-zero vacuum expectation value at zero temperature:
\begin{equation}
\label{eq:vevdef}
v_h(T = 0) = v_{\rm EW} \approx 246.2~\mathrm{GeV} \, , \quad
v_s(T = 0) = 0 \, .
\end{equation}
Allowing for a non-zero singlet vev
$ \phi^{\min}_s  = v_s(T)$
at finite temperature, the cxSM exhibits a 
FOEWPT of the form $(0,v_s(T)) \to (v_h(T),0)$,
where $\phi^{\min}_h = v_h(T)$
is the vev of the Higgs field
(visualized in \cref{fig:phasestructure}).\footnote{
	We find that direct minimization of the loop-improved effective potential can result in a small unphysical non-zero Higgs vev in the high-temperature phase.
	This is further discussed in \cref{sec:thermo}.
	} 

In total, the cxSM has
six scalar degrees of freedom: 
a CP-even scalar~$h$ from the real part of the neutral
component of~$\Phi$ that can be identified with
the detected Higgs boson at 125\,GeV,
three would-be Goldstone bosons $G_{0, \pm}$
from the CP-odd and the charged components
of~$\Phi$,
and two scalar degrees of freedom $s$ and $A$
from the real and imaginary parts of the
singlet field $S$, respectively.
As indicated above,
we consider the case where at zero temperature only
the Higgs field $h$ has a vev, whereas the vev
of the singlet field vanishes.
At
finite temperature
we study FOEWPTs from a U(1)-symmetry
breaking phase in which
the singlet field has a non-zero vev~$v_s$,
into an EW symmetry breaking phase with
a non-zero Higgs vev$~v_h$.
We therefore consider the
two background fields $\phi_h$ and $\phi_s$:
if both $\phi_h \neq 0$ and $\phi_s \neq 0$,
the two CP-even fields $h$ and $s$ mix
with each other, whereas the would-be Goldstone fields
$G_0$ and $G^\pm$ and the 
other
component of
the singlet field~$A$ are mass eigenstates.
The field-dependent mass matrix of the
CP-even states $h$ and $s$ is given by
\begin{equation}
\label{eq:scalarmixmatrix}
M^2_{hs} = \left(
    \begin{matrix}
        \mu_h^2+3\lambda_h \phi_h^2 +\frac{\lambda_{hs} \phi_s^2}{4}  & \frac{\lambda_{hs} \phi_h \phi_s}{2} \\
        \frac{\lambda_{hs} \phi_h \phi_s}{2} & \frac{\mu_s^2}{2}+\frac{3 \lambda_s \phi_s^2}{4} +\frac{\lambda_{hs} \phi_h^2}{4}
\end{matrix}
\right).
\end{equation}
The field-dependent mass of~$A$ is given by
\begin{equation}
\label{eq:Amassvalue}
    M^2_A = \frac{\mu_s^2}{2}+\frac{\lambda_s \phi_s^2}{4} +\frac{\lambda_{hs} \phi_h^2}{4} \, ,
\end{equation}
and for the would-be Goldstone bosons in the Landau gauge, we have
\begin{equation}
    M^2_{G_{0,\pm}} = \mu_h^2+\lambda_h \phi_h^2 +\frac{\lambda_{hs} \phi_s^2}{4} \, .
\end{equation}
Here, it should be noted that the Goldstone boson
masses 
also receive
contributions from the gauge-fixing,
which we do not display here.
We discuss the mass eigenstates of the would-be Goldstone 
and gauge bosons 
in different
gauges in detail in \cref{app:masseigenstates}.

The cxSM has three free parameters in addition to the SM ones.
We choose to treat the singlet mass $m_s$, the singlet quartic
coupling $\lambda_s$ and the portal coupling $\lambda_{hs}$
as free parameters.
For the SM sector, we use the Higgs boson mass $m_h$,
the top quark mass $m_t$, the gauge boson masses $m_W$ and
$m_Z$, and the Fermi constant $G_F$ as input parameters.
Their numerical values are given in \cref{app:parameters},
where we also show the tree-level relations between 
those
input parameters and the Lagrangian parameters.
We note that in our notation, a lowercase $m_p$ refers
to the physical mass of the particle $p$, 
whereas an uppercase $M_p$ refers to the field-dependent tree-level mass parameter.

The 
phenomenology of the
cxSM 
at zero temperature
differs from the SM by the presence of
the degenerate states~$s$ and~$A$, which are stable
due to
the unbroken U(1) symmetry at zero temperature.
In principle, they can be dark matter
candidates, but dark matter direct detection experiments
and searches for the invisible decay of the 125\,GeV
Higgs boson (which are possible in the cxSM if the decays
$h \to ss$ and $h \to AA$ are kinematically
allowed) exclude this possibility
except for the $s$-channel resonance region
with dark matter masses of $m_s \approx
62.5$\,GeV~\cite{Chiang:2017nmu,Biekotter:2022ckj} or for
heavy masses of $m_s \gtrsim 2$~TeV~\cite{Arcadi:2024wwg}.\footnote{The
constraints from direct detection experiments can
be avoided by considering a softly-broken U(1)-symmetry
in which the state~$A$ acts as
pseudo-Nambu-Goldstone dark matter candidate
in the presence of a non-zero 
vev~$v_s$~\cite{Gross:2017dan}.
However, this version of the cxSM is not able
to realize a strong FOEWPT~\cite{Kannike:2019wsn}.}
In our discussion, focusing on the description of
a strong FOEWPT in the early universe, we do not treat the
singlet states as possible dark matter candidates.
Consequently, we do not limit the discussion to the
parameter regions that would be consistent with
experimental limits from (among others)
direct detection experiments. 
We thus assume that the states $s$ and $A$ are not
stable on cosmological timescales, either due to
tiny $\mathbbm{Z}_2$-breaking couplings facilitating
the decays of $s$ and $A$ into SM particles,
or due to the presence of an extended dark sector
with particles lighter than $s$ and $A$ 
into which
the latter can decay.

In such a situation, and assuming $m_s > 62.5$\,GeV such
that the presence of $s$ and $A$ does not give rise
to exotic or invisible decays of the 125\,GeV Higgs
boson $h$, the cxSM is very challenging to 
experimentally
distinguish from the SM. The possibility of realizing a strong FOEWPT
in this model has consequently
been called a \textit{nightmare scenario}~\cite{Curtin:2014jma},
especially since
the 
modifications of the Higgs sector giving rise to the 
phase transition would not lead to
detectable traces at the LHC~\cite{Ashoorioon:2009nf}.
However, in the upcoming decade, the LISA experiment
is expected to be launched. LISA is sensitive to
GWs in the frequency band that coincides with the
peak frequencies of primordial GW backgrounds
produced during an EW phase transition.
Hence, LISA will offer a new and potentially unique
window to test the EW phase transition in the
U(1)-symmetric cxSM. It is therefore vital to
provide accurate predictions for the parameters
characterizing the transition with a precision that
makes it possible to determine the parameter space
regions that predict potentially detectable GW signals.

\section{Effective action and effective potential}
\label{sec:effpot}

For the study of the vacuum structure 
and possible phase transitions 
we use the one-particle-irreducible (1PI) effective action formalism~\cite{Jona-Lasinio:1964zvf} which we will briefly describe here.
We define the effective action $\Gamma[\phi]$ as the Legendre transform of the generating functional of connected Green's functions, $W[J]$, with respect to the source $J(x)$ that couples to the background field $\phi(x)$ 
\begin{equation}
\label{eq:effaction}
\Gamma[\phi] = W[J] - \int d^4x J(x) \phi(x).
\end{equation}
In this 
section $\phi$ represents all the fields within the theory.
The effective action as defined in \cref{eq:effaction} is the generating functional for 1PI correlation functions.

Importantly, the true vacuum of the theory (with $J = 0$) satisfies
\begin{equation}
    \frac{\delta \Gamma[\phi(x)]}{\delta \phi (x)} = 0,
\end{equation}
allowing us to determine the vacuum structure by simply extremizing the effective action.
$\Gamma [\phi]$
can be expanded in powers of derivatives of $\phi(x)$ as
\begin{equation}
\label{eq:effaction_expand}
    \Gamma[\phi] = \int d^4x \left[ -V(\phi) + \frac{1}{2} Z(\phi) (\partial_\mu \phi \,\partial^\mu \phi) + \cdots \right] \, ,
\end{equation}
where $V(\phi)$ is the effective potential and $Z(\phi)$ is a field renormalization constant. 
Under the assumption of homogeneity of the background field $\phi$, the effective potential becomes a density of the effective action;
simply put, the effective potential encodes the quantum corrections to the classical potential
(as defined in \cref{eq:vtree} for the cxSM).

We evaluate the effective potential and
the effective action in a perturbative expansion
$V(\phi) = V_0(\phi) + V_1(\phi) + V_2(\phi) + \dots$
within the so-called background field method~\cite{Coleman:1973jx}.
The tree-level effective potential $V_0$ coincides with the classical
scalar potential of the theory.
The one loop effective potential $V_1(\phi)$ reads~\cite{Jackiw:1974cv}
\begin{equation}
    V_1(\phi) = - \frac{i}{2} \sum_{i \in \text{fields}} \eta_i \int_p \log \det i G^{-1}_i(p,\phi),
\end{equation}
where $G^{-1}_i(p,\phi)$ are the $\phi$-dependent inverse propagators of the quantum fields in the theory, and $p$ denotes their momentum.
The normalization factors are $\eta_i = 1$ for bosons and $\eta_i = -2$ for fermions and ghosts.
This expression can be written as
\begin{equation}
\label{eq:effpot1loop}
    V_1 = \sum_{i\in V,S,f,c} n_i \, I(M_i^2) \, ,
\end{equation}
in which we sum over the index $i$ labeling the
background-field-dependent mass eigenstates $M_i$ contained in the theory,
arising from the determinant of the inverse propagator.
The factors $n_i$ are $n_i = d-1$ for
gauge fields (V), $n_i = -4N_c$ for Dirac fermions (f), $n_i = -2$ for ghosts (c),
and $n_i = 1$ for scalar fields (S);\footnote{
We note that (depending on the gauge) the factors $n_i$ do not generally correspond to the number of physical degrees of freedom.
}
here, $d$ is the number of space-time dimensions.
The loop integrals $I(M_i^2)$ are divergent in $d=4$, but can be renormalized in the \msbar-scheme.
Then, the integral takes the form
\begin{equation}
\label{eq:CWpotential}
     I^{4D}(M_i) = \left.\frac{1}{2}\int_p \log(p^2+M_i^2) \right|_{\msbar} = \frac{1}{64\pi^2}M_i^4\left(\log\frac{M_i^2}{\mu^2}-c_i\right) \, ,
\end{equation}
where the superscript $4D$ refers to the four-dimensional approach used here.
The constants $c_i$ are $c_i = 5/6$ for transverse and longitudinal polarizations of vector bosons and $c_i = 3/2$ for all other degrees of freedom \cite{Coleman:1973jx}.
In $d=4$ dimensions, the \msbar renormalized
one-loop (bubble) integral $I^{4D}$
explicitly depends on the renormalization scale $\mu$.
The one-loop part $V_1$ of the effective potential
expressed in terms of $I^{4D}$ given in \cref{eq:CWpotential}
is known
as the Coleman-Weinberg potential.

The one-loop effective potential in the
dimensionally reduced EFT, 
in $d = 3$ Euclidean space, reduces to the same form 
as in
\cref{eq:effpot1loop}, with
\begin{equation}
	\label{eq:I3D}
    I^{3D}(M_i) = - \frac{M_i^{3/2}}{12\pi}.
\end{equation}
Accordingly, the effective potential in $3D$ does not explicitly depend on the renormalization scale at the one-loop level.
However, it does show an explicit $\mu_{3D}$-dependence
starting from two loops~\cite{Farakos:1994kx}, see
the discussion in \cref{sec:DR}.
The benefit of evaluating the effective potential in 3D
compared to the effective potential in 4D
will become apparent when considering temperature corrections to the effective potential in~\cref{sec:temp} and 
the related
scale uncertainties.

\begin{figure}
    \centering
    \includegraphics[width=0.6\linewidth]{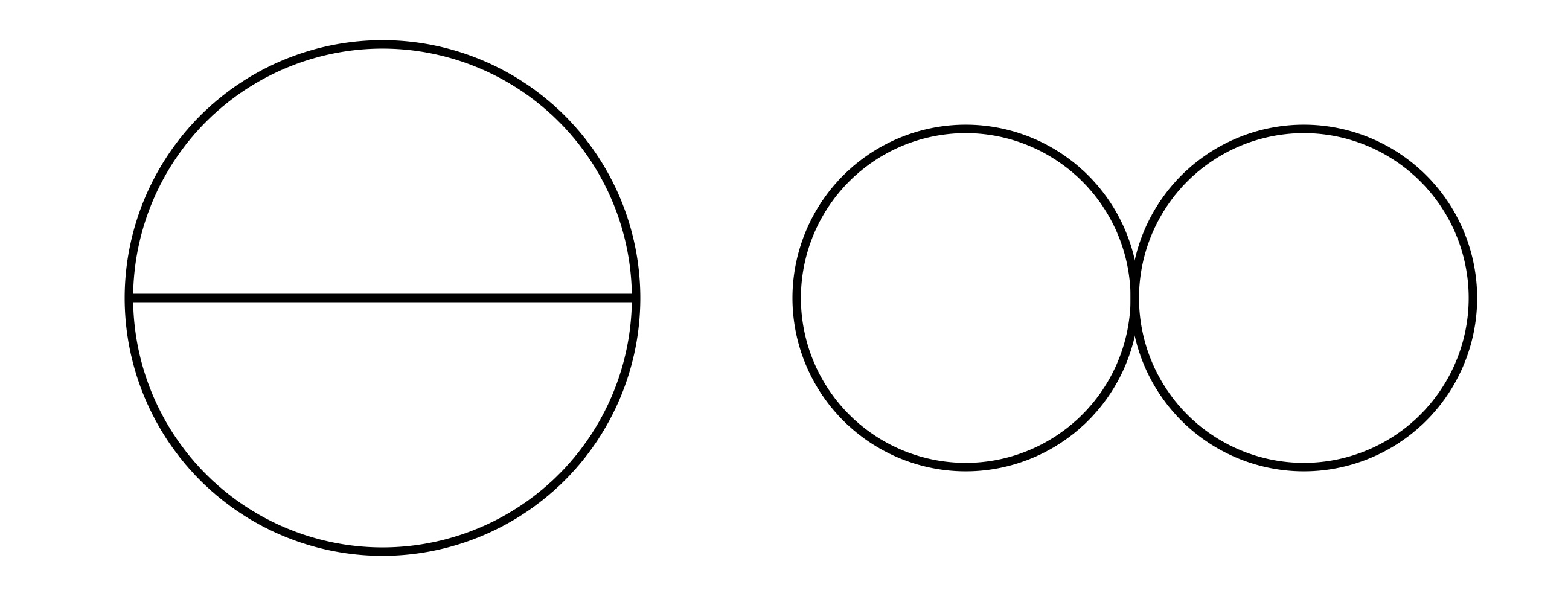}
    \caption{Topologies of prototype diagrams with background field dependent propagators entering the effective potential calculation at the two-loop level.}
    \label{fig:twoloop}
\end{figure}

The higher-loop effective potential requires the evaluation of vacuum diagrams with dressed propagators (\textit{prototype} graphs~\cite{Coleman:1973jx}).
For instance, at two-loop order, the effective potential receives contributions both from factorizable figure-8 (double-bubble) diagrams and genuine two-loop irreducible diagrams, as shown in~\cref{fig:twoloop}.
The resulting two-loop piece of the potential
can schematically be written as:
\begin{equation}
\label{eq:twolooppotential}
    V_2 = V_{VVV}+V_{VVS}+V_{VSS}+V_{SSS}+V_{ffV}+V_{ffS}+V_{ccV}+V_{ccS}+V_{VV}+V_{SS}+V_{VS} \, .
\end{equation}
Here, the terms with three indices represent
contributions from genuine two-loop diagrams, whereas
the terms with two indices represent contributions
from figure-8 diagrams, and the indices
($V$, $S$, $f$, $c$) denote
the particle type inserted into the propagators
of the diagrams.
The explicit form of the two- and three-loop
effective potential in terms of
master integrals has been computed in $4D$ in
Refs.~\cite{Martin:2017lqn,Martin:2018emo}
using the background field method.
We follow this method to evaluate the two-loop effective potential in the dimensionally reduced $3D$ theory in $\rxi$-gauge, see \cref{sec:DR} for details.
In three dimensions, the evaluation of the two-loop potential is more straightforward compared to four dimensions 
since the two-loop integrals 
evaluate
to simple logarithmic expressions.

In the following, we investigate the dependence on a gauge choice and the renormalization scale dependence of the effective potential in more detail.
It is worth noting that many of these results can be generalized to the effective action, which plays a crucial role in the study of non-equilibrium thermodynamics. 

\subsection{Gauge dependence} 
\label{sec:gauge}

The effective action shown in \cref{eq:effaction}
is defined for a non-zero source term, which violates
the Ward-Takahashi identities \cite{Fukuda:1975di}.
As a consequence, 
the effective action and thus also the
effective potential are intrinsically
gauge-dependent quantities.

The gauge dependence of the effective potential is governed by the Nielsen identity~\cite{Nielsen:1975fs},
\begin{equation}
\label{eq:nielsen}
    \xi \frac{\partial  }{\partial \xi} \Veff(\phi, \xi ) = C(\phi, \xi ) \frac{\partial}{\partial \phi}\Veff(\phi, \xi ),
\end{equation}
where $\xi$ represents a set of gauge-fixing parameters.
This equation follows 
from 
Ward identities
with a non-zero background current~\cite{Nielsen:1975fs,
Fukuda:1975di}. The functions $C(\phi, \xi )$ 
can be determined
order-by-order in perturbation theory
independently 
of $\Veff$ (see, for instance,
Refs.~\cite{Metaxas:1995ab,Alexander:2008hd}).

The Nielsen identity shows that the effective
potential is 
gauge-independent at its extremal
points $v$,
\begin{equation}
	\label{eq:nielsen0}
	\xi \frac{\partial  }{\partial \xi} \Veff(v, \xi ) = 0\, .
\end{equation}
The extremizing field value itself, however, is gauge-dependent (which is not 
a problem, as it is not a physical observable, and occurs as a consequence of the spontaneous breaking of the gauge symmetry), 
\begin{equation}
	\label{eq:nielsenv}
    \frac{\partial v }{\partial \xi} = C(v, \xi ).
\end{equation}
Away from the extrema, \cref{eq:nielsen} shows that the gauge dependence can be expressed in terms of field redefinitions governed by the 
functions $C(\phi, \xi )$.

Since the values of the effective potential at its stationary points are gauge-independent, the physical observables that depend solely on these values are gauge-independent as well.
For instance, the critical temperature $T_c$, which is defined as the temperature at which two local minima are degenerate (see \cref{sec:equilibrium} for a detailed discussion), is gauge-independent.
Therefore, it follows from~\cref{eq:nielsenv} that the
ratio
$\Delta v_h(T_c)/T_c$, where $\Delta v_h(T_c)$ denotes the
difference of $v_h$ between two phases (local minima)
at $T_c$, used in the context of EW baryogenesis in the baryon number preservation condition $\Delta v_h(T_c)/T_c > 1$, is gauge-dependent~\cite{Patel:2011th}.

\Cref{eq:nielsen} 
incorporates contributions to all orders
and thus needs careful treatment when truncating a calculation at finite loop-order.
Schematically, at one-loop order, we have
\begin{equation}
	\label{eq:nielsen-1-loop}
	\xi\frac{\partial}{\partial \xi}\Veff_1 =  C_1 (\phi,\xi) \frac{\partial}{\partial \phi}  \Veff_0,
\end{equation}
where we indicate the respective loop orders by the subscripts.
Notice that different extremizing field values $v$ are relevant for the l.h.s.\ and r.h.s.\  of \cref{eq:nielsen-1-loop}.
This lies at the heart of the so-called \emph{$\hbar$-expansion}~\cite{Fukuda:1975di,Patel:2011th} which we will briefly discuss here.

In order to ensure gauge-independence of physical observables, a loop expansion of both the effective potential and its extremizing field is crucial.
We write both $V(\phi)$ and $v$ as a formal expansion\footnote{We
only keep the powers of $\hbar$ explicitly in this section for clarity and leave them implicitly in the rest of the paper.}
in the loop-counting parameter $\hbar$,
\begin{align}
    V(\phi)&= V_0(\phi)+\hbar V_1(\phi)+\hbar^2 V_2(\phi)+\mathcal{O}(\hbar^3) \; ,\\
    v&= v_0+\hbar v_1+\hbar^2 v_2+\mathcal{O}(\hbar^3) \; .
\end{align}
Extremizing the potential and collecting 
contributions in terms of powers of $\hbar$ result in 
\begin{equation}
	\label{eq:hbar1}
    \begin{split}
    	0 &=  \partial_\phi V_0\big|_{v_0} + \hbar \left[\partial_\phi V_1 + v_1 \partial_\phi^2 V_0\right]\bigg|_{v_0}\\&+\hbar^2 \left[\partial_\phi V_2 + v_2 \partial_\phi^2 V_0+v_1 \partial_\phi^2 V_1+\frac{1}{2}v_1^2\partial_\phi^3 V_0\right]\Bigg|_{v_0} + \mathcal{O}(\hbar^3).
    \end{split}
\end{equation}
This defines the tree-level extremum
\begin{equation}
\partial_\phi V_0 \big|_{v_0} = 0 \, ,
\end{equation}
the one-loop extremum
\begin{equation}
v_1 = \left.-\frac{\partial_\phi V_1}{\partial_\phi^2 V_0}\right|_{v_0} \, ,
\end{equation}
and the two-loop extremum
\begin{equation}
	\label{eq:minima}
v_2 = \left.\frac{-2\partial_\phi V_2 (\partial_\phi^2 V_0)^2 +2\partial_\phi V_1 \partial_\phi^2 V_0\partial_\phi^2 V_1-(\partial_\phi V_1)^2\partial_\phi^3 V_0}{2(\partial_\phi^2 V_0)^3}\right|_{v_0} \, .
\end{equation}
The potential at the extremum is then given by
\begin{equation}
	\label{eq:hbarexp}
	V(v)=V_0(v_0)
	+\hbar V_1(v_0)
	+\hbar^2 \left(V_2(v_0) -\frac{1}{2}\left.\frac{(\partial_\phi V_1)^2}{\partial_\phi^2 V_0}\right|_{v_0} \right) +\mathcal{O}(\hbar^3),
\end{equation}
where each term is evaluated at the tree-level
extremum $v_0$.
In this expansion,
the dependence on the gauge-fixing parameters cancels order by order according to \cref{eq:nielsen},
whereas the naive (numerical) minimization
of the $n^{\text{th}}$-loop improved potential would lead
to an uncancelled gauge-dependence in $V(v)$.
However, in the presence of parametric enhancement of loop contributions --- for instance, in the
high-temperature regime (see~\cref{sec:temp}) 
or for radiatively generated potential
barriers~\cite{Metaxas:1995ab} --- the $\hbar$-expansion 
could lead to results that significantly deviate from the direct minimization method~\cite{Patel:2011th}.
The fact that
the potential at its
extremum given in \cref{eq:hbarexp} 
is gauge-independent order by order
(unlike the minimizing field value)
has the noteworthy 
consequence that problematic behaviors
of specific gauges---for instance, IR-divergences caused by massless Goldstone
bosons in the Landau gauge---have to cancel out
term by term in \cref{eq:hbarexp}~\cite{Martin:2014bca}.

For the cxSM, we have to minimize along two field directions
$\phi_h$ and $\phi_s$. The corresponding $\hbar$-expansion
of the potential values at the extremum then reads
\begin{align}
&V(v_h,v_s) =
  V_0(v_{h,0},v_{s,0}) +
  \hbar \, V_1(v_{h,0},v_{s,0}) +
  \hbar^2 \Bigg(V_2(v_{h,0},v_{s,0}) - \\
&\frac{1}{2} \frac{
  (\partial_{\phi_s}^2 V_0) \left(
  \partial_{\phi_h} V_1\right)^2
  + (\partial_{\phi_h}^2 V_0)
  \left(\partial_{\phi_s} V_1\right)^2
  - 2(\partial_{\phi_h} \partial_{\phi_s} V_0)
  \left(\partial_{\phi_h} V_1\right)
  \left(\partial_{\phi_s} V_1\right)} {
  (\partial_{\phi_h}^2 V_0)(\partial_{\phi_s}^2 V_0) -
  (\partial_{\phi_h} \partial_{\phi_s} V_0)^2 }
  \Bigg|_{v_{h,0},v_{s,0}} \Bigg) +
  \mathcal{O}(\hbar^3) \, , \notag
\end{align}
where all quantities are evaluated at the tree-level
minimum $\phi_h = v_{h,0}$ and $\phi_s= v_{s,0}$.\footnote{However, it is not necessary to use the $\hbar$-expansion along the singlet field direction, because it is not charged under any gauge group and is therefore not affected by the gauge fixing.}

\subsection{Renormalization scale dependence}
\label{sec:scale_variation}

When working in a perturbative expansion, due to the truncation of the perturbative series at some loop order, one inevitably introduces a dependence of the effective potential on 
either the choices made for on-shell-type renormalizations or on
the renormalization scale $\mu$.
Focusing here on the latter case, there is both an explicit (logarithmic) dependence
on the scale~$\mu$ and an implicit scale dependence of the couplings and the fields.
The full $\mu$-dependence of the perturbatively evaluated effective potential is governed by the renormalization group equation\footnote{Note that this is just expressing the fact that $\dd \Veff/ \dd \mu = 0$ 
if all loop orders are taken into account.
}
\begin{equation}
	\label{eq:rge}
	\left(\mu \frac{\partial }{\partial \mu} - \gamma \phi \frac{\partial}{\partial \phi} + \beta_i \frac{\partial}{\partial g_i} \right)\Veff = 0.
\end{equation}
This equation shows that the explicit scale dependence of the effective potential cancels against
the respective contributions from the scale dependence of the couplings $g_i$ and the field $\phi$, expressed in terms of the
$\beta$-functions $\beta_i$
and the anomalous dimension $\gamma$.\footnote{There is an additional scale dependence of the effective potential: it depends on the initial scale $\mu_\text{in}$ at which it is first determined before evolving it to another scale $\mu$. In our case, we have $\mu_\text{in} = m_Z$ (see \cref{app:renormalization}). 
This additional scale dependence is governed by the RGE-like equation~\cite{Andreassen:2014eha} 
$$
\mu_\text{in} \frac{\partial}{\partial \mu_\text{in}} \Veff(\phi; \mu_\text{in}, \mu) = \gamma \phi \frac{\partial}{\partial \phi}  \Veff(\phi; \mu_\text{in}, \mu).
$$
A complete study of this effect goes beyond the scope of our current work and we leave it for future studies.}

\Cref{eq:rge} holds order by order in perturbation theory.
For instance, the explicit scale dependence of the one-loop potential is canceled by the running of parameters and fields (which appears starting from one loop, i.e.\ at $ \mathcal{O}(\hbar)$) in the  tree-level potential: 
\begin{equation}
\label{eq:rge1}
\mu \frac{\partial}{\partial \mu}\Veff_1 = \left( \gamma_1 \phi \frac{\partial}{\partial \phi} - \beta_{i,1} \frac{\partial}{\partial g_i} \right)\Veff_0,
\end{equation}
where $\gamma_1$ and $\beta_{i,1}$ are the one-loop anomalous dimension and beta functions, respectively.
The inclusion of the running of couplings and fields can therefore serve as a consistency check for the loop-improved potential.
If one includes
the running of parameters also in the \emph{one-loop} effective potential, the residual, uncancelled scale dependence will be of two-loop order, \ie
\begin{equation}
\label{eq:rge2}
\left(\mu \frac{\partial }{\partial \mu} - \hbar\gamma_1 \phi \frac{\partial}{\partial \phi} + \hbar\beta_{i,1} \frac{\partial}{\partial g_i} \right)\left(\Veff_0+\hbar\Veff_1\right) = \mathcal{O}(\hbar^2),
\end{equation}
where we explicitly write out the powers of $\hbar$ to
indicate
the respective loop orders for each term.
This residual scale dependence
propagates
to the predictions for 
physical observables derived from the loop-improved potential,
providing an estimate for the size
of missing $\mu$-dependent higher-loop terms.
Comparing the $\mu$-dependence of 
the prediction for an observable at increasing loop orders can be used 
to study the behavior
of the perturbative expansion, and therefore
serves as an important indicator of its
validity when truncated at a certain loop order.

As an aside, we note that the potential may contain scale-dependent, but field-independent terms (\emph{cosmological constant} or \emph{vacuum energy} terms) whose running is not canceled by using loop-improved couplings.
Such scale dependencies can be dealt with by including the running of a vacuum energy term in~\cref{eq:rge} (see Ref.~\cite{Bando:1992np}).
Since field-independent terms in the potential are not
relevant for the vacuum structure of the theory, we do not
include the running of the vacuum energy in our numerical analysis.

\subsection{Temperature corrections}
\label{sec:temp}

In thermal field theory, the effective potential at non-zero temperature can be evaluated using the imaginary time formalism.
For instance, at one-loop order, the integral in \cref{eq:CWpotential} is modified to
\begin{equation}
\label{eq:I4dT}
    I^{4D}(M_i, T) = \frac{1}{2}\SumInt_p \log(p^2+M_i^2) =I^{4D}(M_i)^{T=0}\mp T\int_{\vec{p}} \log(1+n_{B/F}(E_{i,\vec{p}},T)),
\end{equation}
where we
split the one-loop contributions into vacuum and temperature-dependent parts, 
while $E_{i,\vec{p}}=\sqrt{M_i^2+\vec{p}^{\,2}}$
and $n_{B/F}$ refers to Bose-Einstein and Fermi-Dirac distributions for bosons and fermions, respectively.
The sum-integral is taken over Matsubara modes $\SumInt_p \equiv T \sum_{n} \int_{\vec{p}}$, and the four-momenta become $p = (\omega_n, \vec{p})$.
For bosons, we have $\omega_n = 2n\pi T$ and for fermions $\omega_n = 2(n+1/2)\pi T$.
In the second term on the right-hand side of the equation, 
the integral is evaluated in $d=3$ momentum space,
and it is commonly written in terms of thermal functions $J_{B/F}$,
\begin{equation}
    \begin{split}
        \mp T\int_{\vec{p}} \log(1+n_{B/F}(E_{\vec{p}},T))&=\mp\frac{T^4}{2\pi^2}\int_0^{\infty}dxx^2\log\left(1\pm \exp\left(-\sqrt{x^2+\frac{M_i^2}{T^2}}\right)\right)\\&\equiv \frac{T^4}{2\pi^2}J_{B/F}\left(\frac{M_i^2}{T^2}\right) \, .
    \end{split}
\end{equation}
For a phase transition to occur, the temperature-induced corrections to the effective potential must be large enough to qualitatively alter its structure.
Therefore, the leading-order thermal corrections should be parametrically of the same order as the tree-level potential.
This, in turn,  
implies that the usual loop expansion 
in general is not applicable, and a
careful power counting is required.

For instance, in Ref.~\cite{Gould:2021oba} it was shown that due to the 
differences between
loop and coupling expansions for thermal phase transitions, in order to cancel the leading order (one-loop thermal)  renormalization scale dependence at high temperatures, one has to add two-loop thermal mass contributions, so that the residual dependence is sub-leading.\footnote{Similar
issues can be present at zero temperature for a radiatively generated barrier, where the two-loop contribution is required~\cite{Andreassen:2014eha}.}
This is in contrast to the zero temperature case, 
where the leading renormalization scale dependence of the couplings is canceled by the one-loop effective potential.
Similarly, ensuring 
the cancellation of gauge-dependent contributions
poses additional challenges
at finite temperature, as the standard $\hbar$-expansion
is no longer sufficient and must be replaced by a
perturbative expansion formulated in terms
of a (thermal) coupling.

In theories with gauge interactions, leading thermal corrections arise from one-loop self-energy diagrams and scale as $\mathcal{O}(g^2 T^2)$, where $g$ is a generic gauge coupling in the theory.
Under the assumption of perturbativity, these corrections can still be significant at sufficiently high temperatures due to the quadratic dependence on $T$.
In this case, non-zero Matsubara modes are special;
their effective thermal masses are of order $\sim \pi T$ (\textit{hard} modes), while zero modes, which could acquire a thermal screening mass $m_D$, are at most $\sim gT$ (\textit{soft} modes).
This hierarchy of scales 
has important implications
for perturbative calculations,
since the vacuum diagrams of soft modes with additional petals of hard modes will all contribute at the same order, \ie $\mathcal{O}(g^3T^2)$.
Accordingly, one must
resum these diagrams in order
to go beyond the leading order of $\mathcal{O}(g^2T^2)$.

To this end, we can write the one-loop thermal effective potential integral for bosons as a sum of contributions from zero and non-zero Matsubara modes in the imaginary time formalism
\begin{equation}
    I^{4D}(M_i, T) = \frac{1}{2} T \int_{\vec{p}} \log(\vec{p}^{\,2}+M_i^2) +  \frac{1}{2}\SumInt_{p}' \log(p^2+M_i^2),
\end{equation} where the sum goes over all but zero Matsubara modes $\SumInt_p' \equiv T \sum_{n\neq0} \int_{\vec{p}}$ .
The soft modes then can be resummed as $M_i^2 \to M_i^2 + \Sigma_i^T$, where $\Sigma_i^T \sim g^2 T^2$ are the thermal self-energies of the soft modes 
that are
summed.
This leads to 
\begin{equation}
\label{eq:daisy-AE}
    \begin{split}
        I^{4D}_{\text{resummed}}(M_i, T) &= \frac{1}{2} T \int_{\vec{p}} \log(\vec{p}^2+M_i^2+\Sigma^T_i) +  \frac{1}{2}\SumInt_{p}' \log(p^2+M_i^2)\\
        &= I^{4D}(M_i, T) - \frac{T}{12\pi} \left((M_i^2+\Sigma^T_i)^{3/2}- (M_i^2)^{3/2} \right).
    \end{split}
\end{equation}
The last term is an additional contribution from the resummation, 
called a
\textit{daisy} term.
The resummation of soft modes only, as in \cref{eq:daisy-AE}, is usually coined Arnold-Espinosa (AE) resummation~\cite{Arnold:1992rz}.
An alternative method where one resums both, hard and soft modes, is the Parwani-type resummation~\cite{Parwani:1991gq}.\footnote{
Compared
to AE resummation, Parwani resummation
leads to a stronger temperature dependence of the
scalar potential~\cite{Biekotter:2021ysx,Bahl:2024ykv}
and potentially significantly larger predictions for the
strength of
the transition~\cite{Kainulainen:2019kyp,
Basler:2016obg,Bittar:2025lcr}.
By inserting the thermally corrected
masses also into the radiative corrections from non-zero Matsubara modes, 
the Parwani method
mixes IR and UV dynamics.
We checked that in the cxSM, the differences
between AE and Parwani resummation are much below
the theoretical uncertainties from residual
gauge and renormalization scale
dependence.
We will therefore only consider the AE
method in our paper.
}
Similar resummations can also be performed at higher-loop order~\cite{Laine:2017hdk,Bahl:2024ykv}.
An alternative to the diagrammatic resummations is to solve the gap equation~\cite{Boyd:1993tz}.
In our work, we determine the temperature-dependent potential at one-loop order with AE-type diagrammatic daisy resummation in various gauges (see \cref{sec:gauge}) and compare those predictions with the ones from a high-temperature EFT.

\subsection{High-temperature EFT}\label{sec:DR}

The deficiency of the usual loop expansion at high temperature is not surprising---it simply indicates a large separation of the scales at play.
At high temperatures, those are: \textit{hard} ($\sim \pi T$), \textit{soft} ($\sim m_D\sim g T$) and \textit{ultrasoft} ($\sim g^2 T$) contributions.\footnote{
	For radiatively generated potential barriers, an additional \textit{softer} (also called \textit{supersoft}) scale
	($\sim g^{3/2} T$) appears, which characterizes the nucleation scale~\cite{Gould:2021ccf}.
	In the cxSM, the barrier is not radiatively induced,
	such that the \textit{softer} scale is not relevant.
	}
These originate from non-zero Matsubara modes, temporal components of gauge fields and spatial gauge modes together with scalar fields, respectively.

Instead of performing a power counting and resummations by hand, the large separation between scales of the relevant physics can be taken into account using methods of effective field theories (EFTs).
The dynamics of thermally driven phase transitions
can be described via the so-called High-Temperature EFT~\cite{Farakos:1994kx,Kajantie:1995dw}.
It allows to systematically integrate out hard and soft modes, which include all fermionic degrees of freedom. This leaves us with the field theory for the nucleating field, where large logarithms are resummed into effective couplings of the theory.
This matching is often called \textit{dimensional reduction}, as the effective theory is three-dimensional after integrating out the hard modes.

Schematically, the matching of mass parameters ($\mu_{\phi}$), scalar self-couplings ($\lambda_{\phi}$) and gauge ($g$) couplings  at the \textit{hard} scale can be represented as
\begin{equation}
    \begin{split}
        \label{eq:matching}
            \mu_{\phi,3D}^2 &=
        \mathrlap{\color{gray}\overbrace{\phantom{\;\; \mu_{\phi}^2 + \# g^2T^2}}^{\mathcal{O}(g^2)}}
        \underbracket{\:\mu_{\phi}^2}_{\text{tree}}
        + \underbracket{\# g^2T^2 + \# g^2\mu_{\phi}^2}_{\text{1-loop}} \mathrlap{\color{gray}\kern-30pt\overbrace{ \phantom{\; \# g^2\mu_{\phi}^2+\# g^4T^2}}^{\mathcal{O}(g^4)}}
        + \underbracket{\# g^4T^2}_{\text{2-loop}} +\mathcal{O}(g^6) \, ,\\ \quad \lambda_{\phi,3D}&={\color{gray}\overbrace{\color{black}\underbracket{\; \lambda_{\phi}T}_{\text{tree}}}^{\mathcal{O}(g^2)}}+ {\color{gray}\overbrace{\color{black}\underbracket{\# g^4T}_{\text{1-loop}}}^{\mathcal{O}(g^4)}}  +  \mathcal{O}(g^6) \, ,\\ \quad g_{3D}^2&={\color{gray}\overbrace{\color{black}\underbracket{\:g^2T}_{\text{tree}}}^{\mathcal{O}(g^2)}}+ {\color{gray}\overbrace{\color{black}\underbracket{\# g^4T}_{\text{1-loop}}}^{\mathcal{O}(g^4)}}  +  \mathcal{O}(g^6) \, ,
    \end{split}
\end{equation}
where $\mu_{\phi,3D}$, $\lambda_{\phi,3D}$ and
$g_{\phi,3D}$ are the corresponding
parameters
in the dimensionally reduced $3D$ EFT.
Here, with underbraces we indicate which loop order contributes to the matching while overbraces show the respective coupling order when assuming the scaling $\mu_{\phi}^2 \sim g^2T^2$, $\lambda_{\phi} \sim g^2$, as is customary for strong thermal phase transitions~\cite{Gould:2021oba}.
Expressions of the form $\#g^n$ denote dimensionless 
factors at 
$\mathcal{O}(g^n)$, which generally depend on the scalar self-couplings and gauge charges.

\begin{figure}[h]
    \centering
    \includegraphics[width=\linewidth]{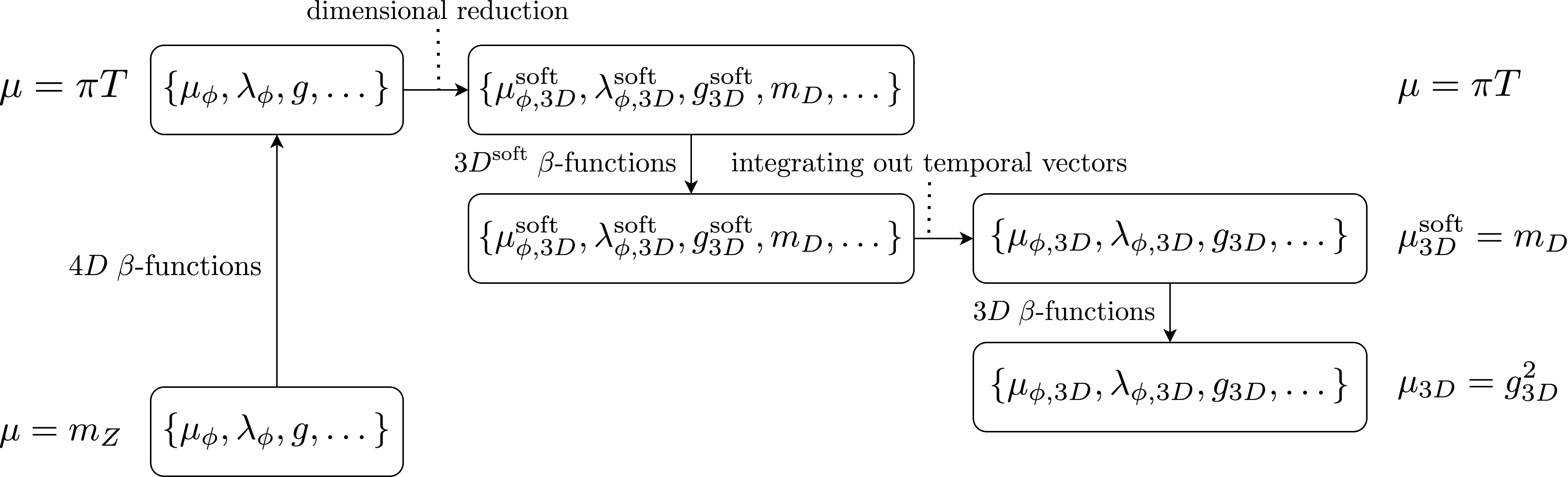}
    \caption{
    	Illustration of the procedure for systematically integrating out relevant thermal scales within the dimensional reduction approach.
    	Starting from the initial scale ($\mu\sim m_Z$), Lagrangian parameters---determined from physical input parameters (see \cref{app:parameters})---are evolved to the \textit{hard} scale ($\mu\sim\pi T$), where the full theory is matched to the dimensionally reduced EFT.
    	The effective couplings are then evolved to the \textit{soft} scale ($\mu_{3D}^{\rm soft}\sim m_D$), where temporal vector modes are subsequently integrated out, yielding the final EFT used to compute thermodynamic observables at the \textit{ultrasoft} scale, characterized by the gauge coupling ($\mu_{3D}\sim g_{3D}^2$).}
    \label{fig:dimred}
\end{figure}

The truncation of the matching at a given loop order introduces
a residual dependence on the \textit{hard}  matching scale
$\mu = \pi T$.
The relations in \cref{eq:matching} are modified when integrating out the \textit{soft} scale $\mu_{3D}^{\rm soft} = m_D$.
This gives rise to contributions with odd powers of $g$, starting with $\mathcal{O}(g^3)$.
The high-temperature EFT matching procedure has been automated in \texttt{DRalgo}~\cite{Ekstedt:2022bff} up to  $\mathcal{O}(g^4)$ for generic models. In \cref{fig:dimred} we present a schematic depiction of the matchings and subsequent running of parameters within the dimensional reduction approach as applied here to study the EW phase transition in the cxSM. 
Within the dimensionally reduced $3D$ theory, only the mass parameters $\mu_{\phi, 3D}$ run, starting at two-loop order~\cite{Niemi:2021qvp}.

For the cxSM,
the mass parameters are $\mu_{\phi} = \{\mu_h, \mu_s\}$, and the quartic couplings $\lambda_{\phi} = \{\lambda_h, \lambda_s, \lambda_{hs}\}$.
Dimensional reduction for the SM
with an additional singlet was studied extensively in the past \cite{Schicho:2021gca,Gould:2021dzl,Niemi:2021qvp}, and the matching relations 
that are relevant in this work can be found in \cite{Schicho:2022wty}. For illustration, here we explicitly 
give
the leading one-loop
matching conditions for the mass parameters,
\begin{equation}
    \begin{split}
        \label{eq:explicitmatching}
            \mu_{h,3D}^2 &= \mu^2_h + \frac{1}{48} T^2 \left(3 g'^2 + 9 g^2 + 12 y_t^2 + 2 \lambda_{hs} + 24 \lambda_h \right),\\
            \mu_{s,3D}^2 &=\mu_{s}^2 + \frac{1}{6} T^2 \left(\lambda_s + \lambda_{hs} \right) \, ,
    \end{split}
\end{equation}
where the $\mathcal{O}(T^2)$ terms coincide with the
leading finite-temperature corrections to the classical
scalar potential in the high-temperature expansion.
We explore the impact of the matching accuracy on thermodynamic observables
and the description of the EW phase transition
in the cxSM in detail
in~\cref{sec:thermo}.

The validity of the high-temperature EFT, which formally breaks down if $ m_{h,s} / (\pi T) \gtrsim 1$, can be investigated through the 
impact of
including higher-dimensional operators in the EFT
on the obtained results.
To assess the 
validity
of truncating the EFT expansion at dimension-four operators, we perform the matching of dimension-six scalar operators in the cxSM in the Landau gauge with the help of \texttt{DRalgo}~\cite{Ekstedt:2022bff}. 
Parametrically appearing at $\mathcal{O}(g^6)$ in the matching, they appear in the EFT tree-level potential as:
\begin{equation}
\label{eq:dim6oper}
    V_0^{3D} \supset c_{h^6} (\Phi^\dagger \Phi)^3 + c_{h^4 s^2}(\Phi^\dagger \Phi)^2 |S|^2 + c_{h^2 s^4} \Phi^\dagger \Phi |S|^4 + c_{s^6} |S|^6,
\end{equation}
and the Wilson coefficients read:
\begin{equation}
    \label{eq:c6}
    \begin{split} 
    c_{h^6} &= \frac{\zeta(3)}{6144 \pi^4}\left( 3 g'^6 + 9 g'^4 g^2 + 9 g'^2 g^4 + 9 g^6 + 2 \left( -336 y_t^6 + \lambda_{hs}^3 + 960 \lambda_{h}^3 \right) \right) , \\
    c_{h^4 s^2} &= \frac{\zeta(3)}{512 \pi^4} \lambda_{hs} \left( \lambda_{hs} (\lambda_s + \lambda_{hs}) + 12 \lambda_{hs} \lambda_{h} + 48 \lambda_{h}^2 \right) , \\
    c_{h^2 s^4} &= \frac{\zeta(3)}{1024 \pi^4}\lambda_{hs} \left( 5 \lambda_s^2 + 6 \lambda_s \lambda_{hs} + 2 \lambda_{hs} (\lambda_{hs} + 6 \lambda_{h}) \right), \\
    c_{s^6} &= \frac{\zeta(3)}{1536 \pi^4}\left( 7 \lambda_s^3 + \lambda_{hs}^3 \right).
    \end{split}
\end{equation}
We use this subset of dimension-six operators to assess the validity of the EFT expansion. For this purpose, we do not take into account the gauge-dependence of these matching relations~\cite{Chala:2025aiz,Bernardo:2025vkz},
further operators at this order which include derivatives~\cite{Chala:2024xll,Chala:2025oul} or corrections to the RGE running from the higher-dimensional operators~\cite{Chala:2025cya}.
The impact of the operators in \cref{eq:dim6oper} on thermodynamic observables is discussed in \cref{sec:highdim}.
It should be noted that in the dimensionally reduced theory, scalar fields have mass dimension $1/2$, hence the Wilson coefficients in \cref{eq:c6} are dimensionless. However, we refer to them as dimension-six operators, as they correspond to operators with mass dimension six in 4D.

In the $3D$ EFT resulting from dimensional reduction at high temperature, temperature is no longer a dynamical variable but enters only through the effective parameters of the theory. After matching, the effective potential can be calculated within the EFT in the usual perturbative manner.
At one-loop order, this corresponds to defining the effective potential of~\cref{eq:effpot1loop} in terms of the one-loop integrals of~\cref{eq:I3D}, evaluated in three dimensions. At the next order, one
takes into account the contributions from
the two-loop prototype diagrams (see~\cref{eq:twolooppotential,fig:twoloop}). 

\begin{figure}
    \centering
    \includegraphics[width=0.7\linewidth]{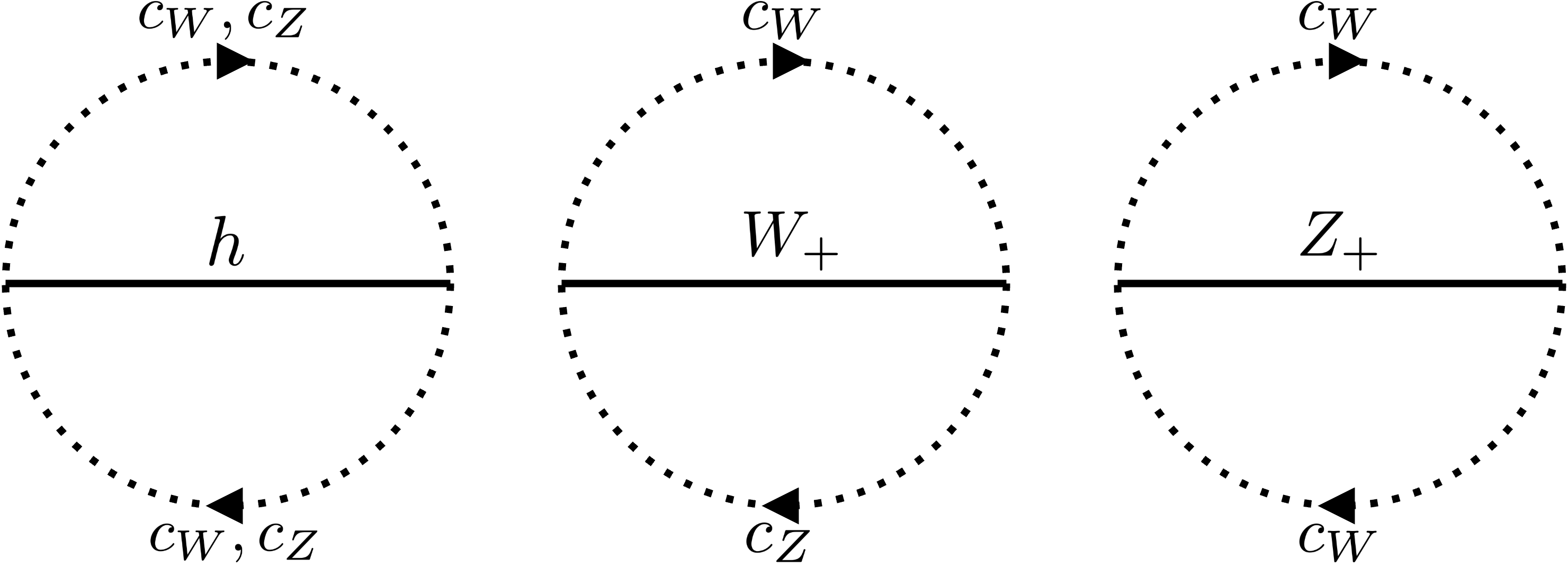}
    \caption{Two-loop ghost--ghost--scalar prototype diagrams that appear in the effective potential in $\rxi$-gauge, but are absent in the Landau gauge. See \cref{app:masseigenstates} for the gauge-dependent mass eigenvalues used in their evaluation.}
    \label{fig:ccStwoloop}
\end{figure}

In the present work, we calculate the effective potential within the dimensionally reduced theory in $\rxi$-gauge up to two-loop order.
Compared to the algorithmically generated effective potential in \texttt{DRalgo}~\cite{Ekstedt:2022bff}, which is
computed in the Landau gauge, the two-loop potential in $\rxi$-gauge contains additional diagrams, namely ghost--ghost--scalar vacuum prototype graphs (see \cref{fig:ccStwoloop}), and the mass eigenvalues of vectors, Goldstone scalars and ghosts are altered (see \cref{app:masseigenstates} for details).
After incorporating the contributions from these
additional diagrams and the $\rxi$-dependent masses, we
explicitly checked that the expression for the potential evaluated in $\rxi$-gauge becomes 
gauge-independent
in the $\hbar$-expansion at
the tree-level minima and that it reproduces the potential in the Landau gauge in the $\xi\to 0$ limit.
This serves as a highly non-trivial check of the newly introduced contributions to the potential and the inherent intricate gauge cancellations at the two-loop level.

As mentioned previously, the explicit dependence on the EFT renormalization scale $\mu_{3D}$ appears in the effective potential at the two-loop level. We explore the dependence of thermodynamic observables on $\mu_{3D}$ in \cref{sec:equilibrium}
in order to 
assess
the related theoretical uncertainties
in the description of the EW phase transition.
We compare the uncertainty related to the dependence on
the renormalization scale 
with
the one resulting from
residual gauge dependence (if the $\hbar$-expansion is
not applied).

\section{Thermodynamics of the First Order Phase Transition}
\label{sec:thermo}

In this section, we
numerically investigate
the effects of thermal corrections on phase transition dynamics in a comparison of various possible approaches to the problem. 
As discussed in \cref{sec:cxsm},
we focus on the case of
the flip-flop scenario in the cxSM.
We study parameter regions where the
EW phase transition can be of first order and would potentially be strong enough to produce detectable
primordial stochastic GW backgrounds.
The thermodynamics of a FOEWPT
can be formally split into two parts: equilibrium thermodynamics, which is characterized by the critical temperature and latent heat, and nucleation thermodynamics, which describes the nucleation of the new phase.

\subsection{Critical temperature and latent heat}
\label{sec:equilibrium}

In studies of phase transition thermodynamics, one of the central observables is the critical temperature, $T_c$.
It is defined as the temperature at which the pressures --- given by $p = -V^{4D}$ in each minimum in the $d = 4$ theory and
by $p = -T V^{3D}$ in the dimensionally reduced theory, see \cref{sec:DR},
where $V^{4D}$ and $V^{3D}$ are the potential values at the minima
--- at two coexisting local minima are equal.

To characterize the strength of the phase transition,
we compute the latent heat at the critical temperature,
$L_c$, which quantifies the difference in energy density,
$\rho = T (\partial p/\partial T) -p$,  between two phases:
\begin{equation}
    L_c \equiv L(T_c) = \Delta\rho\big|_{T=T_c}= T_c \frac{d\Delta p}{dT} \bigg|_{T=T_c} = -T_c \frac{d\Delta \Veff}{dT} \bigg|_{T=T_c}.
    \label{eq:Lc}
\end{equation}
Here, $\Delta V=-\Delta p$ is the difference in the effective potential between high and low temperature phases. In our notation, this corresponds to $\Delta V=\Delta V^{4D}$ or $\Delta V=T\Delta V^{3D}$ in the $3D$ EFT approach.
For a FOEWPT,
$L_c$ represents the discontinuous release of energy as the system moves from a metastable phase to a stable broken symmetry phase.
A large latent heat corresponds to a strong FOEWPT,
which is essential for scenarios like EW baryogenesis, where a significant departure from thermal equilibrium is necessary to generate and maintain the baryon asymmetry.
Additionally, sufficiently
strong FOEWPTs
can produce GWs (see \cref{sec:gravwave}) with peak amplitudes in reach
of future space-based GW 
observatories.

In this section, we will discuss the scale and gauge dependence of $T_c$ and $L_c$ for the EW phase transition
predicted in the cxSM using the following set of approaches:
\begin{enumerate}
	\item[(a)] the {\boldmath$4D$} \textbf{approach} with the effective potential at one-loop order, including the AE-type daisy resummation of~\cref{eq:daisy-AE} (unless stated otherwise) and two different sets of relations between physical input and Lagrangian parameters at the one-loop level: \msbar- and OS-like scheme (denoted \os), see \cref{app:renormalization} for details. 
    We evaluate the effective potential as
    \begin{align}
    \label{eq:Veff4d}
        V^{4D}(\phi_h,\phi_s,T;\mu,\xi) &= V_0(\phi_h,\phi_s)+V^{T=0}_1(\phi_h,\phi_s;\mu,\xi) \notag \\
        &+V^{T\neq0}_1(\phi_h,\phi_s,T;\xi)+V_\text{daisy}(\phi_h,\phi_s,T;\xi),
    \end{align}
    where we construct the one-loop temperature-dependent potential using \cref{eq:effpot1loop,eq:I4dT} and add the contribution from daisy resummed diagrams as in \cref{eq:daisy-AE}.
    For the latter, we only 
    use
    the leading-order thermal screening masses.
    In the \msbar-scheme, we use one-loop improved Lagrange parameters (\cref{eq:one-loop-rels}) in each part of the effective potential~\cref{eq:Veff4d}.
    In the \os-scheme, a UV-finite counterterm piece $V_{\rm CT}$ is added to the potential, which shifts the parameters such that the first and second derivatives of the potential $V^{4D}$
    in the tree-level minimum $(\phi_h,\phi_s) = (v_h,0)$ are
    the same as the ones of the tree-level potential $V_0$.
    This ensures that the scalar vevs and masses derived from $V^{4D}$
    remain at their tree-level values within
    the \os prescription based on the effective
    potential, which, however, does not account for
    momentum-dependent radiative corrections
     (see \cref{eq:OS-conditions} for details). 
	\item[(b)] the {\boldmath$3D$} \textbf{approach} based on the high-temperature EFT using dimensional reduction (see \cref{sec:DR} for details) and \msbar-relations.\footnote{
        To apply the \os-scheme in the $4D$ approach,
        one makes use of the fact that the effective potential
        can be separated into a $T = 0$ part
        (first line on the right-hand side
        of \cref{eq:Veff4d})
        and a part that
        explicitly depends on the temperature
        (second line of \cref{eq:Veff4d}).
        The counterterm potential is then defined using only the
        $T = 0$ part of the potential.
        In the $3D$ EFT approach, the explicit temperature 
        dependence enters into all terms of the potential
        from the matching conditions of the
        effective potential parameters, such that one
        cannot split the potential into a $T=0$ part
        and a part that explicitly depends on~$T$.
        We therefore do not apply an \os-scheme
        in the $3D$ approach.}
	In this case, we evaluate the effective potential up to two-loop order within the effective theory
    \begin{equation}
    \label{eq:Veff3d}
        V^{3D}(\phi_h,\phi_s\;\mu_{3D},\xi) = V^{3D}_0(\phi_h,\phi_s)+V^{3D}_1(\phi_h,\phi_s;\xi)+V^{3D}_2(\phi_h,\phi_s;\mu_{3D},\xi).
    \end{equation}
    The temperature dependence of the potential arises
    implicitly through 
    the effective couplings
    as shown in \cref{eq:matching}.
    The dependence on the \textit{hard} matching scale
    $\mu$ (see \cref{fig:dimred} for a schematic depiction of the matching procedure) also arises implicitly through the effective couplings of the theory, whereas the dependence on
    the \textit{ultrasoft} scale $\mu_{3D}$ is explicit in $V_2^{3D}$.
    Unless specified otherwise, in the $3D$ approach
    we apply $\mathcal{O}(g^4)$ matching and integrate
    out \textit{hard} and \textit{soft} scales. 
\end{enumerate}

In this section, we use a
reference parameter point with the singlet mass fixed
to $m_s = 100\,\text{GeV}$ and the singlet self-coupling
fixed to $\lambda_s=1$, while varying the remaining 
free parameter of the cxSM in the range
$\lambda_{hs}\in [1,1.3]$.
In this parameter range, we find a thermodynamic behavior
compatible with strong FOEWPTs.
For larger values of the portal coupling $\lambda_{hs}$
(with the other parameters kept fixed),
we find that the low-temperature global minimum is not the EW one,
preventing the onset of an EW
phase transition during the evolution of the universe.
On the contrary,
for smaller values of $\lambda_{hs}$,
there is no strong first-order EW phase transition.
We further analyze parameter planes in which $\lambda_s$ is
kept fixed and both $m_s$ and $\lambda_{hs}$ are varied
in \cref{sec:gravwave},
including a phenomenological discussion of the resulting
GW signals and the experimental prospects for their
detection at future experiments.
Throughout the numerical analysis, we 
vary $\lambda_{hs}$ and keep $\lambda_s = 1$ fixed
to streamline the discussion.
In general, larger values of $\lambda_s$ with $m_s$
and $\lambda_{hs}$ kept fixed reduce the strength
of the transition, whereas smaller values make the
transition stronger (irrespective of the perturbative
approach used in the analysis).

In the remainder of this section,
unless specified otherwise (\textit{i.e.}\ when using the $\hbar$-expansion), we directly minimize the real part of the effective potential to 
determine the relevant thermodynamic quantities.
\footnote{The imaginary part of the effective potential is related to the decay rate of the homogeneous state~\cite{Weinberg:1987vp}.}

\subsubsection{Scale dependence}
\label{sec:scale_dependence}

\begin{figure}[t]
\centering
\includegraphics[width=0.48\textwidth]{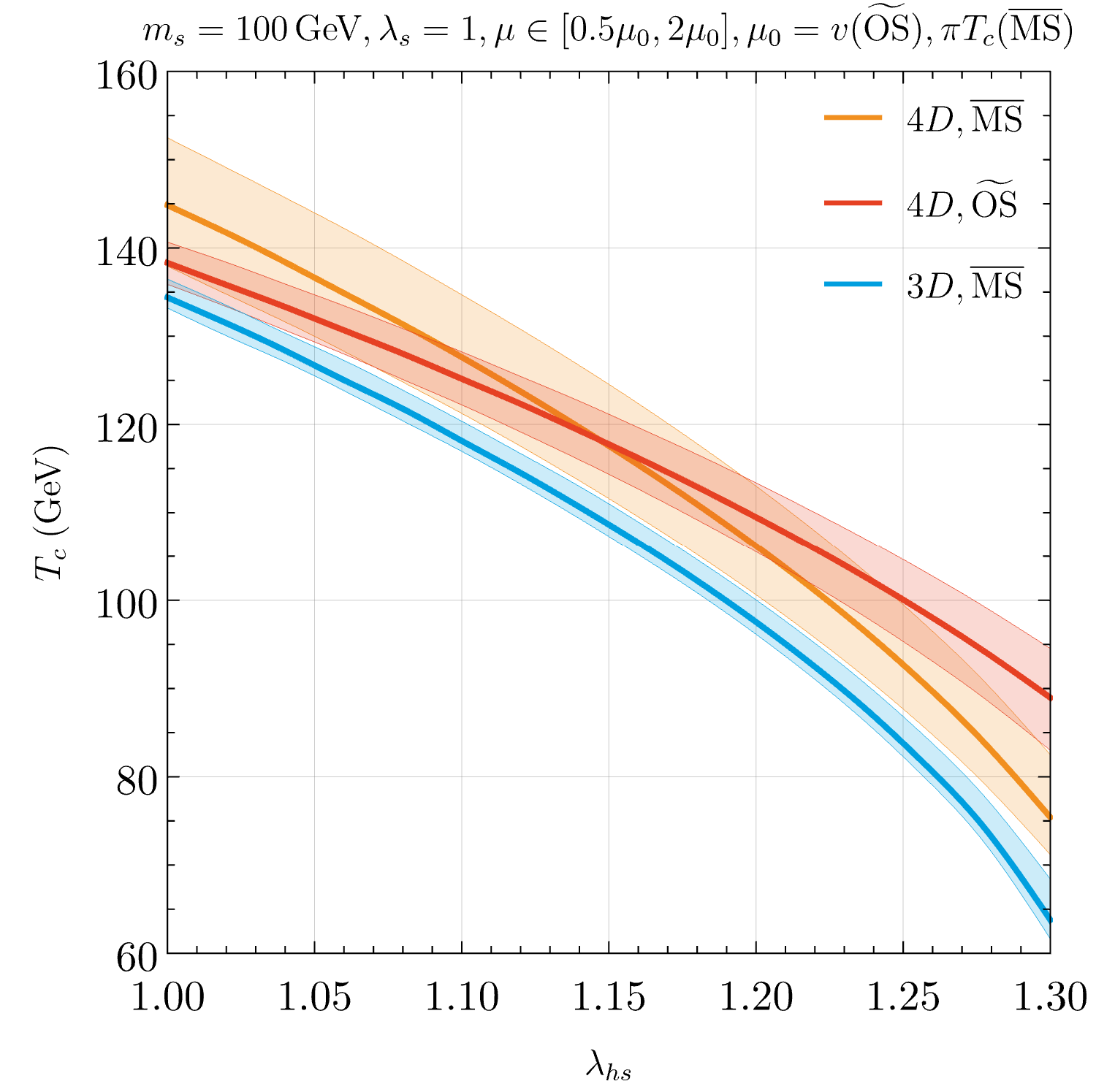}
~
\includegraphics[width=0.48\textwidth]{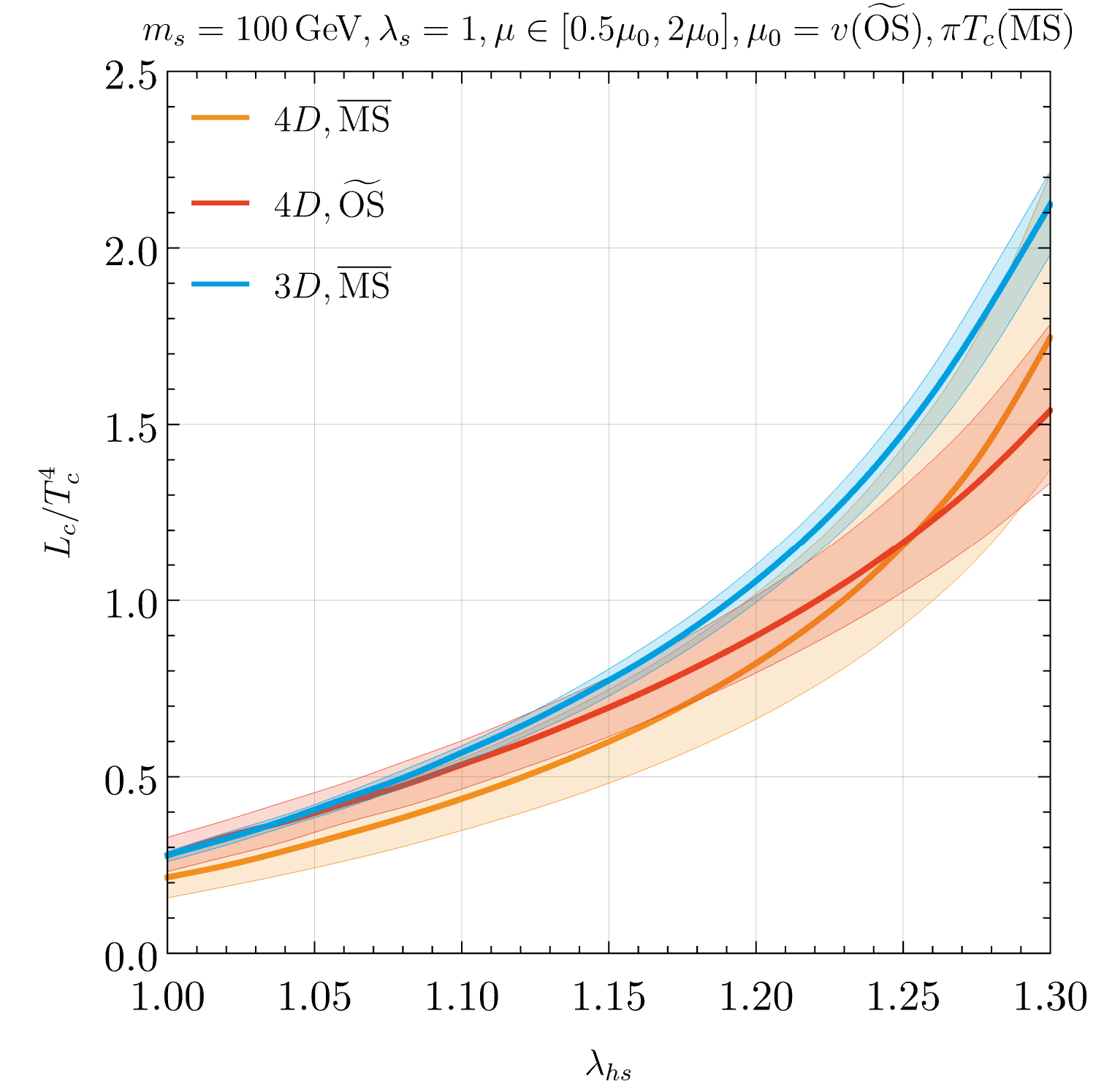}
\caption{
Critical temperature (\textit{left}) and latent heat (\textit{right}) for $m_s=100\,\text{GeV}$, $\lambda_s=1$, $\lambda_{hs}\in[1,1.3]$ determined by direct potential minimization in the Landau gauge.
For the $4D$ approaches,
the shaded bands around the solid lines are obtained
from a variation of the renormalization scale by a factor
of 2,
with the central value being fixed at
$\mu = \pi T$ and $\mu = v$ for \msbar and \os renormalization schemes, respectively
(see text for details).
For the $3D$ approach, the shaded band
results from a variation of the
\textit{hard} matching scale by a factor
of two around the central value $\mu = \pi T$.
}
\label{fig:TcLcScale4d3d}
\end{figure}

We first investigate the scale dependence
in the predictions for the critical temperature $T_c$
and latent heat $L_c$
using the different approaches discussed above.
In the left plot of \cref{fig:TcLcScale4d3d} we show with the
solid lines the predictions for $T_c$ as a function of $\lambda_{hs}$
using the $4D$ approach in the \msbar-scheme and the \os-scheme
with the orange and red lines, respectively,
and using the $3D$ approach
with the blue line. The colored shaded bands around these lines
result from a variation of the renormalization and matching
scale $\mu$ by a factor of 2 around the central values
$\mu_0^{\msbar} = \pi T_c$ and $\mu_0^{\os} = v$ in
the \msbar- and the \os-scheme, respectively.
These results were obtained from the effective
potential in the Landau gauge, and the determination of the degenerate minima was carried out
by numerically minimizing the loop-improved potential
(without applying the $\hbar$-expansion).\footnote{The direct numerical minimization of the two-loop effective potential can 
give rise to a
spurious logarithmic divergence originating from the triple Higgs
sunset diagram \cite{Niemi:2024vzw}. This feature is not problematic for the present investigation, since, for the case of a strong first-order phase transition, the divergence does not occur near the local minima in field space.
}
The widths of the bands illustrate the theoretical
uncertainties related to the renormalization scale
dependence.
Across the entire $\lambda_{hs}$ range,
we find that the renormalization scale dependence
is significantly reduced in the $3D$ approach
compared to the $4D$ approaches.

While the orange and the red bands representing the $4D$ approaches overlap over the whole
range of $\lambda_{hs}$ shown,
their slopes
slightly differ. The \msbar-scheme predicts smaller
values of $T_c$ at the upper end of the
$\lambda_{hs}$-range, and the \os-scheme predicts
smaller values of $T_c$ at the lower end
of the $\lambda_{hs}$-range.
This difference results from the different
origins of the scale dependence.
In the \msbar-scheme, it is determined by the
one-loop RGE running of the couplings,
while for the \os-scheme, it results from
the scale dependence of the one-loop \os\
counterterms (which does not necessarily coincide
with the RGE running).\footnote{For
the $4D$-\msbar approach, our findings agree
with the ones reported in Ref.~\cite{Chiang:2017nmu}.}
Additionally, differences between the two schemes arise due to the different choices for the central scales $\mu_0$.
The \msbar approach uses a variable central scale $\mu_0=\pi T_c$, while the \os one has its central scale fixed at $\mu_0=v_{h,0}$.
Therefore, a stronger parametric dependence for the former approach becomes visible, especially in the parameter regions where $T_c$ changes quickly.
Compared to the blue band representing the $3D$ approach, we observe that it lies below the other two bands, only showing an overlap with the other bands at the smallest values of $\lambda_{hs}$ shown.
This discrepancy between the $3D$ and the
$4D$ approaches suggests that a scale variation
with a factor of two might not be sufficient
for a robust estimation of the theory uncertainty
related to missing higher-order contributions
in the perturbative expansion, and instead
a wider scale variation or other ways to estimate the 
theoretical uncertainties
should be considered.
In addition, it should be taken into account that the $3D$ approach suffers from an additional theoretical uncertainty
from the truncation of the high-$T$ EFT at dimension four,
which is discussed in detail in \cref{sec:highdim}.

In the right plot of \cref{fig:TcLcScale4d3d}, we show the predictions for the latent heat
$L_c$ defined in \cref{eq:Lc} using the three
approaches, where the definitions of the solid lines
and the colored bands are as in the left plot
of \cref{fig:TcLcScale4d3d}.
For the latent heat, we find
a better agreement between the $4D$ and
the $3D$ approaches compared to the predictions
for the critical temperature $T_c$.
As for $T_c$, the smallest theoretical uncertainty
from the scale variation is found using the
dimensionally reduced EFT.
For the strongest transitions at the upper end
of the depicted $\lambda_{hs}$-range, the blue
band for the $3D$ approach lies entirely above
the red band corresponding to the $4D$-\os-approach,
but it still overlaps with the orange band
corresponding to the $4D$-\msbar-approach.
This suggests that in particular
for the $4D$-\os-approach the renormalization scale
variation by a factor of 2 does not fully capture
the theory uncertainty from missing
higher orders. 
For completeness, we note that with increasing
values of $\lambda_s$ the uncertainty bands
shown in both plots in \cref{fig:TcLcScale4d3d} move
to larger values of $T_c$ and smaller values of $L_c / T_c^4$.
The relative widths of the different bands do
not change significantly, such that our observations and
conclusions about the theoretical uncertainties of the
different approaches remain unchanged also for
different values of $\lambda_s$.

\begin{figure}[t]
\centering
\includegraphics[width=0.48\textwidth]{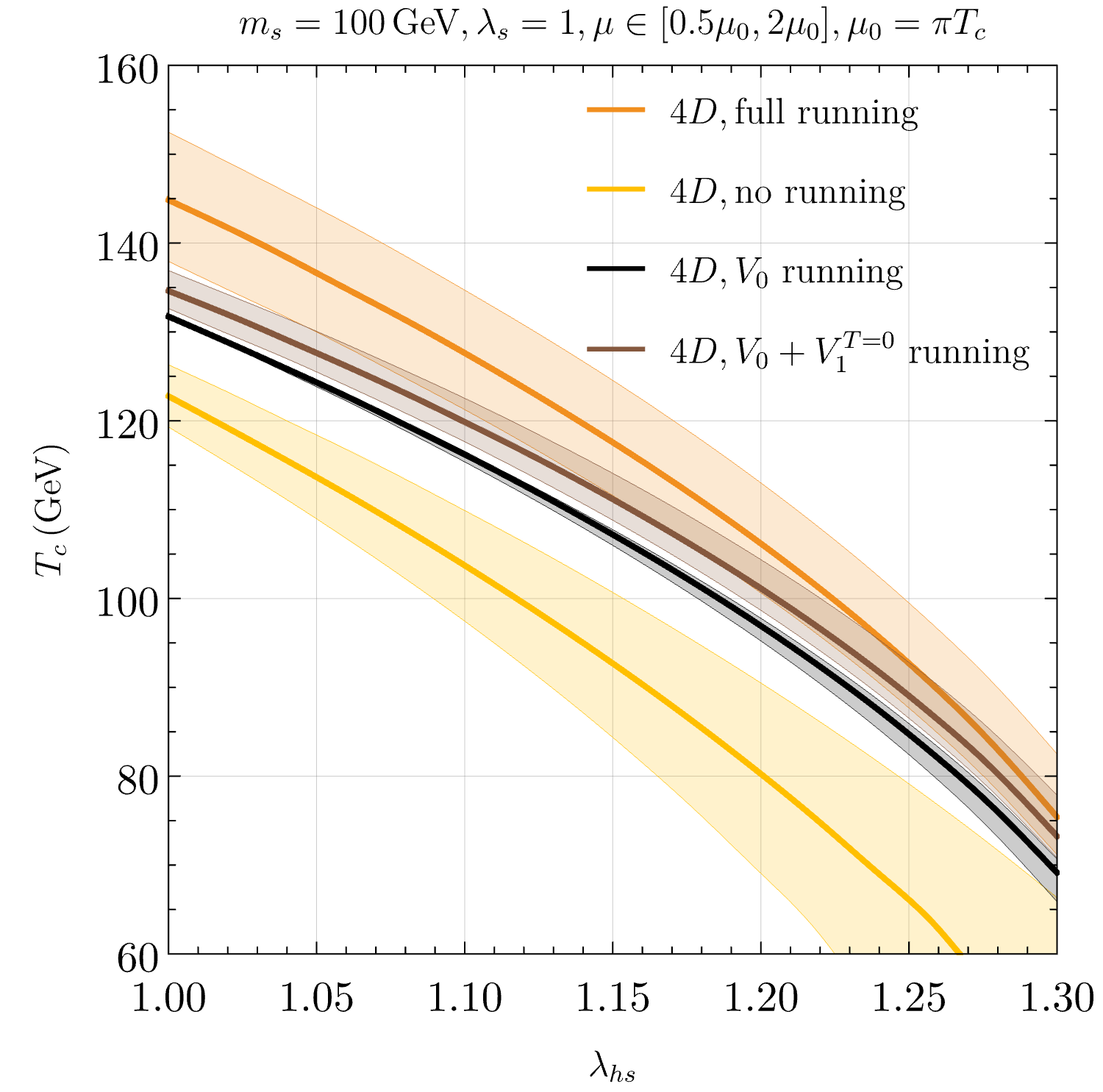}
~
\includegraphics[width=0.48\textwidth]{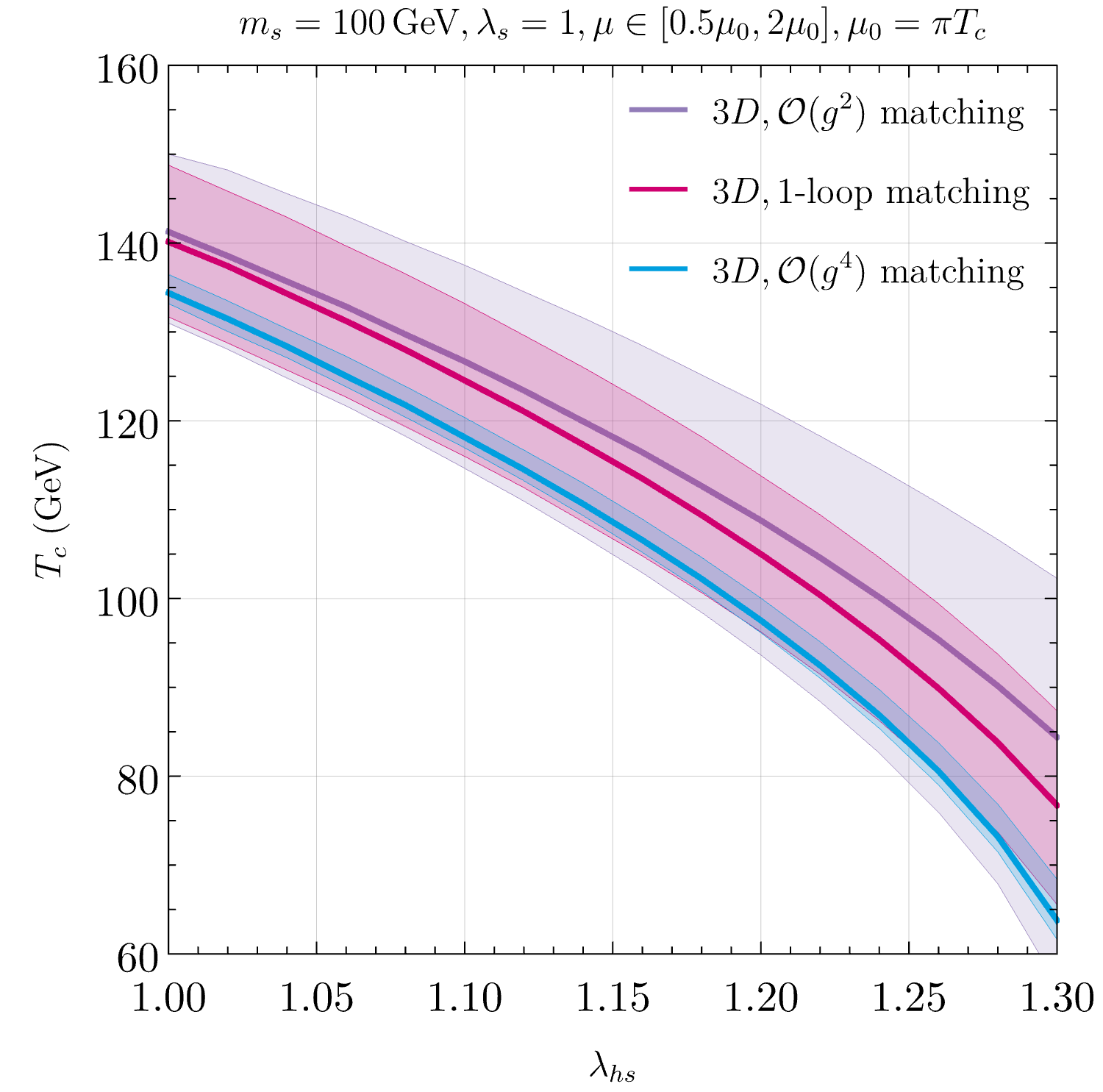}
\caption{
Scale-dependence of the critical temperature when using the \msbar-input scheme in different settings:
including the RGE running of Lagrange parameters in different parts of the effective potential in the $4D$ approach (\textit{left}) and using different matching orders in the $3D$ approach (\textit{right}).}
\label{fig:TcScaleplot2}
\end{figure}

As discussed in \cref{sec:scale_variation},
the renormalization
scale dependence in the $4D$ approach has different
origins. The effective potential exhibits both an
explicit scale dependence and an implicit scale
dependence from the RGE running of the couplings
and fields.
In order to investigate the main sources
of the uncertainty stemming from the scale variation,
we show in the left plot \cref{fig:TcScaleplot2} 
the predictions for $T_c$ in the $4D$-\msbar\ approach
with the RGE running effects included in different
parts of the effective potential.
As before, the solid lines show the predictions for
the central choice $\mu = \pi T_c$, while the
shaded bands indicate the uncertainty in $T_c$
from a variation of $\mu$ by a factor of 2.
When including no RGE running at all (yellow),
we find a sizable impact of the scale variation which
can be attributed solely to
the explicit scale dependencies in $V_{1}^{T=0}$.
As expected from~\cref{eq:rge},
including the running only in the tree-level part
almost completely cancels out the scale dependencies, which
gives rise to
the very small black uncertainty band.
However, this should not be interpreted as an indication
of small theory uncertainty, since the potential size
of missing higher-order corrections is largely
encoded in the RG-improved one-loop contributions.
As such, the apparent insensitivity to the renormalization
scale in this setup does not provide a reliable
estimate of the theory uncertainty from the two-loop
corrections or of the 
behavior
of the perturbative expansion.

Including the running also in the zero-temperature
one-loop piece of the effective potential
$V_{1}^{T=0}$ leads to a slightly larger
uncertainty band (brown). 
Since the one-loop RGE running included in $V_{1}^{T=0}$ is formally of two-loop order, this dependence on $\mu$
can be 
used
as an estimate for missing higher-order corrections in the $T=0$ part of the potential.
The main scale dependence, however, comes from the inclusion of the parameter running in the thermal part of the effective potential, which dominates for higher temperatures (orange, showing the full result, which is the same as in the left plot of \cref{fig:TcLcScale4d3d}).
This is apparent by the scale variation
becoming more pronounced in this case,
despite the scale dependence of $V_{1}^{T=0}$
being cancelled.
Moreover, as a consequence of the growing
temperature corrections with increasing
temperature, the orange band width grows
in size for higher temperatures,
while the opposite is true 
for the case where no
running effects are included (yellow).

As discussed in \cref{sec:DR}, in the $3D$ approach
the theory uncertainty from missing higher orders
can be estimated from the dependence on the hard
matching scale $\mu$ (the dependence on
the renormalization scale $\mu_{3D}$ is discussed below).
In the 
right
plot
of \cref{fig:TcScaleplot2}
we show the scale dependence of the predictions
for the critical temperature at different matching
orders in the dimensionally reduced EFT.
As before, the solid lines indicate the predictions
for $T_c$ with the scale set to the central choice
$\mu = \pi T_c$, and the shaded bands indicate the
uncertainty originating from a variation of the
matching scale by a factor of 2.
For the sake of separating uncertainties originating from various sources, we keep the effective potential at two-loop order for all matching orders.
We observe that a matching at the leading
$\mathcal{O}(g^2)$ accuracy (violet) gives rise to a
significant scale dependence and a large theory
uncertainty which surpasses the uncertainty from
the renormalization scale dependence in the $4D$
approach shown in the left plot of \cref{fig:TcScaleplot2}.
The scale dependence in the $3D$ approach is reduced
at the lower end of the $T_c$ values depicted in the plot
if a matching at the full one-loop level is carried out
(magenta),
which contains the corrections of $\mathcal{O}(g^2)$
and partially accounts for corrections of
$\mathcal{O}(g^4)$ (see \cref{eq:matching}).\footnote{Our
one-loop results within the three-dimensional framework are in agreement with those presented in Ref.~\cite{Schicho:2022wty}.}
However, for the largest values of $T_c$ shown in
the plot, the dependence on the hard matching scale
is barely reduced at one-loop matching accuracy.
The blue band indicates the results when
including two-loop contributions and a
matching at full NLO $\mathcal{O}(g^4)$ is carried out.
This is the same result as in the left plot of \cref{fig:TcLcScale4d3d}.
We see that the NLO matching reduces the scale
dependence over the whole range of $\lambda_{hs}$
by a factor of about five.
Notably, only at $\mathcal{O}(g^4)$ matching accuracy
the scale dependence in the $3D$ approach is smaller
than the renormalization scale dependence in the
$4D$ approach shown in the left plot
of \cref{fig:TcScaleplot2}. This demonstrates
the necessity of working at $\mathcal{O}(g^4)$
matching accuracy in order to reduce theory
uncertainties by applying dimensional reduction
and working in the high-temperature EFT.

In addition to the dependence on the hard matching
scale $\mu$, the effective potential in the $3D$ EFT
shows a dependence on the renormalization
scale $\mu_{3D}$.
As shown in \cref{eq:Veff3d},
in $3D$ the explicit dependence
on the renormalization scale enters only
at the two-loop level, whereas the dependence in the
tree-level and one-loop pieces of the effective
potential is only implicit via the RGE running
of the parameters. Specifically for the cxSM,
we can write
\begin{equation}
\label{eq:Veff3d-scale}
\Veff^{3D} = V^{3D}_0 \big(\mu^2_{h/s, 3D}(\mu_{3D})\big) + V^{3D}_1 \big(\mu^2_{h/s, 3D}(\mu_{3D})\big) + V^{3D}_2 \big(\mu_{3D}; \mu^2_{h/s, 3D}(\mu_{3D})\big) \, ,
\end{equation}
where the explicit $\mu_{3D}$-dependence appears
in $V_2^{3D}$, and the RGE running of the
mass parameters $\mu_{h,3D}^2$ and $\mu_{s,3D}^2$ gives
rise to the $\mu_{3D}$-dependence in
$V_0^{3D}$ and $V_1^{3D}$. 

\begin{figure}[t]
\centering
\includegraphics[width=0.48\textwidth]{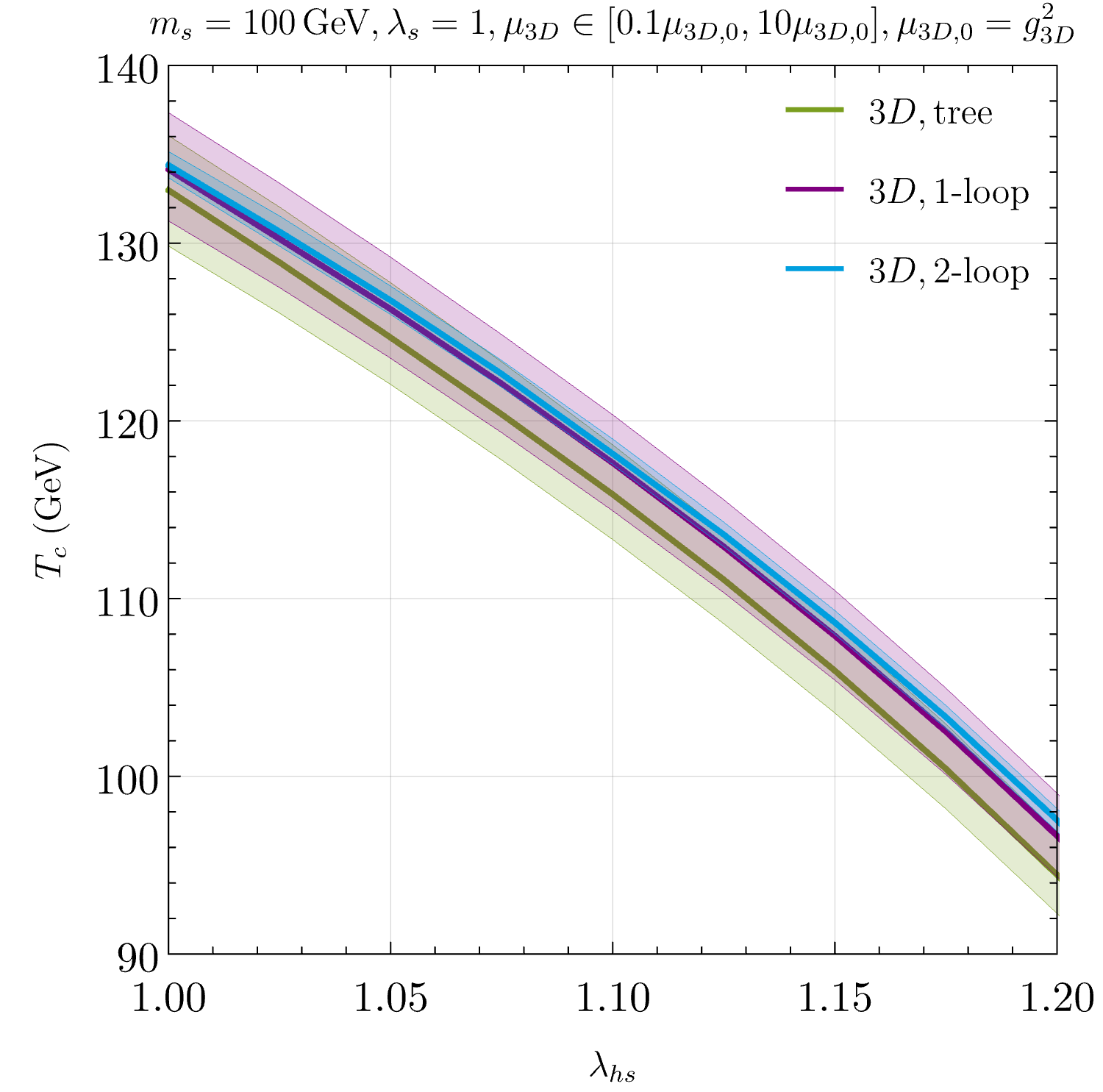}
\caption{Variation of $T_c$ with the $3D$ renormalization scale
$\mu_{3D}\in[\mu_0 /10, 10\mu_0]$
for different loop orders
within the $3D$ approach.}
\label{fig:TcScaleplot3}
\end{figure}

The dependence of the critical temperature $T_c$
on the renormalization scale $\mu_{3D}$
(for fixed $\mu$) within the
EFT is shown in
\cref{fig:TcScaleplot3} for the tree-level, one-loop
and the two-loop effective potential, where we vary the
scale around a central value $\mu_{3D} = g^2_{3D}$,
where $g_{3D}$ is the $\text{SU(2)}$ charge of the effective theory.
The dependence on the EFT scale $\mu_{3D}$ is of far smaller magnitude than the one on the \textit{hard} matching scale $\mu$ (see the discussion above),
even if only 
the tree-level potential is used.
In order for the $\mu_{3D}$-variation to be visible
at all in \cref{fig:TcScaleplot3},
we therefore show a variation
around the central value by a factor of ten,
in contrast to the factor of two for the $\mu$-variation,
and we narrow down the shown range for the
portal coupling $\lambda_{hs}$.
The explicit renormalization scale dependence in $V_2^{3D}$
cancels the implicit scale dependence of the couplings $\mu^2_{h, 3D}$ and $\mu^2_{s, 3D}$ in the tree-level $V_0^{3D}$, while the residual uncanceled scale dependence comes from including the running of couplings in $V_1^{3D}$ and $V_2^{3D}$.
In agreement with this expectation, we find
that the renormalization scale dependence cancels
to a large degree at the two-loop level (blue),
while the results merely receive an overall shift
but no reduction of the scale dependence
when comparing the predictions at
tree-level (green) and one-loop level (purple).
Moreover, the overlap of the three different bands suggests a good loop convergence within the EFT.

Finally, the effective potential in the $3D$
EFT depends on a third \text{soft} scale
$\mu^{\rm soft}_{3D}$, which appears at the
intermediate step of integrating out temporal components
of the gauge fields (see \cref{fig:dimred}).
We find that the residual dependence on
$\mu^{\rm soft}_{3D}$ is quantitatively insignificant.
Accordingly, in the discussion of theoretical
uncertainties of GW predictions in
\cref{sec:gravwave} we include only the
uncertainty associated
with the hard matching scale $\mu$, as it was found to
be substantially larger than the uncertainties arising
from both the renormalization scale $\mu_{3D}$
and the soft scale $\mu^{\rm soft}_{3D}$.
We do, however, include two additional sources of
theory uncertainty from the residual gauge
dependence and the breakdown of the $3D$ high-temperature
EFT at low temperatures, which we estimate by
studying the impact of higher-dimensional
operators in the EFT, as discussed in detail
in \cref{sec:gauge_dep} and
\cref{sec:highdim}, respectively.

\subsubsection{Gauge dependence}
\label{sec:gauge_dep}

Next, we compare the different methods in terms of the residual gauge-dependence of the results.
For this, we choose the following gauge fixing condition for the cxSM in the basis of mass-eigenstates after electroweak symmetry breaking:
\begin{equation}
\label{eq:gaugefixingCXSM}
    \mathcal{L}_{GF} = -\frac{1}{2}\left(F_{\gamma}^2+F_{Z}^2+2F^+_{W}F^-_{W}\right),
\end{equation}
with
\begin{align}
    F_{\gamma}&=-\frac{1}{ (\xi^\gamma_1)^{1/2}}\partial^\mu A_{\mu},\\
    F_{Z}&=-\frac{1}{ (\xi^Z_1)^{1/2}}\left(\partial^\mu Z_{\mu} -  \xi_2^Z \frac{\sqrt{g^2+g'^2}\phi^{GF}}{2}G_0 \right),\\
    F^{\pm}_{W}&=-\frac{1}{ (\xi^W_1)^{1/2}}\left(\partial^\mu W^{\pm}_{\mu} \mp i \xi_2^W \frac{g\phi^{GF}}{2}G_{\pm} \right),
\end{align}
where $\xi_1^{Z,W,\gamma}$, $\xi_2^{Z,W}$ and $\phi_i^{GF}$ are the gauge-fixing parameters. 
We suppress the superscripts $(Z,W,\gamma)$ in the following for better readability.
Some commonly used choices for the gauge-fixing parameters, which we study in the presented work, are
\begin{enumerate}
	\item \textbf{Fermi gauge}: $\xi_1 = \xi$, $\xi_2 = 0$.
	In this case, ghosts become massless, but calculations are complicated by the presence of 
    mixing propagators between vector bosons and the corresponding Goldstone bosons. Moreover, 
    infrared (IR) divergences arise, necessitating additional resummation procedures~\cite{Espinosa:2016uaw}.
	
	\item \textbf{$\rxi$-gauges}:  $\xi_1 = \xi_2 = \xi$.\footnote{
		Note that in general, $\xi_1$ and $\xi_2$ run differently under the RGE flow~\cite{Martin:2018emo}.
        However, this only becomes relevant at the two-loop order in the effective potential in $4D$, which we do not consider in this work.
	}
	These are usually introduced to cancel the aforementioned 
    mixing propagators.
	Two subclasses can be defined here:
	\begin{enumerate}
		\item In the so-called \emph{background} $\rxi$-gauge, one uses $\phi^{GF} = \phi_h$.
		Thence, the vector-Goldstone mixing is canceled for all values of the background field $\phi_h$, and ghosts become massive.
		\item In the \emph{standard} $\rxi$-gauge, one uses $\phi^{GF} = v_h$.
		Here, 
        the vector-Goldstone mixing is canceled only at the minimum field configuration $\phi_{h, \text{min}} = v_h$, and the ghosts are massive.
		Note that in this gauge, the effective potential is not symmetric under $\phi_h\to-\phi_h$.
	\end{enumerate}
	
	\item \textbf{Landau gauge} is a limit of both Fermi and $\rxi$-gauges with $\xi \to 0$.\footnote{This
    is a fixed point of the RGE.} Ghosts are massless and 
    mixing propagators are
    absent.

\end{enumerate}
We list the gauge-dependent mass eigenstates of the cxSM in different gauges explicitly in \cref{app:masseigenstates}.
In our work, we will employ the \emph{background} $\rxi$-gauge and take all $\xi^{W,Z,\gamma} = \xi$, except where explicitly mentioned otherwise, which allows us to analyze the
numerical impact of residual gauge dependence by continuously
varying the gauge-fixing parameter $\xi$.

A few comments on the \emph{background} vs.\ the \emph{standard} $\rxi$-gauge are in order.
The former has the advantage that
propagator
mixing between Goldstone bosons
and vector fields is avoided for all field values, which significantly simplifies the form of the effective potential beyond tree-level.
Though this comes at a price: the effective potential in this gauge cannot be used as a generating functional for all $n$-point correlation functions at zero momentum.
More specifically, starting from two-point functions at one-loop order, there is a mismatch between the regular Feynman-diagrammatic calculation (\eg the scalar self-energy at one-loop) and the second field derivative of the effective potential in the \emph{background} $\rxi$-gauge.
The mismatch appears because taking derivatives of $V$ w.r.t.\ $\phi_h$ does not commute with setting $\phi^{GF}\to \phi_h$.
This is in principle known from Ref.~\cite{Fukuda:1975di} where it was denoted as ``\textit{bad gauge}'',
but this fact 
has not been emphasized very much  
in the literature in this context.
Somewhat accidentally, though, there is no mismatch between one-loop tadpole diagrams and the first field derivatives of $V$~\cite{Fukuda:1975di}.
For the use of
the effective potential and the effective action to describe phase transitions, the difference between the two \rxi-gauges matters, as the variation of the background scalar field value is the essence of the computations.
For the effective potential in the cxSM
we have explicitly verified that only the effective potential in the \emph{standard} $\rxi$-gauge reproduces $n$-point functions calculated perturbatively at zero external momentum.
However, in our analysis below, we also include the results obtained in the \emph{background} $\rxi$-gauge, as it is frequently used in the studies of phase transitions.
This is another non-trivial cross-check of our
implementation of the effective potential in the $R_\xi$-gauge.

Since
we are interested in
estimating the theory uncertainty of the
parameters characterizing 
a FOEWPT arising from their residual gauge-parameter dependence,
we need to define a reasonable variation range for the gauge-fixing parameter.
The upper bound can
be defined by requiring perturbativity, which
schematically gives rise to a limit of
$|\xi|  \lesssim 1/g \sim 10$,\footnote{It was shown in Ref.~\cite{DiLuzio:2014bua} that taking larger values of the gauge-fixing parameter can also lead to spurious Landau poles.}
while for higher values of the gauge-fixing parameters
the 
perturbative behavior of the loop expansion
begins to fail.
We have therefore chosen
the range $\xi \in [-10,10]$ for the Fermi and \emph{background} \rxi gauges and $\xi \in [0,10]$ for the \emph{standard} \rxi\ gauge (see the discussion below).

\begin{figure}[t]
\centering
\includegraphics[width=0.55\textwidth]{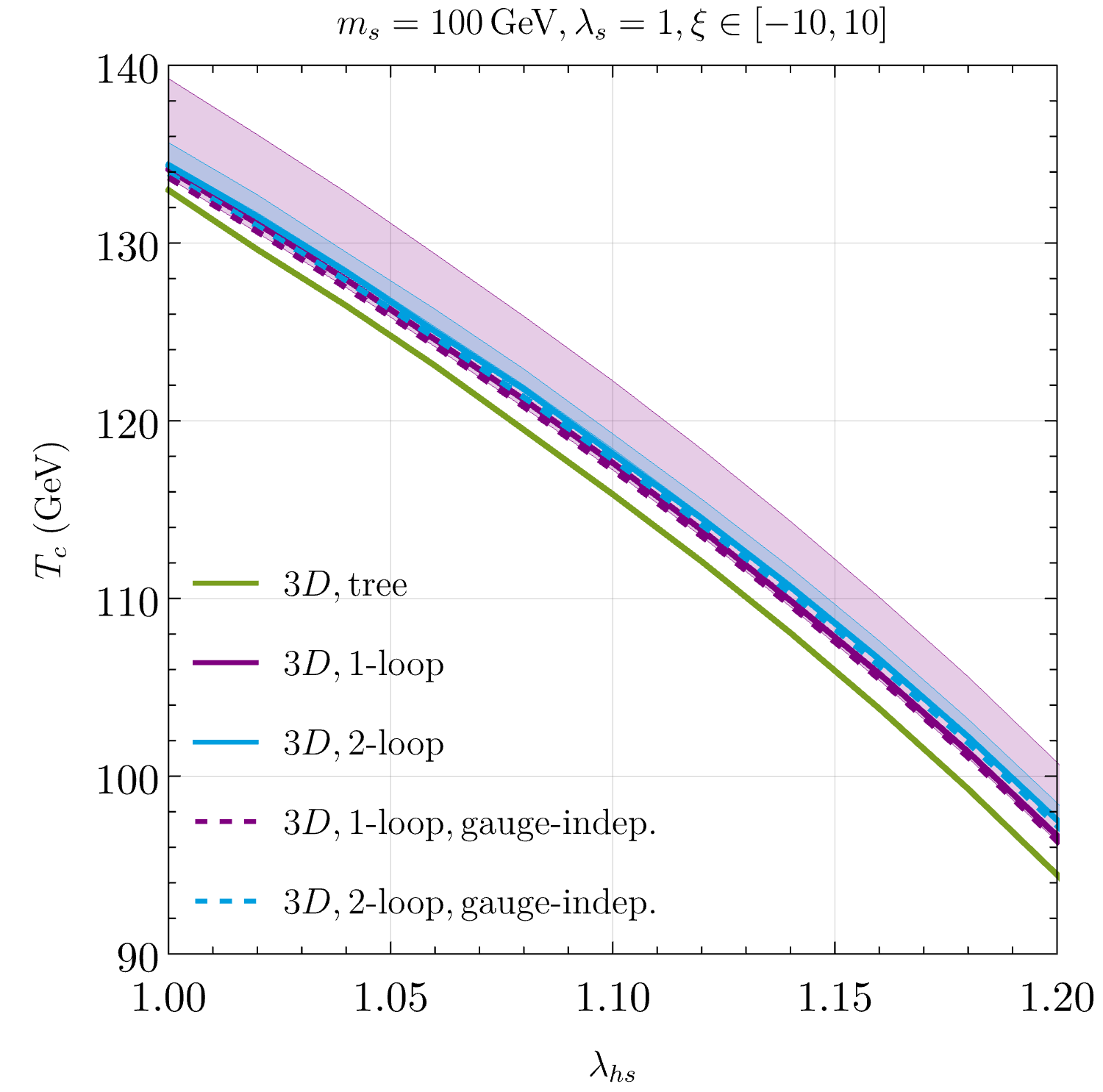}
\caption{
Critical temperature $T_c$ determined for $m_s=100\,\text{GeV}$, $\lambda_s=1$, $\lambda_{hs}\in[1,1.2]$ for different loop orders within the $3D$ approach.
The bands represent variations of the gauge-fixing parameter $\xi \in [-10,10]$ in $\rxi$-gauge.
Dashed lines represent the gauge-independent determination
of $T_c$ in the $\hbar$-expansion.
}
\label{fig:TcScaleplot3right}
\end{figure}

To begin, we show the gauge variation of $T_c$ for different loop orders of the effective potential in the $3D$ approach with the $\mathcal{O}(g^4)$ matching accuracy (see \cref{sec:DR} for details) in
\cref{fig:TcScaleplot3right}.
The solid lines indicate the predictions for $\xi = 0$,
which corresponds to the Landau gauge, and the shaded
uncertainty bands show the variation of $T_c$ when
varying $\xi$ between -10 and 10.
The result for the tree-level potential (green) is
only visible as a solid line because it does not depend on $\xi$.
We find that the gauge dependence is reduced by more than a factor of three when considering
the two-loop effective potential (blue) in comparison to the one-loop potential (purple).
The reduction of the theory uncertainty from the residual
gauge dependence
strongly resembles the reduction of the
scale-dependence when including higher loop orders of the effective potential
as shown in \cref{fig:TcScaleplot3}.
One can explain
this behavior via the Nielsen identity
shown in \cref{eq:nielsen0}.
For the gauge-dependent results shown in \cref{fig:TcScaleplot3right},
we determine the $n$-loop minimum $v^{(n)}=v_0 +v_1+\dots + v_n$ of the potential via direct minimization of the $n$-loop potential, \ie
\begin{equation}
	\frac{\partial}{\partial \phi} \left.(\Veff_0 + \Veff_1 + \dots + \Veff_n)\right|_{v^{(n)}} = 0 \, .
\end{equation}
Now, we insert this into the Nielsen identity for the all-order potential
\begin{equation}
	\label{eq:Nge}
	\left.\xi \frac{\partial }{\partial \xi} (\Veff_0 + \Veff_1 + \dots )\right|_{v^{(n)}} 
	= C(\phi,\xi) \left.\frac{\partial}{\partial \phi}\left(V_{n+1} + V_{n+2} + \dots\right)\right|_{v^{(n)}}.
\end{equation}
Rearranging terms, we find for the order of the gauge dependence of the $n$-loop potential:
\begin{align}
	\left.\xi \frac{\partial}{\partial \xi} (V_0 + V_1 + \dots + V_n)\right|_{v^{(n)}} 
	=
	\left.\left(C(\phi,\xi) \frac{\partial}{\partial \phi} - \frac{\partial}{\partial \xi} \right)
	\left(V_{n+1} + V_{n+2} + \dots\right)\right|_{v^{(n)}} = \mathcal{O}(\hbar^{n+1}) \, .
\end{align}
Thus, the residual gauge dependence of the effective
potential is of the order of $\hbar^{n+1}$, i.e.~the order of the
missing higher-order terms in the potential.
As a consequence, the predictions for 
thermodynamic quantities like $T_c$ 
have a reduced residual gauge dependence
with increasing loop order that is visible
for our results shown
in \cref{fig:TcScaleplot3right}.

We have discussed in \cref{sec:gauge} how a gauge-independent
prediction for the critical temperature can be achieved
by applying the $\hbar$-expansion.
We indicate the predictions for $T_c$ using the
$\hbar$-expansion considering the different loop orders with
the dashed lines in \cref{fig:TcScaleplot3right}.
Both at one- and two-loop order, we find that the
gauge-independent
$\hbar$-expansion results
lie within the shaded uncertainty bands and
are thus captured by the gauge variation in the direct minimization procedure.
We also find that the gauge-independent predictions
lie close to the Landau-gauge results
which are shown with the solid lines.

\begin{figure}[t]
\centering
\includegraphics[width=0.48\textwidth]{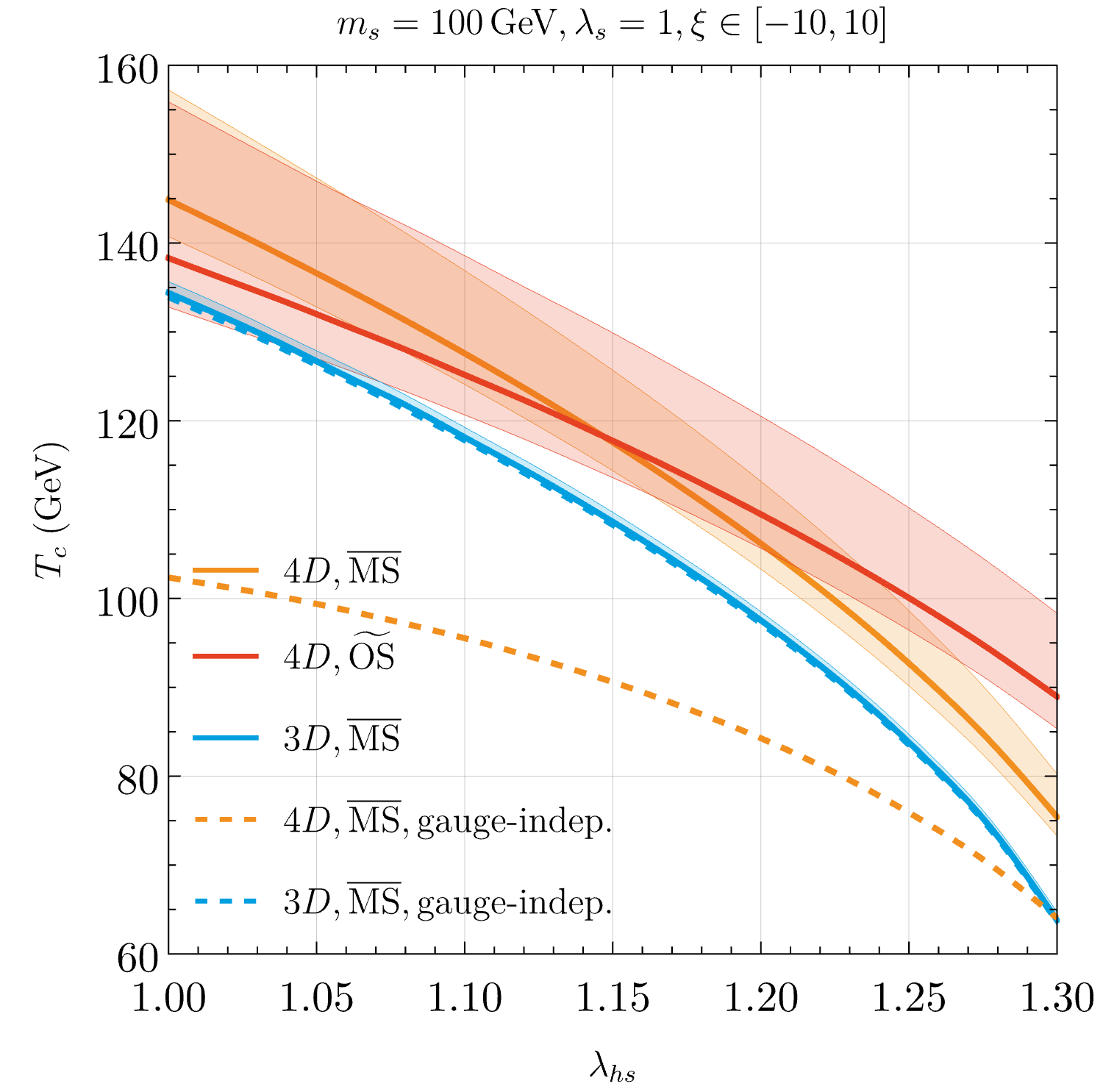}
~
\includegraphics[width=0.48\textwidth]{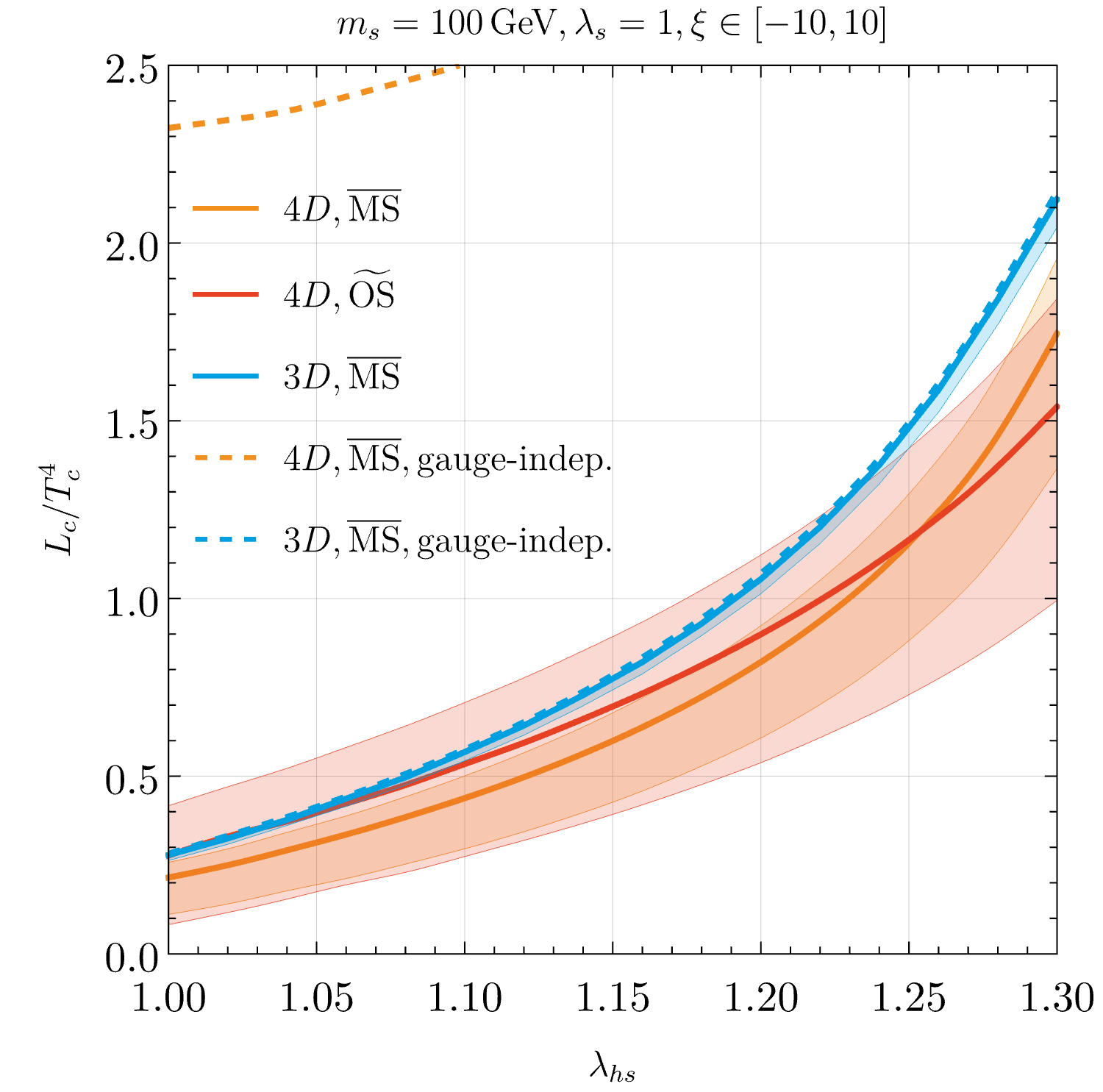}
\caption{Critical temperature (\textit{left}) and latent heat (\textit{right}) determined for $m_s=100\,\text{GeV}$, $\lambda_s=1$, $\lambda_{hs}\in[1,1.3]$.
The bands represent variations of the gauge-fixing parameter $\xi \in [-10,10]$ in $\rxi$-gauge.
Dashed lines represent the gauge-independent approach based on the expansion around the tree-level minimum.
}
\label{fig:TcLcXi4d3d}
\end{figure}

Next, we compare the different approaches
defined
at the beginning of this section ($4D$-\msbar; \ $4D$-\os; \
$3D$-\msbar\ at two-loop level with $\mathcal{O}(g^4)$
matching)
in terms of the numerical impact of the
residual gauge-dependence on the predictions for
the critical temperature $T_c$
and the latent heat $L_c$.
This is depicted in \cref{fig:TcLcXi4d3d} for $T_c$ (\textit{left} plot) and $L_c$ (\textit{right} plot), with
the solid lines showing the result for $\xi = 0$,
and the shaded bands indicate the theoretical uncertainty
from the $\xi$-variation as discussed above. Moreover,
the dashed lines
show the predictions for the manifestly
gauge-independent calculation, as determined using the $\hbar$-expansion about the tree-level minima of the effective potential (\cref{eq:hbarexp}). 
We note that in the $4D$-\os approach the finite counterterms are gauge-dependent (see \cref{app:renormalization}); hence, it is not possible to perform an $\hbar$-expansion
leading to a gauge-independent prediction.
As already visible in \cref{fig:TcScaleplot3right},
we find good agreement between
the gauge-independent results
and the gauge-dependent results based on
the direct potential minimization in the $3D$ case (blue).
On the contrary, there is a
significant 
deviation
between the
gauge-independent results based on the $\hbar$-expansion
and the ones from direct numerical minimization of the
loop-corrected potential in the $4D$ approach, irrespective
of whether the \msbar- or the \os-prescription is 
used.
The gauge-independent predictions for $T_c$ deviate by
more than 50\% from the results where no $\hbar$-expansion
is applied, and the gauge-independent predictions for 
$L_c/T_c^4 \sim 2.5\text{--}4$
deviate by a factor of a few across the parameter space shown on the plot.
The main culprit for these huge discrepancies in
the $4D$ approaches
lies in the fact that the tree-level minimum is
significantly shifted by leading thermal loop corrections.
Furthermore, in the $4D$ approach, the method to extract gauge-independent daisy resummation contributions, following Ref.~\cite{Patel:2011th}, is somewhat \textit{ad hoc}, relying on evaluating different parts of the potential at different minima values.
Taking into account the observations from the discussion
of the renormalization scale dependence
in \cref{sec:scale_variation}, our results
suggest that the inherent theoretical uncertainties of the $4D$ approach
at the one-loop level
are neither captured by a scale variation about a factor of two, nor by the rather large $\xi$ variation 
that we have applied
here.
On the contrary, the proximity of the gauge-independent and the Landau gauge result in the $3D$ approach once again indicates a good 
perturbative behavior
of the effective potential. 
As for the scale dependence discussed in
\cref{sec:scale_variation}, we also note here that the
relative widths of the different uncertainty bands corresponding
to the different perturbative approaches
shown in \cref{fig:TcLcScale4d3d}
remain very similar
also for other values of the quartic coupling $\lambda_s$.
Thus, our conclusion that the $3D$ approach shows a better perturbative
behavior compared to the $4D$ approach holds also for
different values of $\lambda_s$.

\begin{figure}[t]
        \includegraphics[width=0.48\textwidth]{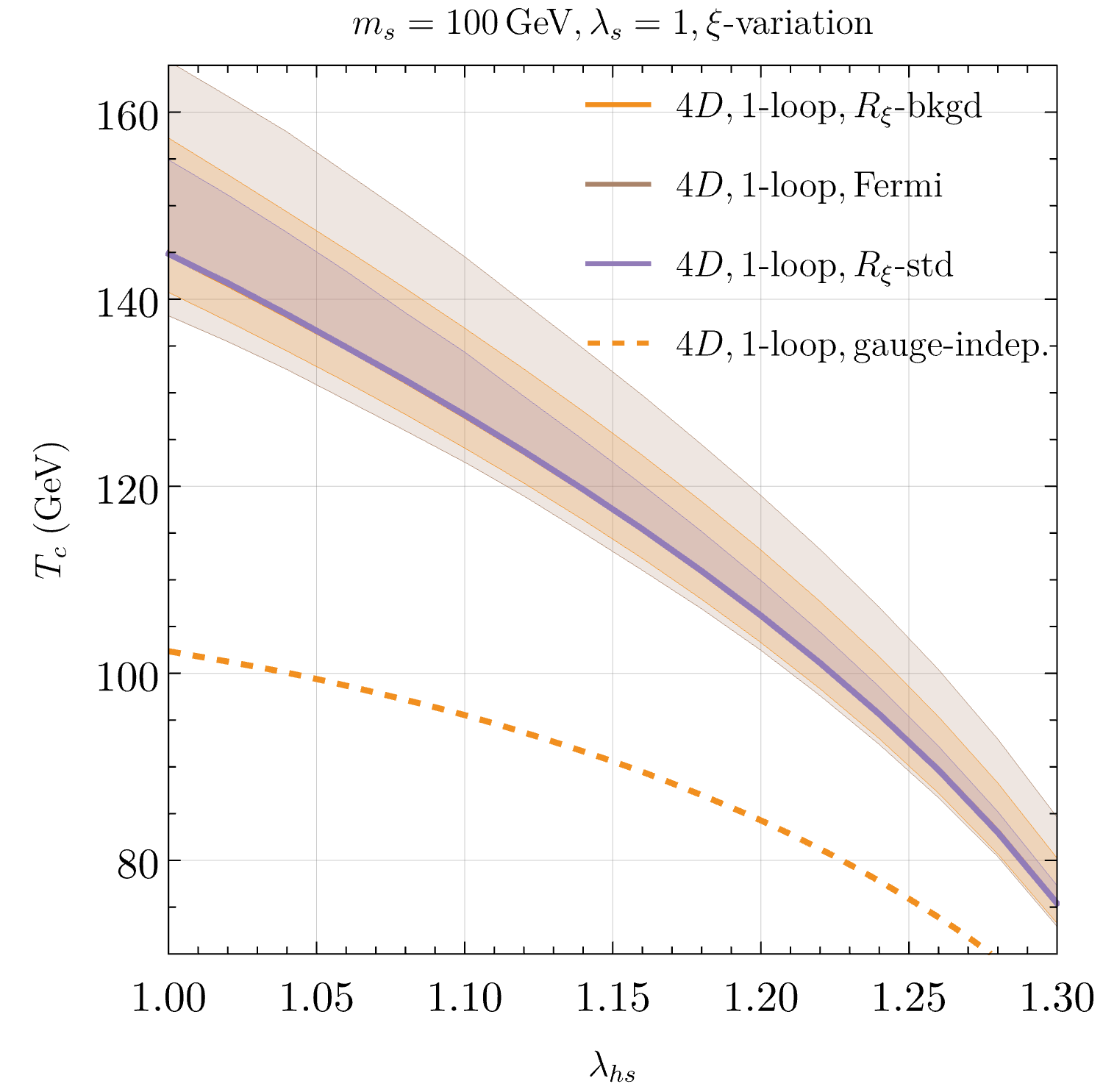}
        ~
         \includegraphics[width=0.48\textwidth]{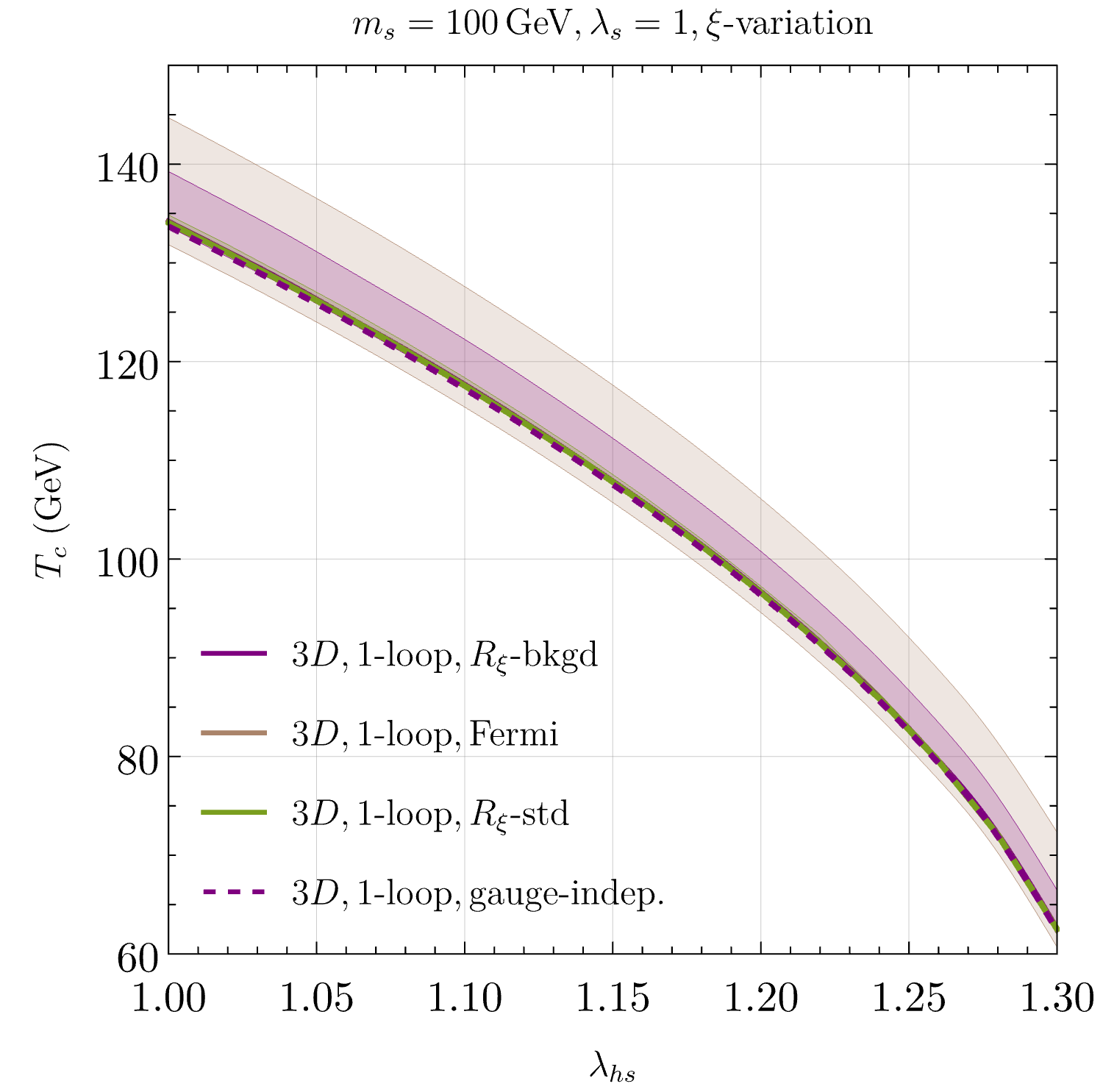}
\caption{
Gauge-parameter induced variation of the critical temperature for Fermi, \textit{standard} and \textit{background} $\rxi$-gauges (see \cref{sec:gauge} for definitions).
The \textit{left} panel shows $T_c$ determined by minimizing the $4D$ effective potential with daisy resummations and input parameters determined in the \msbar-scheme.
The \textit{right} panel shows $T_c$ in different gauges within the $3D$-\msbar approach at one-loop order. For the \textit{background} $\rxi$ and Fermi gauges, the $\xi \in [-10,10]$ variation is shown.
In the \textit{standard} $\rxi$-gauge, $\phi^{GF}$ is set to the tree-level minimum $v_{h,0}$, and the gauge-fixing parameter is varied within $\xi \in [0,10]$.
The solid and dashed lines correspond to results obtained in the Landau gauge and using a gauge-independent approach via the $\hbar$-expansion, respectively.
}
\label{fig:TcDifferentGauges4d3d}
\end{figure}

We proceed by exploring differences in the choice of the gauge-fixing condition. 
A comparison between the critical temperature determined by direct minimization of the effective potential in the Fermi, \textit{background} $\rxi$ and \textit{standard} $\rxi$ gauges,
defined in \cref{sec:gauge}, is presented in \cref{fig:TcDifferentGauges4d3d} for the $4D$-\msbar
(left plot) and $3D$-\msbar (right plot) approaches.
For a better comparison between the $3D$ and the $4D$ approaches,
we use here the one-loop effective potential in the $3D$ approach
which is at
the same loop order as the
effective potential in the $4D$ approach.
For the \textit{background} $\rxi$ and Fermi gauges, we
show uncertainty bands using a $\xi$-variation
in the range $\xi \in [-10,10]$.
In the \textit{standard} $\rxi$-gauge, $\phi^{GF}$ is set to
the tree-level minimum $v_{h,0}$, and the gauge-fixing
parameter is varied 
using only positive values of $\xi$
within the range $\xi \in [0,10]$, see the discussion below.
The variation of the gauge-fixing parameter $\xi \in [-10,10]$ in Fermi gauge results in a larger uncertainty in the $T_c$-determination in both the $4D$ and $3D$ approaches as compared to the \textit{background} $\rxi$-gauge, but 
the qualitative results are similar.
However, we find that positive $\xi$ values
give rise to lower critical temperatures in Fermi gauge,
while the opposite is true in the
$\rxi$-gauges.

For the \textit{standard} $\rxi$-gauge, we use $\phi^{GF} = v_{h,0}$ in the gauge-fixing condition, thus assigning
the local EW symmetry-breaking minimum with positive
value of $v_{h,0}$ as the physical minimum.
This gauge fixing explicitly breaks the $\phi_h \to -\phi_h$ symmetry of the effective potential, which is preserved in the other gauges.
As a consequence, in the standard $\rxi$-gauge the
two EW symmetry breaking minima at $+v_{h}$ and at
$-v_{h}$ are not degenerate anymore, whereas their
degeneracy is preserved in the other gauges.
For negative $\xi$-values, we find that the
minimum at
$\phi_s = 0$ and $\phi_h<0$ becomes the
global minimum at low temperatures.
By tracing the different phases of the potential
at finite temperature and computing the probabilities
for a FOEWPT from the vacuum with $(v_{s} \neq 0, v_{h} = 0)$
into the vacua with $(v_{s} = 0, v_{h} \neq 0)$, one
finds that the transition rates are larger for a transition
into the vacuum corresponding to the global
minimum with $v_{h} < 0$. However, this contradicts
the choice of setting $\phi^{GF} = v_{h,0} > 0$ in the gauge-fixing
conditions, as discussed above.
Due to this inconsistency in the standard $R_\xi$-gauge,
we discard these points 
with negative $\xi$-values
as unphysical, and limit the
variation of the gauge-fixing parameter
in this gauge to positive
values $\xi \in [0, 10]$ where the minimum with $v_{h} > 0$
is the global minimum of the potential at $T = 0$.\footnote{Since
$\xi = 0$ is the lower limit of the $\xi$-variation in the
standard $R_\xi$-gauge, the solid lines indicating the predictions
in the Landau gauge coincide with the lower edge of the
uncertainty bands of the standard \rxi-gauge.}
We have verified that upon performing an $\hbar$-expansion in 
the \emph{standard} \rxi-gauge, we obtain gauge-independent results that agree with the results from other gauges after an $\hbar$-expansion.

The results for the gauge-independent determination of $T_c$ via an $\hbar$-expansion are shown by dashed lines in \cref{fig:TcDifferentGauges4d3d}.
As a consequence of 
the $\hbar$-expansion,
the results are
independent of both the gauge-fixing parameter
and the gauge-fixing condition.
As already discussed above, the
application of the $\hbar$-expansion in the
$3D$ EFT gives rise to a result that lies
within the uncertainty bands from the $\xi$-variation
using the gauge-dependent numerical minimization
of the loop potential. Hence, the $3D$ approach
complemented with the $\hbar$-expansion
provides a numerically precise and gauge-independent
prediction for the critical temperature.
On the contrary, in the $4D$ approach the
application of the $\hbar$-expansion at one-loop
level gives rise to a result that is gauge-independent
but shows large deviations from the results using
the direct minimization of the loop potential,
and only the gauge-dependent results (without
$\hbar$-expansion) are in agreement with the
predictions of the $3D$ EFT.

Interestingly, we find that the loop corrections in some gauge-fixing setups can qualitatively change the vacuum structure depicted in \cref{fig:phasestructure}.
For large $\xi$-values, the high temperature phase
can develop a small non-zero value $v_{h}$ of the Higgs condensate at the minimum in addition to
a non-vanishing
singlet vev $v_s$.
This is analogous to
the EW symmetry
non-restoration at high temperature, a phenomenon
that recently gained 
interest~\cite{Meade:2018saz,Baldes:2018nel,
Glioti:2018roy,
Carena:2021onl,Matsedonskyi:2021hti,
Biekotter:2021ysx,Bittar:2025lcr}.
For instance, in the $\textit{background } \rxi$ gauge, for $\lambda_{hs}=\lambda_{s}=1, m_s=100\, \text{GeV}$ at $T=T_c$, this happens for $\xi \gtrsim7$ in the $4D$-\msbar approach.
We find
non-vanishing Higgs vevs in the high-$T$ phase, both
in $\rxi$ and Fermi gauges, while the vacuum structure
with $v_h = 0$ in the high-$T$ phase
is preserved in the Landau gauge.
In addition,
the high-temperature phase does not develop a
vev for the
Higgs field if the $\hbar$-expansion is performed.
Hence, 
we regard the occurrence of non-vanishing Higgs vevs in the high-$T$ phase
as an unphysical artifact of the direct minimization of the effective potential that can arise in certain
gauges if the vacuum structure of the potential
is not analyzed in a strictly gauge-independent way.

\subsection{Nucleation thermodynamics}
\label{sec:nucleationthermo}

As a next step
towards predictions for 
GW
signals from cosmological phase transitions
we now
study
the dynamics of the associated bubble nucleation processes that induce the onset of the phase transition.
One important quantity in this context is the nucleation rate at non-zero temperatures $\Gamma(T)$, which describes the probability per unit time and unit volume that a bubble of a new phase will nucleate.
It is the defining quantity for the temperatures at which bubble nucleation and percolation takes place, which for the EW phase transition
can be regarded
as the temperatures for the
on-set and the completion of the
phase transition, respectively.
The nucleation rate in the semi-classical approximation is given by \cite{Coleman:1977py,Linde:1980tt,Affleck:1980ac}
\begin{equation}
\label{eq:classicalrate}
\Gamma(T) \approx A e^{-S_B/T} \, ,
\end{equation}
where the prefactor $A = A_{\rm stat} A_{\rm dyn}$
contains statistical $A_{\rm stat}\approx T^3 (S_B/2\pi T)^{3/2}$ and dynamical components $A_{\rm dyn}\approx T$, which capture equilibrium and non-equilibrium effects, respectively \cite{Ekstedt:2022tqk}.
The bounce action $S_B$ is the
Euclidean action evaluated at the bounce solution $\phi^B_i$,
\begin{equation}
  S_B \equiv S\left[\phi^B_i\right] =
  4\pi\int\limits_0^\infty
  d\rho\rho^2 \left[
    \frac{1}{2}\left
      (\frac{d\phi^B_i(\rho)}{d\rho}
    \right)^2 +
  V^{4D}_0(\phi^B_i(\rho))\right] \, ,
    \label{eq:bounceAction}
\end{equation}
where $\phi^B_i$ are the field profiles that solve the equations of motion
subject to
specific boundary conditions, corresponding to the $O(3)$ symmetric profile of the bubble of the \textit{true} vacuum in the space of \textit{false} vacuum configurations,
\begin{equation}
    \frac{d^2\phi}{d\rho^2} + \frac{2}{\rho} \frac{d\phi}{d\rho} - \nabla_\phi V^{4D}_0(\phi)= 0, \enspace \text{with} \enspace
    \phi^B(\infty) = \phi_\textrm{false}, \; \left.\frac{d\phi^B}{d\rho}\right|_{\rho=0} = 0 \, . \label{eq:bounceEquation}
\end{equation}
Loop corrections of higher order
to~\cref{eq:classicalrate} 
arise from fluctuations around the bounce profile of~\cref{eq:bounceEquation}. 
At one-loop level, they would appear as a functional determinant prefactor~\cite{Linde:1980tt}.
They are formally
of higher order, but can be compatible in
size to the leading order bounce action~\cite{Gould:2021ccf,Ekstedt:2021kyx,Ekstedt:2023sqc}.

A complication arises from the fact that the phase transition is induced by thermal fluctuations, which themselves are loop effects.
Hence, the lowest order bounce action, as defined in \cref{eq:bounceAction}, often does not exist.
A common solution is to calculate the bounce action of~\cref{eq:bounceAction} by replacing the tree-level potential with the effective potential, $V^{4D}_0 \to \Veff^{4D}$ \cite{Linde:1980tt,Linde:1981zj}, which includes radiative corrections by construction.\footnote{In practice, one only 
incorporates the real part of the effective potential $V^{4D}_0 \to \mathrm{Re}[\Veff^{4D}]$. The presence of a non-vanishing imaginary part again signals the shortcomings of this approach~\cite{Weinberg:1992ds,Croon:2020cgk}.}
This simple replacement is what is usually
carried out in the $4D$ approach in order
to account for radiative corrections.
However, it does not correspond to a proper perturbative expansion of the bounce action,
as higher-loop corrections to other terms in the derivative expansion (similar to \cref{eq:effaction_expand}) could be just as important \cite{Weinberg:1992ds, Bodeker:1993kj, Surig:1997ne}.
Another drawback is that the effective potential calculation relies on the homogeneity of the background field, which is not the case for the bounce configuration.
It also leads to the double-counting of fluctuations, in both the calculation of the effective potential and in the prefactor to the bounce. 
Moreover, the nucleation rate calculated in this ad hoc approach is gauge-dependent, as a remnant of the gauge dependence of the effective potential
at non-stationary points.
Nevertheless, in many cases, the dominant contributions to the
action arise from the leading thermal corrections
already captured by the effective potential,
making this approach a reasonable first
approximation 
if
more rigorous methods are
computationally too expensive.
Consequently,
despite the issues raised above,
this method remains widely
used, as it often provides a qualitatively reliable
estimate of the phase transition dynamics, and
it is often sufficient
for identifying parameter regions where a 
strong
FOEWPT occurs
(see \cref{sec:gravwave}).

The calculation of the nucleation rate
in the $3D$ approach proceeds along the same lines as
in the $4D$ approach discussed above.
After taking into account the scalar field redefinition,
the nucleation rate takes the form
\begin{equation}
\label{eq:classicalrate3d}
\Gamma(T) \approx A e^{-S^{3D}_B} \, ,
\end{equation}
with
\begin{equation}
  S^{3D}_B =
  4\pi\int\limits_0^\infty
  d\rho\,\rho^2 \left[
    \frac{1}{2}\left
      (\frac{d\phi^B_i(\rho)}{d\rho}
    \right)^2 +
  V^{3D}_0(\phi^B_i(\rho))\right] \, ,
    \label{eq:bounceAction3D}
\end{equation}
and the field profiles $\phi^B_i$ of the
bounce solution satisfy the
corresponding
equations of motions as shown in
\cref{eq:bounceEquation} with $V_0^{4D}$ replaced
by $V_0^{3D}$.
A theoretically consistent treatment
of the bounce action that includes radiative
and thermal corrections can be achieved
in a high-temperature EFT by integrating out the leading thermal fluctuations, absorbing them into an effective leading-order bounce action~\cite{Gould:2021ccf, Lofgren:2021ogg, Hirvonen:2021zej}.
Then, loop corrections to the bounce action within the EFT can be systematically accounted for \cite{Ekstedt:2023sqc}.
The loop-improved nucleation action derived in this way is gauge-independent, as it is defined by
the solution of the tree-level bounce equation of motion~\cite{Fukuda:1975di, Metaxas:1995ab, Metaxas:2000cw}.
As mentioned above, adding a complete set of one-loop corrections to the bounce action would require the computation of the functional determinant of the fluctuations around the bounce configuration, which goes beyond the scope of this work.

The dynamics of bubble nucleation 
implies that
the nucleation temperature $T_n$ and the percolation temperature $T_p$~\cite{Huber:2007vva}
are important quantities in this context.
The former is defined as the temperature at which the probability of a single bubble nucleating in a Hubble volume becomes significant,
\begin{equation}
\label{eq:Tn}
\int_{t(T_c)}^{t(T_n)} dt \frac{\Gamma(t)}{H(t)^3} = \int_{T_n}^{T_c} \frac{dT}{T} \frac{\Gamma(T)}{H(T)^4} \approx 1 \, ,
\end{equation}
where $T_c$ is the critical temperature at which the two minima are degenerate.
For the Hubble parameter $H(T)$, characterizing the expansion rate of the universe,
we can assume that the universe is
radiation dominated, such that
\begin{equation}
	H(T)^2=\frac{\rho_{\text{rad}}(T)}{3 M_{p}^2} \, .
\end{equation}
Here,
$\rho_{\text{rad}}$ is the radiation energy density given by $\rho_{\text{rad}} = g_* T^4 \pi^2/30$, $M_p = 2.4 \times 10^{18}\, \text{GeV}$ is the reduced Planck mass, and $g_* = g_*(T)$ is the effective number of relativistic degrees of freedom.

The percolation temperature $T_p$ is the temperature at which bubbles of the new phase have grown sufficiently to form a connected network that stretches over a significant fraction of the universe~\cite{Guth:1981uk}.
At this point, the phase transition effectively completes, and we expect most of the
GW produced during the transition
to originate from this moment~\cite{Ellis:2018mja,Athron:2023rfq},
which occurs when the fraction of the extended volume occupied by the new phase is around 34\%.\footnote{This corresponds to the probability $P_f(T_p) = e^{-0.34} \approx 71\% $ that a random point of space is still in the false vacuum.
The extended volume double counts the overlapping regions of the bubbles, which is removed by exponentiating it~\cite{Athron:2023xlk}.}
The defining relation for $T_p$ is
\begin{equation}
\label{eq:Tp}
\frac{4\pi H(T_p)^3}{3}\int_{T_p}^{T_c} \frac{dT}{T} \frac{\Gamma(T)}{H(T)^4} R(T_p,T)^3 \approx 0.34,
\end{equation}
where
$R(T_p,T)=v_w \left(1 - T/T_p\right) /H(T)$ is the radius of a bubble nucleated at $T$ which expands with the velocity $v_w$ until $T_p$ is reached~\cite{Anderson:1991zb}.
In our work, we make use of the fact that
most of the contributions to the integral shown
in \cref{eq:Tn} come from temperatures near $T_n$.
This allows one to apply the saddle-point
approximation to the integral, and the relevant
temperatures $T_n$ and $T_p$
can be derived from a simple
Taylor expansions of action.
Specifically, we use the
resulting approximations for
$T_n$ and $T_p$ as 
given in Ref.~\cite{Huber:2007vva},
which provides a convenient method for their numerical
evaluation.\footnote{Ref.~\cite{Huber:2007vva} defines
$T_f$, which is the temperature at which the phase transition
completes, given by \cref{eq:Tp}, but requiring the right-hand
side of this equation to be equal to unity.
Ref.~\cite{Huber:2007vva} then derives an approximate analytic
expression to determine $T_f$.
In our analysis, we evaluate this approximate expression and use the resulting value as an estimate for the percolation temperature 
$T_p$, assuming that $T_p \approx T_f$. We have checked
explicitly that the numerical difference between 
$T_p$ computed according to \cref{eq:Tp} and following
the approximation from Ref.~\cite{Huber:2007vva}
is at the percent level within the parameter space investigated
in this work.}
Numerically, we find that the difference between $T_n$ and $T_p$ is small within the parameter region of the cxSM under consideration, reaching at most
a few GeV.

The duration of the phase transition is characterized by the dimensionless quantity $\beta/H$, where $\beta$ is the time derivative of the nucleation rate at the time when the phase transition becomes significant:
\begin{equation}
\beta = \left.\frac{d}{dt} \log\Gamma(t) \right|_{t=t(T_p)} = \begin{cases}\left.H(T)T\dfrac{d}{dT} \left( \dfrac{S_B(T)}{T} \right) \right|_{T=T_p}, & 4D \textrm{ approach,} \\[10pt]
 \left.H(T)T\dfrac{d S^{3D}_B(T)}{dT}  \right|_{T=T_p}, & 3D \textrm{ approach.}
\end{cases}
\label{eq:betaH}
\end{equation}
The ratio $\beta/H$ measures the rapidity of the phase transition, where large values indicate fast transitions.
It is an important parameter for the predictions
of the GW signals, entering both the peak frequency
and the peak amplitude of the GW spectra,
see the discussion in \cref{sec:gwspec} below.

Another parameter
that is relevant for the GW predictions
is $\alpha$---the ratio of energy density liberated during the phase transition over the radiation energy density of the universe,
given by~\cite{Athron:2023xlk}
\begin{equation}
\alpha =\frac{\Delta\theta}{\rho_{\text{rad}}} =\begin{cases}
\dfrac{1}
{\rho_{\text{rad}}}\left( \Delta V^{4D}-\dfrac{1}{4}T\dfrac{d\Delta V^{4D}}{dT}\right)\Big|_{T=T_p}, & 4D \text{ approach,} \\[10pt]
\dfrac{1}
{\rho_{\text{rad}}}\left( \dfrac{3}{4}T\Delta V^{3D}-\dfrac{1}{4}T^2\dfrac{d\Delta V^{3D}}{dT}\right)\Big|_{T=T_p},  & 3D \text{ approach.}
\end{cases} 
\label{eq:alpha}
\end{equation} 

As before, $\Delta$ 
refers to
the difference between high and low temperature phases.
Here, we assumed that the
energy density released is 
related to the change of trace anomaly, $\theta = (\rho-3p)/4$, between phases. Conversely, it could be simply taken as the 
difference of the
energy density, $\rho$, 
between two phases.
The parameter $\alpha$ determines the strength of the phase transition, where larger values correspond to stronger transitions. 

To calculate the bounce action at the given temperature, we use the public code \texttt{CosmoTransitions}~\cite{Wainwright:2011kj}
to solve the bounce equation shown
in \cref{eq:bounceAction}.
The evaluation of the bounce action may suffer from numerical instabilities, which typically arise in the case of 
relatively weak
phase transitions. 
We find that the determination of $\beta/H$ is the most sensitive to these numerical instabilities in our calculations, whereas the computation
of the characteristic temperatures and of the
parameter $\alpha$ is only marginally affected.
Although these numerical instabilities
are present in the bounce solution,
they are sufficiently small as not to alter our
conclusions about the predicted GW spectra discussed below.

The final parameter that is relevant for the
prediction of the GW spectra is the bubble wall
velocity $v_w$, as used already in \cref{eq:Tp}.
Its computation requires a real-time
simulation of the expanding bubbles
in the surrounding plasma.
Performing such simulations
is a non-trivial
and computationally costly task~\cite{Moore:1995si}
which involves solving coupled equations of motion for scalars
and Boltzmann equations for the surrounding plasma~\cite{Ekstedt:2024fyq,Laurent:2022jrs}.
Explicit results have been obtained for the SM~\cite{Dorsch:2018pat},
singlet extensions of the SM~\cite{Lewicki:2021pgr,Krajewski:2024gma,
Branchina:2025jou},
and extensions of the SM by a second SU(2)
Higgs doublet~\cite{Dorsch:2016nrg,
Jiang:2022btc,Ekstedt:2024fyq,
Branchina:2025jou}.
Nevertheless,
there are still large uncertainties~\cite{vandeVis:2025plm},
and the predictions for $v_w$ based on different
approaches do not always agree. 
Moreover, higher-loop corrections to
$v_w$ have been accessed recently~\cite{Dashko:2024anp} and were found to
decrease the wall velocity in a model with
a real scalar singlet field, and
contributions from additional friction terms
beyond $1 \to 1$ particle transmissions
have been shown to be missing in the kinetic approach
to compute $v_w$~\cite{Ai:2025bjw}.
Taking these uncertainties into account,
for the computation of GW signals in realistic
BSM theories
the bubble wall velocity is often treated as a free input parameter (for instance, see~\cite{Biekotter:2022kgf}), or kept fixed at $v_w = 1$.
We leave a study of
uncertainties in the description
of the GW signals related to the imprecise
knowledge of the bubble wall velocity
for future studies and instead focus on the
impact of other sources of theory uncertainty
related to the perturbative description
of the phase transition.
Hence, for simplicity, we assume that
the expanding bubbles reach a terminal
velocity of $v_w = 1$.
This choice is supported by recent results from
real-time lattice simulations, which show 
a ultra-relativistic
detonation solution with $v_w \approx 1$ for
SFOEWPT (with GW signals potentially in reach
of LISA) for which prior computations using
analytic methods based on a steady-state
expansion of the bubble predicted substantially smaller
sub-sonic bubble walls~\cite{Krajewski:2024gma,Krajewski:2024zxg}.

Similarly to~\cref{sec:equilibrium}, we compare  
the different approaches for computing
the parameters characterizing the
thermodynamics of bubble nucleation
and the onset of the SFOEWPTs.
In particular, we are interested in the
renormalization scale dependence and the
gauge dependence in the predictions for
the parameters $T_p$, $\beta / H$  and $\alpha$,
in comparing the $4D$ approaches and the
dimensionally reduced $3D$ EFT.
In the $4D$ approach,
we substitute $V^{4D}_0 \to \Veff^{4D}$ of
\cref{eq:Veff4d} in the bounce equation, \cref{eq:bounceAction,eq:bounceEquation},
to compute $S_B$ in the temperature
range in which the false and the true vacua
coexist.
In the $3D$ approach, we calculate the nucleation rates (and observables which depend on those)
by solving the bounce equation at leading order.
In addition,
by substituting the loop-improved potential of \cref{eq:Veff3d} in the bounce determination (see \cref{eq:bounceAction3D}),
$V^{3D}_0 \to \Veff^{3D}$,
we estimate the impact of
higher-order corrections.

\begin{figure}[t]
    \centering
    \includegraphics[width=0.32\textwidth]{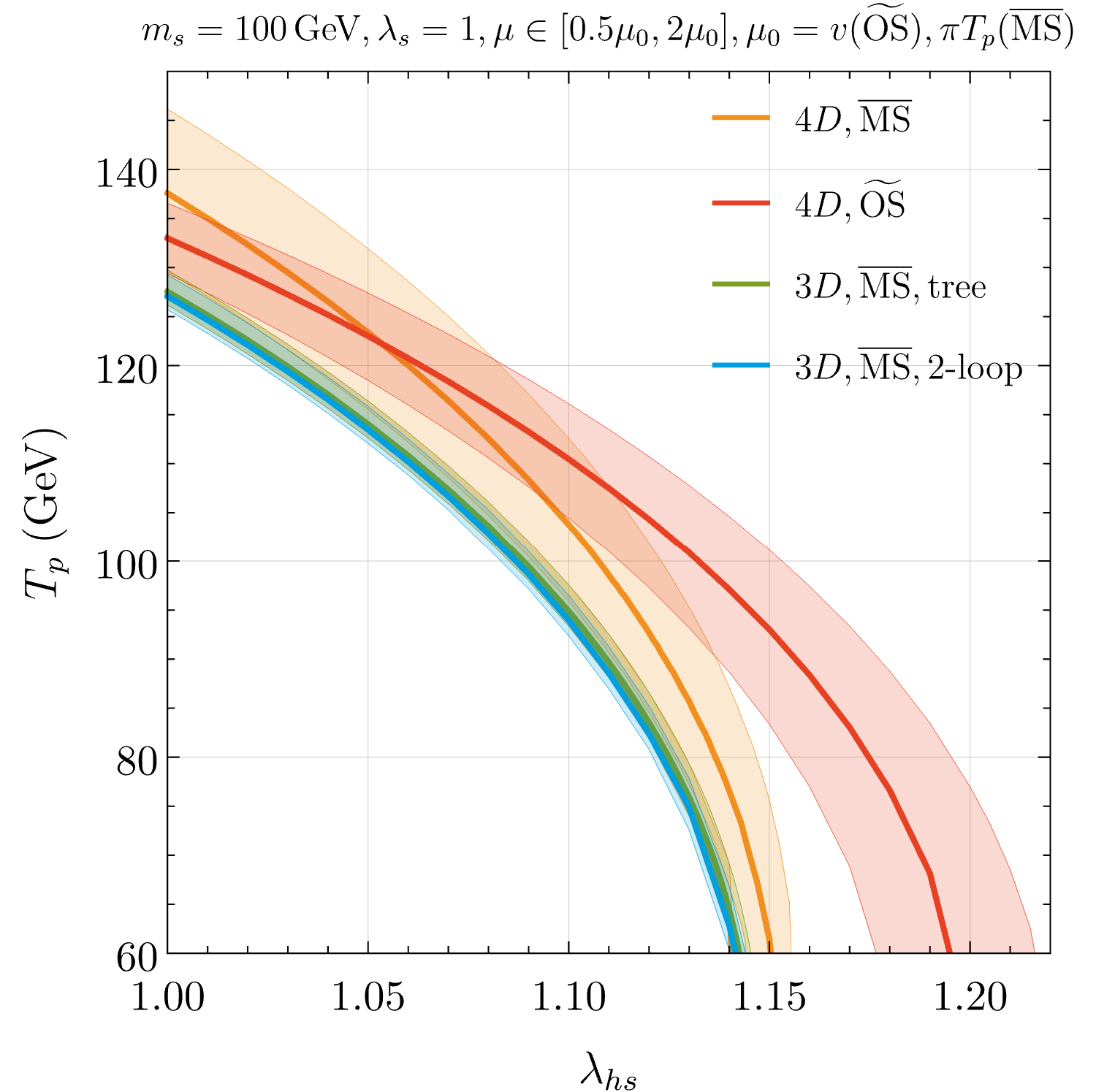}
    \includegraphics[width=0.32\textwidth]{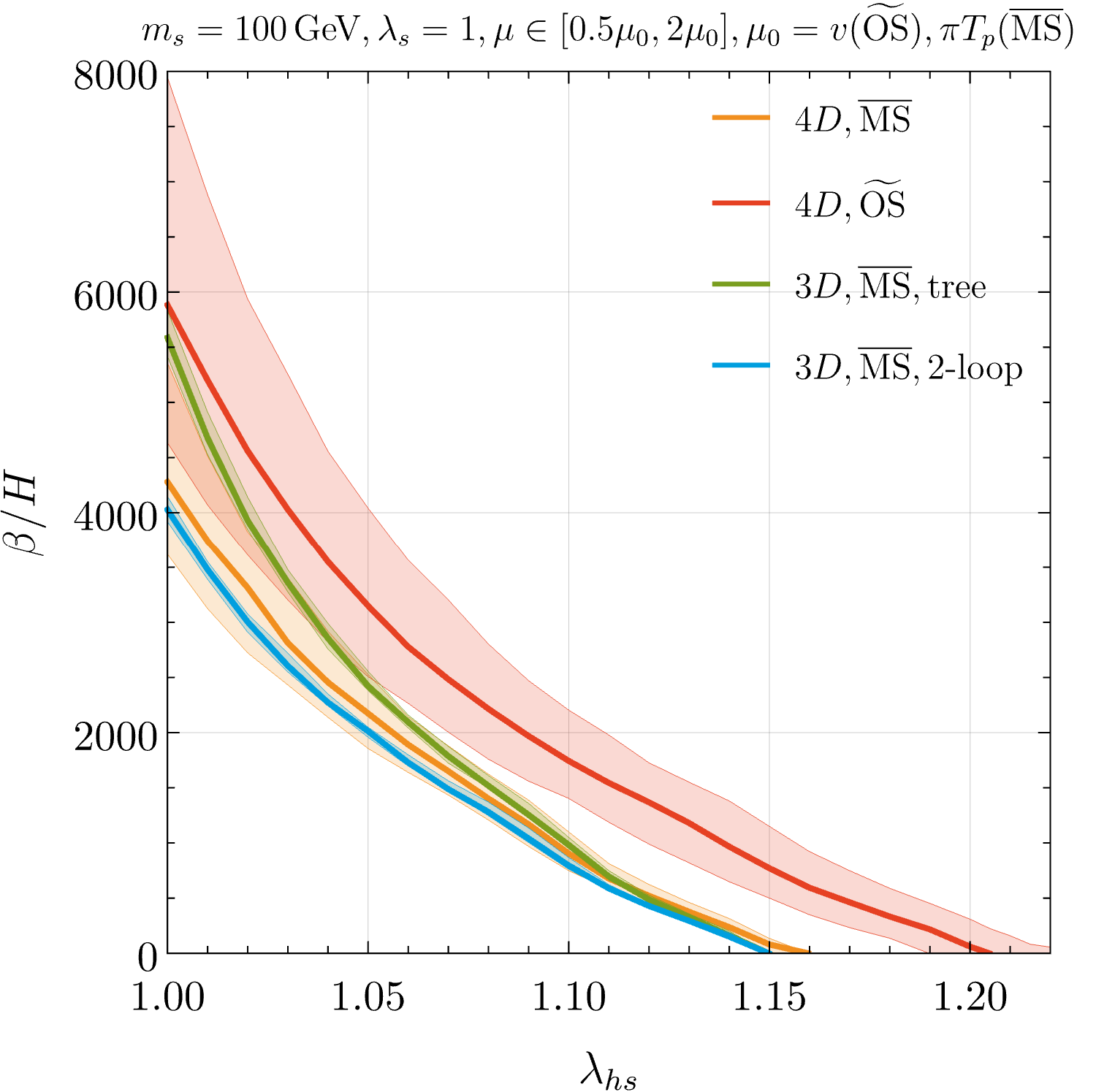}
    \includegraphics[width=0.32\textwidth]{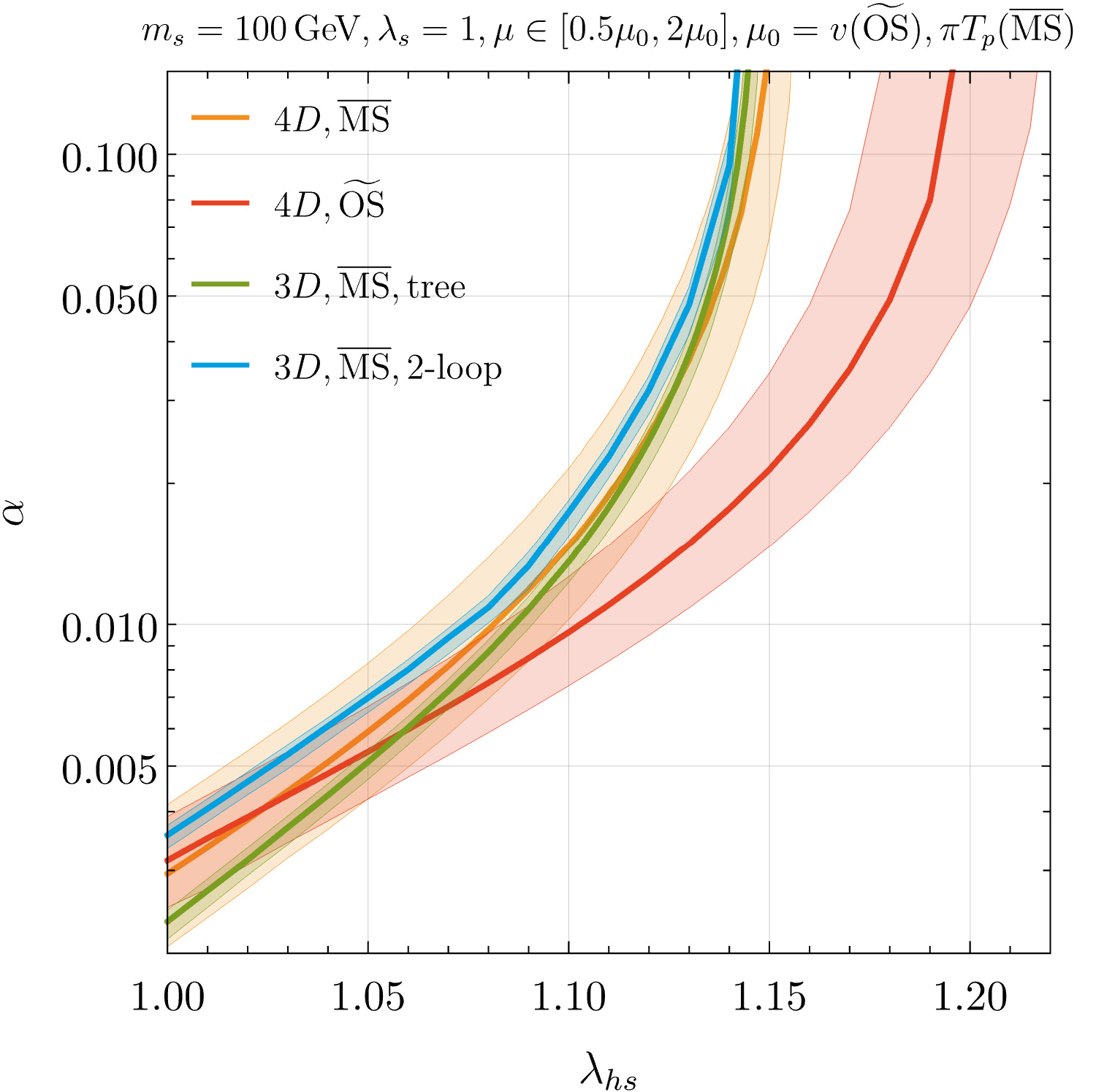}
    \caption{
    	Scale-dependence of the percolation temperature $T_p$ (\textit{left}), phase transition duration $\beta/H$ (\textit{center}) and strength $\alpha$ (\textit{right}) determined for $m_s=100\,\text{GeV}$, $\lambda_s=1$, $\lambda_{hs}\in[1,1.2]$ by direct potential minimization in the Landau gauge.
    	The bands represent the renormalization scale variation around the central values at $\mu = \pi T$ for the \msbar and $\mu = v$ for the \os renormalization schemes, respectively.
    	For the $3D$ approach, the variance of the matching scale
        is shown.
        }
    \label{fig:Tn4d3dmu}
\end{figure}

In \cref{fig:Tn4d3dmu}, we illustrate the renormalization scale dependence of the percolation temperature $T_p$ (left plot),
the inverse of the duration of the
phase transition
$\beta/H$ (middle plot)
and the strength of the
phase transition
$\alpha$ (right plot) calculated within the different approaches. 
As for the critical temperature, the scale dependence is considerably reduced in the $3D$ approach (blue) as compared to the $4D$ ones
($4D$-\msbar\ in orange and $4D$-\os\ in red)
for all three parameters.
We also observe that the $4D$-\msbar\ prediction
aligns more closely with the $3D$-\msbar\ prediction,
while the $4D$-\os\ approach predicts larger values
of $T_p$ (and consequently smaller values of $\alpha$)
for the largest values of $\lambda_{hs}$.
Here, one should take into account the different
choices for the central values $\mu_0$ in
the \msbar- and the \os-schemes, as discussed
already in \cref{sec:scale_dependence}.
For the $3D$ approach,
we show both results using the tree-level (green)
and the loop-improved (blue) potential in the
computation of the bounce action.
The predictions using the loop-improved
potential agree very well with the ones using
the tree-level potential.
This confirms that the loop corrections do
not lead to a significant shift in the predictions
for the parameters characterizing the phase transition,
even though we 
note that
using the loop-corrected potential in \cref{eq:bounceAction} is not a fully consistent loop improvement of the bounce action.

\begin{figure}[t]
\centering
\includegraphics[width=0.32\textwidth]{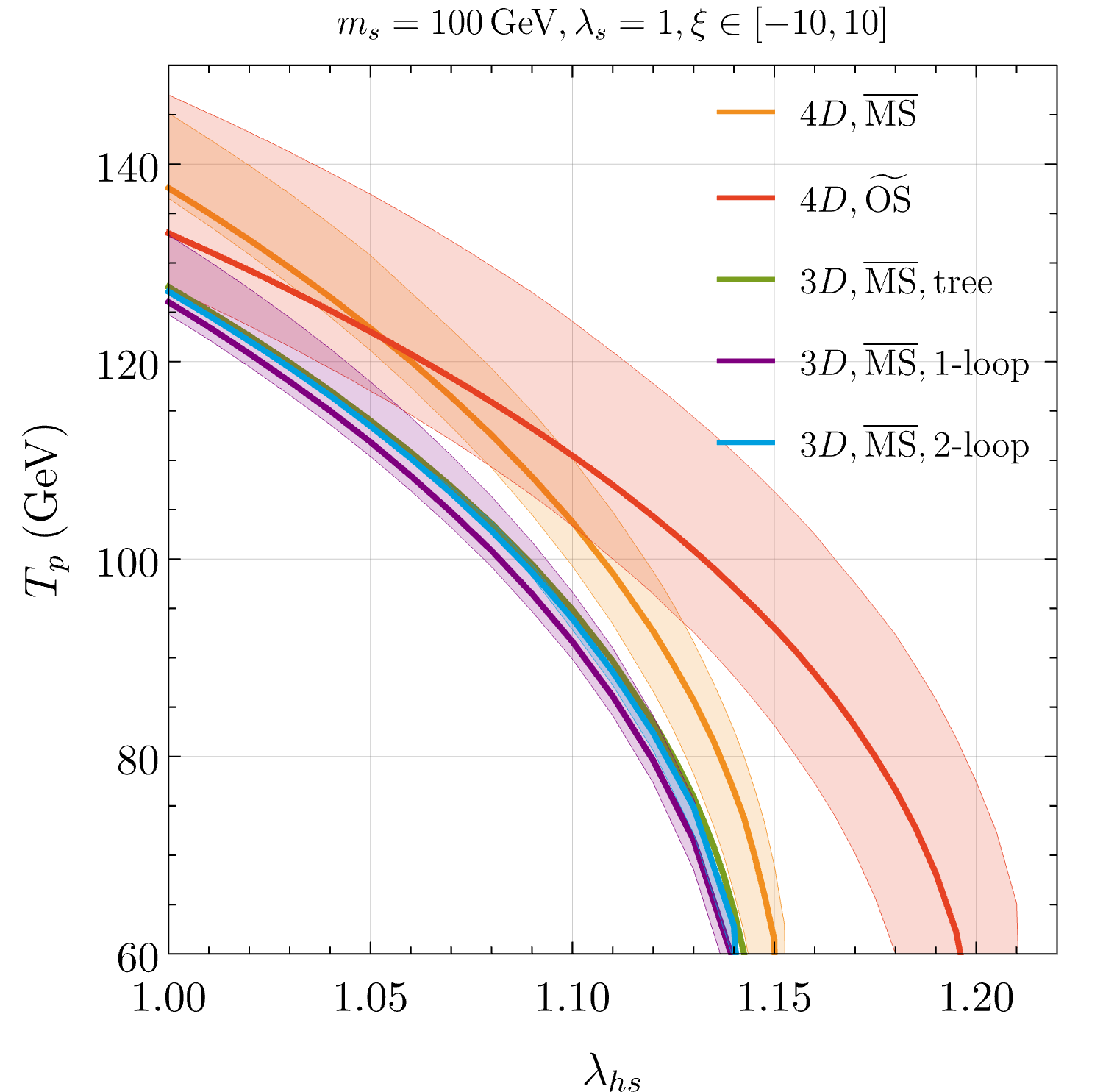}
\includegraphics[width=0.32\textwidth]{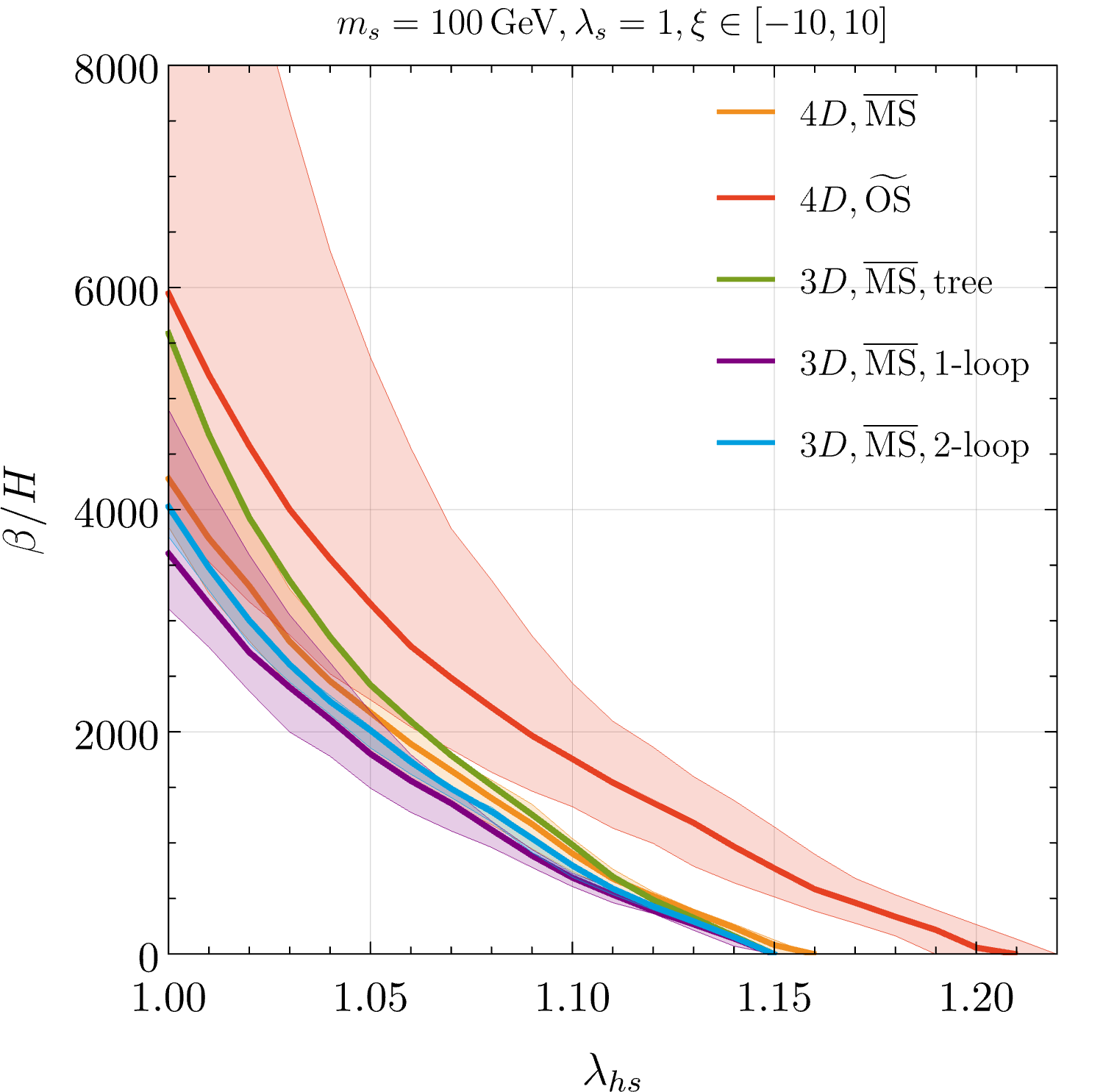}
\includegraphics[width=0.32\textwidth]{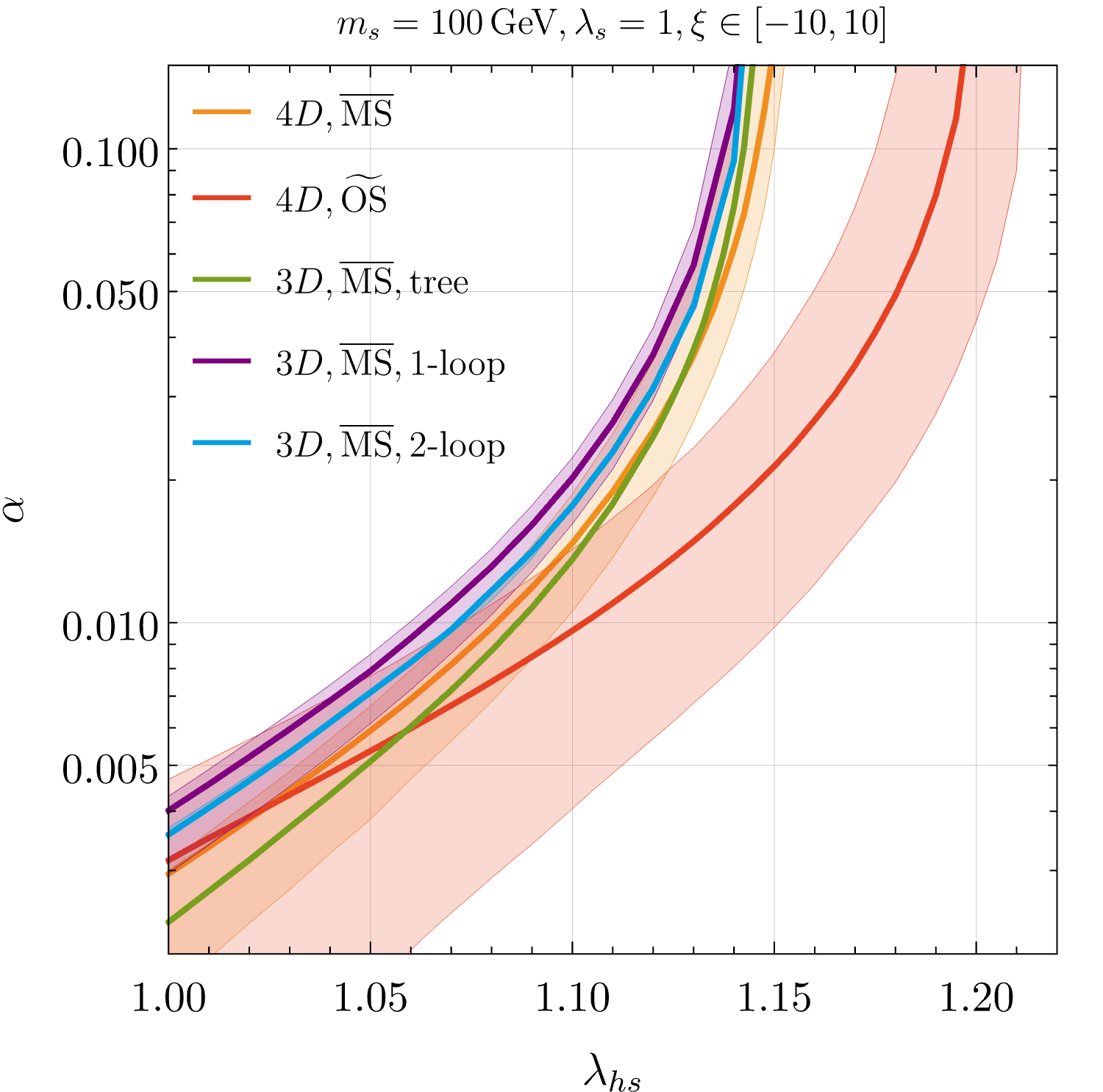}
\caption{Gauge-parameter dependence of the percolation temperature 
$T_p$ (\textit{left}), phase transition duration $\beta/H$ (\textit{center}) and strength $\alpha$ (\textit{right}) for $m_s=100\, \text{GeV}, \lambda_s=1, \lambda_{hs}\in[1,1.2]$ in $R_\xi$-gauge.
The bands 
indicate
the gauge parameter variation of $\xi \in [-10,10]$. 
}
\label{fig:Tn4d3dxi}
\end{figure}

Next, we analyze
the gauge dependence of $T_p$, $\beta/H$ and $\alpha$,
which is shown in \cref{fig:Tn4d3dxi}.
The results are similar to the ones we
observed for the scale dependence, with the
$3D$ EFT result exhibiting a substantially smaller
gauge dependence in all three parameters.
One can therefore expect that the predictions
for the GW signals will be significantly 
less affected by theoretical uncertainties
in the dimensionally reduced EFT compared
to the $4D$ approaches.
In particular,
the $4D$-\os approach exhibits a notably larger dependence on the gauge-fixing parameter compared
to the $4D$-\msbar and the $3D$ approaches.
The larger uncertainty from the gauge dependence
using the \os-prescription can be attributed to
additional uncanceled gauge dependencies
in the finite counterterms added
to the potential, which are derived from taking
derivatives of the one-loop piece of the
potential with respect to the fields
(see the discussion in \cref{sec:equilibrium}
and \cref{app:renormalization}).
As for the renormalization scale variation, we show (within the $3D$ EFT approach) 
results from the tree-level bounce (which are inherently gauge-independent) 
as well as
from the bounce calculation, where the tree-level potential is replaced by the effective potential.
This substitution introduces a gauge dependence of the nucleation observables, but we observe that the resulting shifts are of only moderate magnitude.
Similar to what we observed in \cref{sec:equilibrium}, we observe a gauge dependence reduction when transitioning from the one-loop to the two-loop effective potential.

\subsection{EW baryogenesis and the sphaleron washout criterion}
\label{sec:sphaleron_crit}

In addition to the discussion of gravitational wave production,
which is the main focus of our paper, in this section
we briefly comment on
the theoretical uncertainties in the determination of the order parameter that governs the sphaleron-induced washout of a baryon asymmetry that can be generated during a FOEWPT
catalysed by CP-violating interactions
(see Ref.~\cite{vandeVis:2025efm} for a recent review).
In the conventional treatment, a transition strong enough to avoid the washout is characterized by the criterion $\Delta v_h(T_c)/T_c \gtrsim 1$, where $\Delta v_h(T_c)$ is the difference of the Higgs field value between the two phases at the critical 
temperature~\cite{Dimopoulos:1978kv,Kuzmin:1985mm,Bochkarev:1990gb}. 
As already emphasized in \cref{sec:gauge}, and as follows from the Nielsen identity~\cite{Nielsen:1975fs,Fukuda:1975di,Patel:2011th}, this ratio is gauge-dependent: only the values of the effective potential at its minima are gauge-independent, while the extremizing field value is not.
Here we quantify the associated theoretical uncertainty by comparing the gauge dependence of $\Delta v_h(T_c)/T_c$ across the different perturbative approaches.
Moreover, we study the scale dependence of the predictions
for $\Delta v_h(T_c)/T_c$ by varying the renormalization scale
in the $4D$-approaches and the matching scale in the $3D$ approach.

\begin{figure}[t]
\centering
\includegraphics[width=0.48\textwidth]{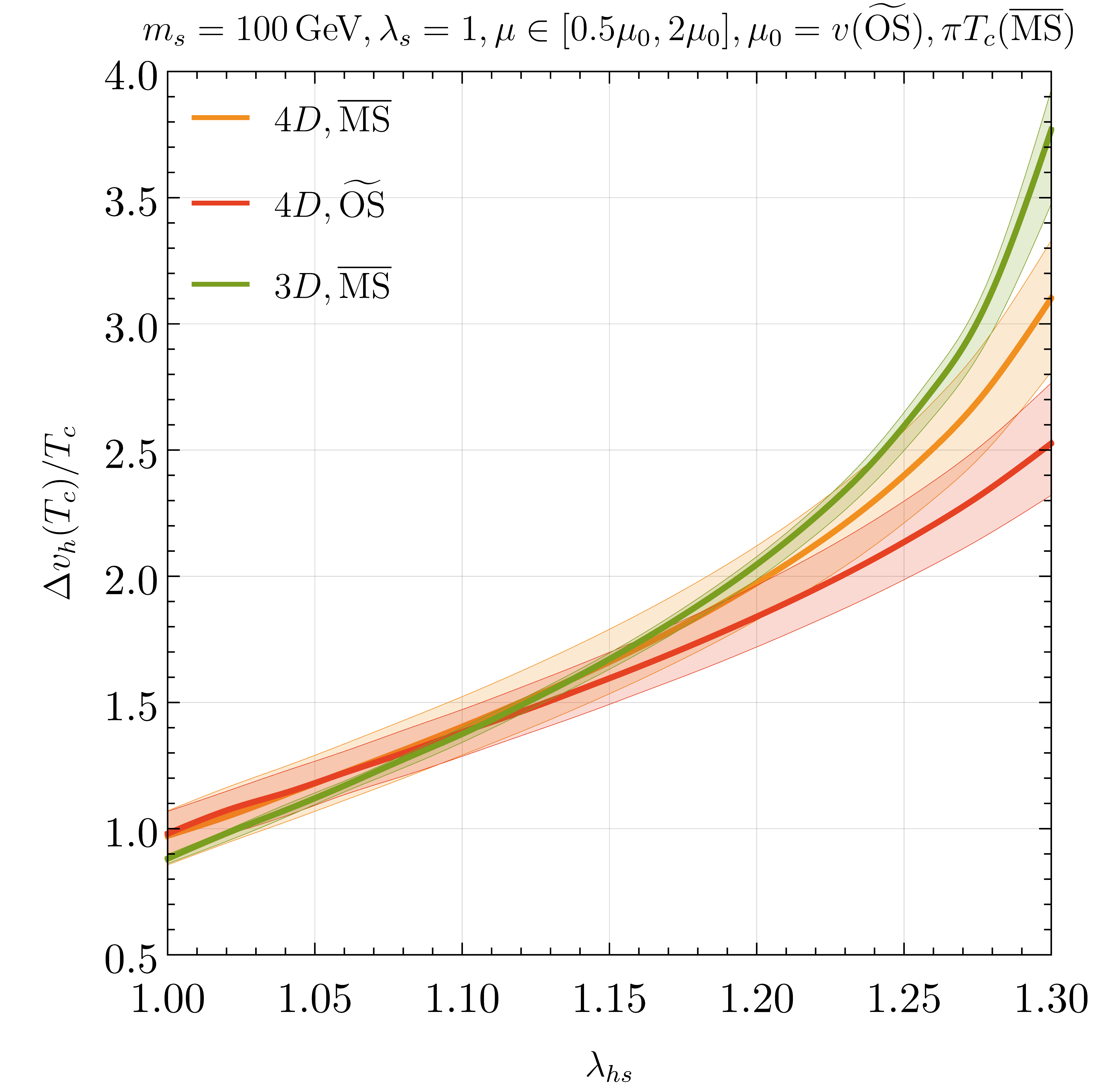}
~
\includegraphics[width=0.48\textwidth]{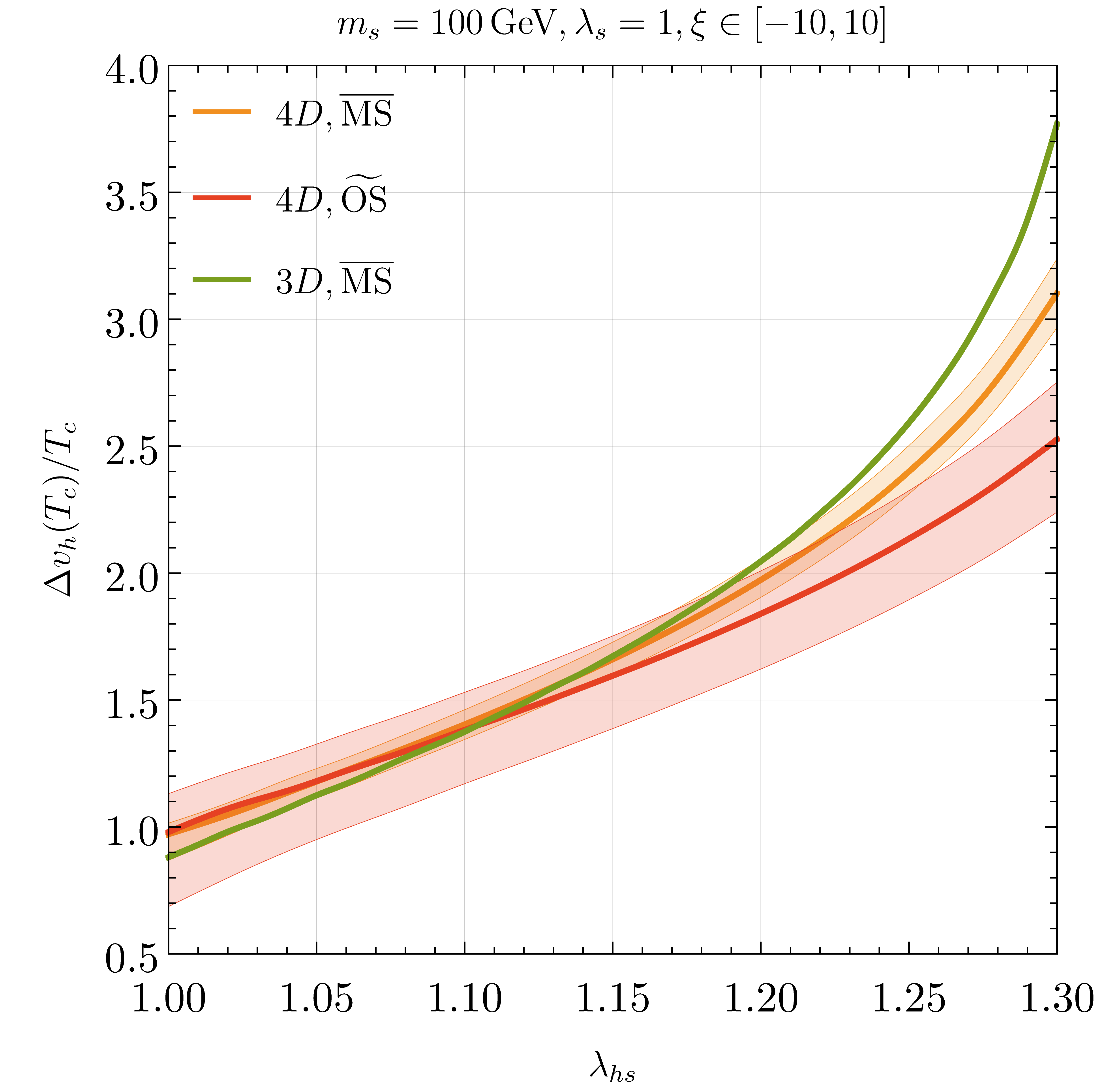}
\caption{Order parameter $\Delta v_h(T_c)/T_c$ relevant for the sphaleron washout criterion, determined for $m_s=100\,\text{GeV}$, $\lambda_s=1$ and $\lambda_{hs}\in[1,1.3]$ in the $4D$-\msbar, $4D$ \os\ and $3D$ approaches.
Renormalization-scale dependence, with the bands obtained from a variation of the scale by a factor of two around the central values $\mu = \pi T_c$ (\msbar) and $\mu = v$ (\os), and of then matching scale in the $3D$ approach, is shown on the \textit{left} panel. Gauge dependence, with the bands representing the variation of the gauge-fixing parameter $\xi \in [-10,10]$ in $\rxi$-gauge is shown in the \textit{right} panel.
In the $3D$ approach, the order parameter is extracted from the minimum of the leading-order effective potential of the dimensionally reduced theory and is gauge-independent.}
\label{fig:dvTc4d3d}
\end{figure}

The results are shown in \cref{fig:dvTc4d3d}. As in the analysis of the nucleation thermodynamics in \cref{sec:nucleationthermo}, in the $3D$ approach we extract the order parameter from the minimum of the leading-order (tree-level) potential of the dimensionally reduced theory, whereas in the $4D$-\msbar\ and $4D$-\os\ approaches, we minimize the loop-corrected effective potential (see \cref{sec:equilibrium}).
Across almost the entire $\lambda_{hs}$ range, all approaches predict $\Delta v_h(T_c)/T_c \gtrsim 1$, such that the qualitative conclusion of a strong
FOEWPT for a given value of $\lambda_{hs}$ is relatively robust against both the scale dependence (left plot)
and the gauge dependence (right plot). The precise value
of $\Delta v_h(T_c)/T_c$, however, differs substantially between the
approaches for the strongest transitions: at  $\lambda_{hs} = 1.3$ we find
$\Delta v_h(T_c)/T_c \approx 3.8$ in the $3D$ approach, while the $4D$-\msbar\
and $4D$-\os\ approaches yield $\approx 3.1$ and $\approx 2.5$, respectively.
This spread is comparable to the one observed for the latent heat in \cref{sec:equilibrium}.

The uncertainty bands from the renormalization-scale dependence are shown in the left panel of
\cref{fig:dvTc4d3d}.
As for the other thermodynamic quantities discussed above, the residual scale dependence is reduced in the $3D$ approach relative to the $4D$-\msbar\ and $4D$-\os\ results.
We note that the $3D$ band originates solely from the variation of the matching scale, whereas in the $4D$ approaches, the scale dependence enters directly through the loop-corrected potential used in the minimization.

The gauge dependence is shown in the right panel of \cref{fig:dvTc4d3d}. The
$4D$-\os\ approach exhibits the largest $\xi$-dependence, 
which, as discussed in
\cref{sec:equilibrium,sec:nucleationthermo}, can be traced back to the additional uncanceled gauge dependence of the finite counterterms (see \cref{app:renormalization}), while the $4D$-\msbar\ approach shows a smaller removed part
gauge dependence that grows towards stronger transitions.
The $3D$ result, in contrast, is manifestly gauge-independent: the order parameter is extracted from the minimum of the tree-level $3D$ potential, whose parameters follow from the gauge-invariant dimensional reduction,
and thus independent of $\xi$ by construction.
This mirrors the gauge-independence of the tree-level bounce action discussed in \cref{sec:nucleationthermo}. 
The $3D$ approach thus provides a gauge-invariant measure of the transition strength, whereas the gauge dependence of the conventional $4D$ criterion constitutes an inherent theory uncertainty of the washout condition.

We stress, finally, that $\Delta v_h(T_c)/T_c$ is only a proxy for the physical,
gauge-invariant washout condition, which is properly formulated in terms of the
broken-phase sphaleron rate~\cite{Patel:2011th,Li:2025kyo}. 
Within the $3D$ EFT, the leading-order sphaleron action is proportional to the gauge-invariant condensate at the minimum of the tree-level $3D$ potential~\cite{Li:2025kyo}. 
As for the nucleation observables in \cref{sec:nucleationthermo}, this leading-order result can be systematically improved by including higher-order corrections within the $3D$ EFT perturbation theory, in analogy to the loop corrections to the bubble nucleation rate.
A dedicated gauge-invariant computation of the sphaleron rate for the cxSM
that include higher-order corrections along these lines
is beyond the scope of the present work.

\section{Gravitational wave signals}
\label{sec:gravwave}

Gravitational waves (GW) produced during a cosmological first-order phase transition might
provide a crucial probe of early universe dynamics
in the
future~\cite{Caprini:2015zlo,Caprini:2019egz}.
Among the primary mechanisms generating these signals, sound waves in the plasma play a dominant role in our model, which does not exhibit strongly supercooled phase transitions, rendering GW production from
runaway bubble
collisions irrelevant.
An additional source of GW production is
magnetohydrodynamic turbulence in the plasma.
Its contribution to the GW signal in an EW phase transition with weak or intermediate strength
of $\alpha \lesssim 0.1$ (see \cref{fig:parscanGWmudep} below) is subdominant compared to the sound wave
contribution~\cite{Auclair:2022jod}.
Moreover, the resulting spectral shape of the
GW signal 
caused by the turbulence
depends on the fraction of energy
converted into turbulent motion in the plasma,
which has not yet been determined robustly
in numerical simulations and would
have to be treated as an additional
free parameter~\cite{Caprini:2024hue}.
Therefore, we do not
include this contribution in our analysis in order to keep
the analysis focused on the theory uncertainty
resulting
from the perturbative description of the SFOEWPT.

\subsection{Spectral density}
\label{sec:gwspec}
The spectral energy density of
GWs
from sound waves
can be approximated as \cite{Hindmarsh:2017gnf,Caprini:2019egz,Guo:2020grp} 
\begin{equation}
    \Omega_{\text{GW}}(f) h^2 \approx \Omega_{\text{GW},\text{peak}} h^2 \left( \frac{f}{f_{\text{peak}}} \right)^3 \left( \frac{7}{4 + 3 (f / f_{\text{peak}})^2} \right)^{7/2},
\end{equation}
where $f_{\text{peak}}$ is the peak frequency of the GW spectrum today, given by  
\begin{equation}
    f_{\text{peak}} \approx 26\,\mu\text{Hz} \times \left( \frac{1}{H_* R_*} \right) \left( \frac{T_*}{100 \,\text{GeV}} \right) \left( \frac{g_*}{100} \right)^{1/6}.
\end{equation}
The peak amplitude of the GW signal today can be approximated by 
\begin{equation}
    \Omega_{\text{GW},\text{peak}} h^2 \approx  4.1\times10^{-7} R_* H_*\left(1-\frac{1}{\sqrt{1+2\tau_{\text{sw}}H_*}}\right) \times \left(\frac{\kappa_{\text{sw}}\alpha}{1+\alpha}\right)^2\left(\frac{100}{g_*}\right)^{1/3},
    \label{eq:peakampli}
\end{equation}
where $T_*$ is the temperature of the plasma after the phase transition.
It
is related to the percolation temperature as $T_* = T_p(1+\alpha)^{1/4} $ and 
takes into account
reheating effects in the early universe. 
Given that $T_* \approx T_p$,
the inverse
phase transition duration $\beta / H$ and strength
$\alpha$ evaluated at the percolation temperature $T_p$
can be used for computing the GW signals.
Moreover,
$\kappa_{\text{sw}}$ is the efficiency factor
for the transformation of vacuum energy released
during the transition into bulk 
kinetic energy,
which we determine as a function of $\alpha$
from numerical fits given in Ref.~\cite{Espinosa:2010hh}.
The mean bubble radius is given by
$R_*H_* \approx (8\pi)^{1/3}(\beta/H)^{-1}$, assuming $v_{w}=1$ as discussed above,
and the sound wave duration 
is related to the  root-mean-square four-velocity
of the plasma $\bar{U}_f$ via
$\tau_\text{sw}H_* = R_*H_*/\bar{U}_f$,
with $\bar{U}_f \approx \sqrt{3\alpha\kappa_{\text{sw}} /4(1+\alpha)}$~\cite{Ellis:2020awk}.

\begin{figure}[t]
\centering
\includegraphics[width=0.48\textwidth]{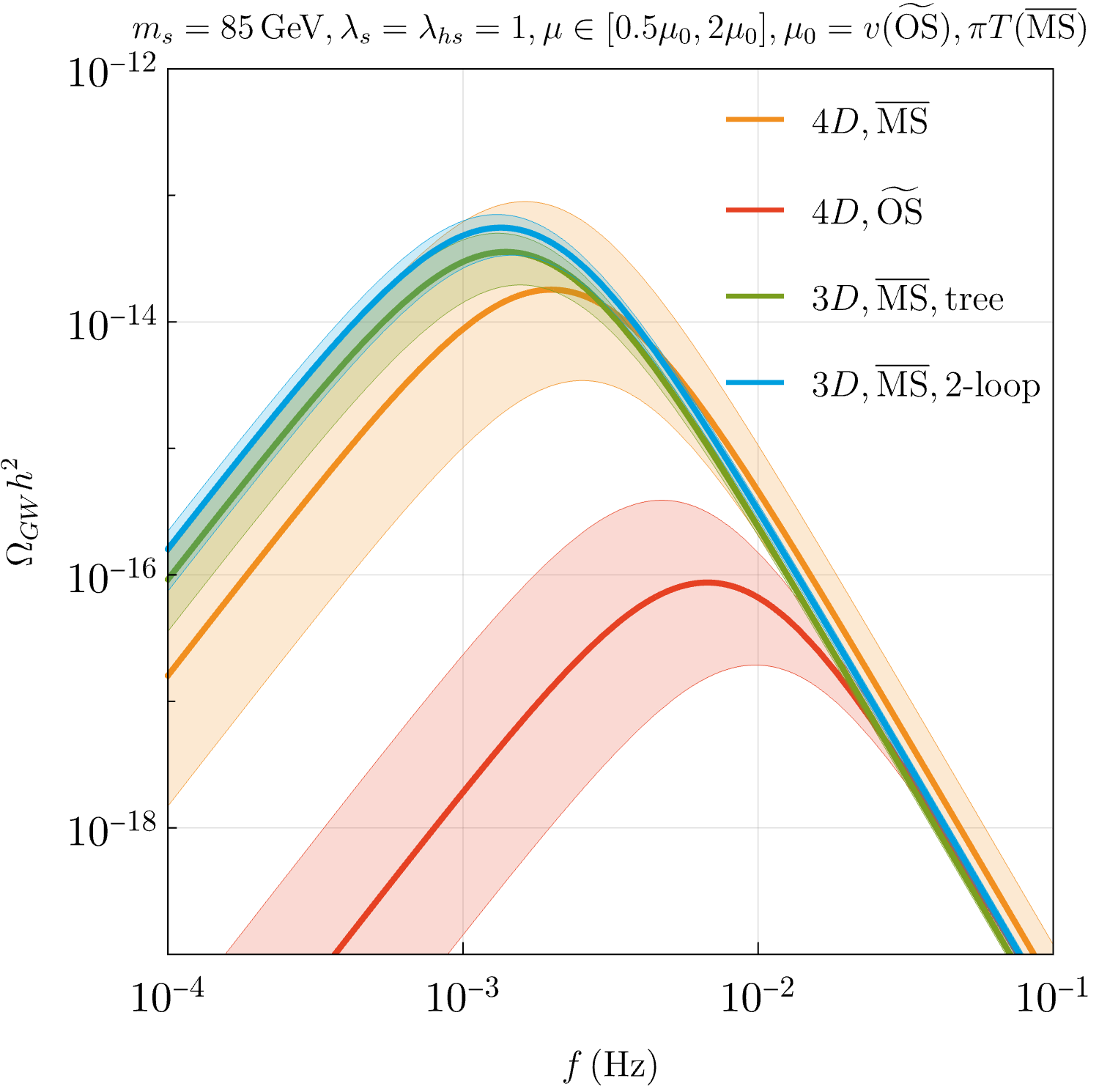}
\includegraphics[width=0.48\textwidth]{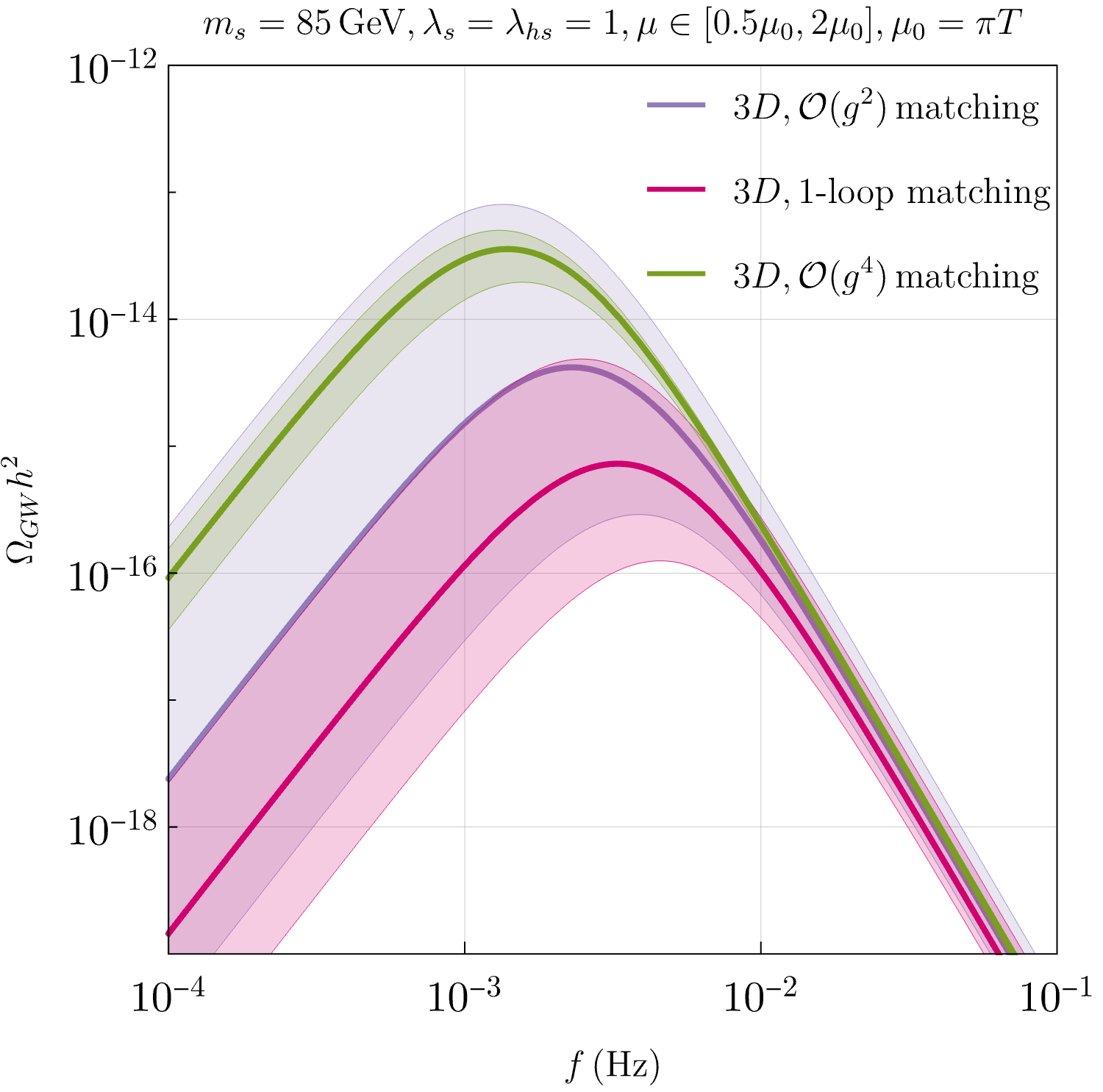}
\caption{
Gravitational wave spectrum for $m_s=85\,\text{GeV}$, $\lambda_s=\lambda_{hs}=1$ evaluated in the Landau gauge.
The bands 
indicate the
renormalization scale variation around the central value $\mu = \pi T$ and $\mu = v$ for the \msbar and \os renormalization schemes, respectively.
For the $3D$ approach, the variance of the hard matching 
scale is shown.
In the \textit{left} panel, we compare the $4D$ and $3D$ approaches, and the \textit{right} panel shows the 
predictions for
different matching orders in the $3D$ approach.}
\label{fig:GWspectrum-mu}
\end{figure}

In the same manner as in the previous sections, we compare different methods for the determination of $\Omega_{\text{GW}}(f) h^2$ in terms of theory
uncertainties from residual scale dependence
and gauge dependence.
In the left plot of
\cref{fig:GWspectrum-mu} we show the
predicted GW signals using the different approaches
introduced in \cref{sec:equilibrium}, where
the shaded bands indicate the theoretical uncertainty
resulting from a variation of the renormalization
scale in the $4D$ approaches and the hard matching
scale in the $3D$ approach. The solid line indicates
the prediction for the central scale choices
as shown on top of the plots.
All results shown are obtained using the Landau gauge,
and we show two predictions for the $3D$-\msbar\
approach using the tree-level effective potential 
(green) or the two-loop level potential (blue)
for the determination of the bounce solution.
The differences between these two predictions
are found to be within the uncertainty resulting
from the scale variation.
In line with the observations from
\cref{sec:nucleationthermo},
we find that the scale dependence of
the predicted GW signals is significantly smaller
using the dimensionally reduced EFT compared
to the $4D$ approaches.
While the $4D$-\msbar\ approach predicts a GW
signal that within its uncertainties is in
agreement with the predictions from the
$3D$ EFT, the $4D$-\os\ approach predicts
a GW signal with a peak amplitude that is roughly
three orders of magnitude smaller, and the
peak frequencies differ by roughly an order
of magnitude.

The discrepancy between the \os\
and the \msbar\ results mainly 
arises
from the
small parameter shifts between the physical input and Lagrange parameters using $\os$ vs.\ \msbar relations  together with the strong parametric dependence of the gravitational wave spectra 
(see e.g.\ Refs.~\cite{Biekotter:2022kgf,Biekotter:2023eil}).
While this discrepancy between the different
renormalization prescriptions appears alarming
at the level of individual benchmark points,
it is important to note that GW predictions
from SFOEWPTs are in many cases extremely sensitive to the
underlying model parameters.
As a result, when comparing broader regions of
parameter space with the goal of identifying
regions of parameter space that allow for potentially
detectable GW signals, the differences between the
approaches are less severe, since the parametric
dependence of the GW signal tends to dominate over
the scheme-related discrepancies
(see \cref{sec:parscans} for further discussion).

In the right plot of \cref{fig:GWspectrum-mu}
we show the relevance of the matching order within the $3D$ approach.
As in the left plot, the shaded uncertainty
bands result from a variation of the matching
scale, while the solid lines indicate the
predictions for the central scale choice,
and all results were obtained in the Landau gauge.
The violet, magenta and green bands belong
to $\mathcal{O}(g^2)$, one-loop and
$\mathcal{O}(g^4)$ matching accuracy, respectively.
In line with the predictions for $T_c$ at different
matching orders as shown in \cref{fig:TcScaleplot2},
we observe a significant reduction of the scale
dependence at $\mathcal{O}(g^4)$ matching accuracy
also for the GW signal,
whereas a matching at lower accuracy does not
improve the theoretical uncertainty from the scale
dependence compared to the $4D$ approach. This
once again indicates the importance of the full $\mathcal{O}(g^4)$ matching in the
dimensional reduction approach.

\begin{figure}[t]
\centering
\includegraphics[width=0.48\textwidth]{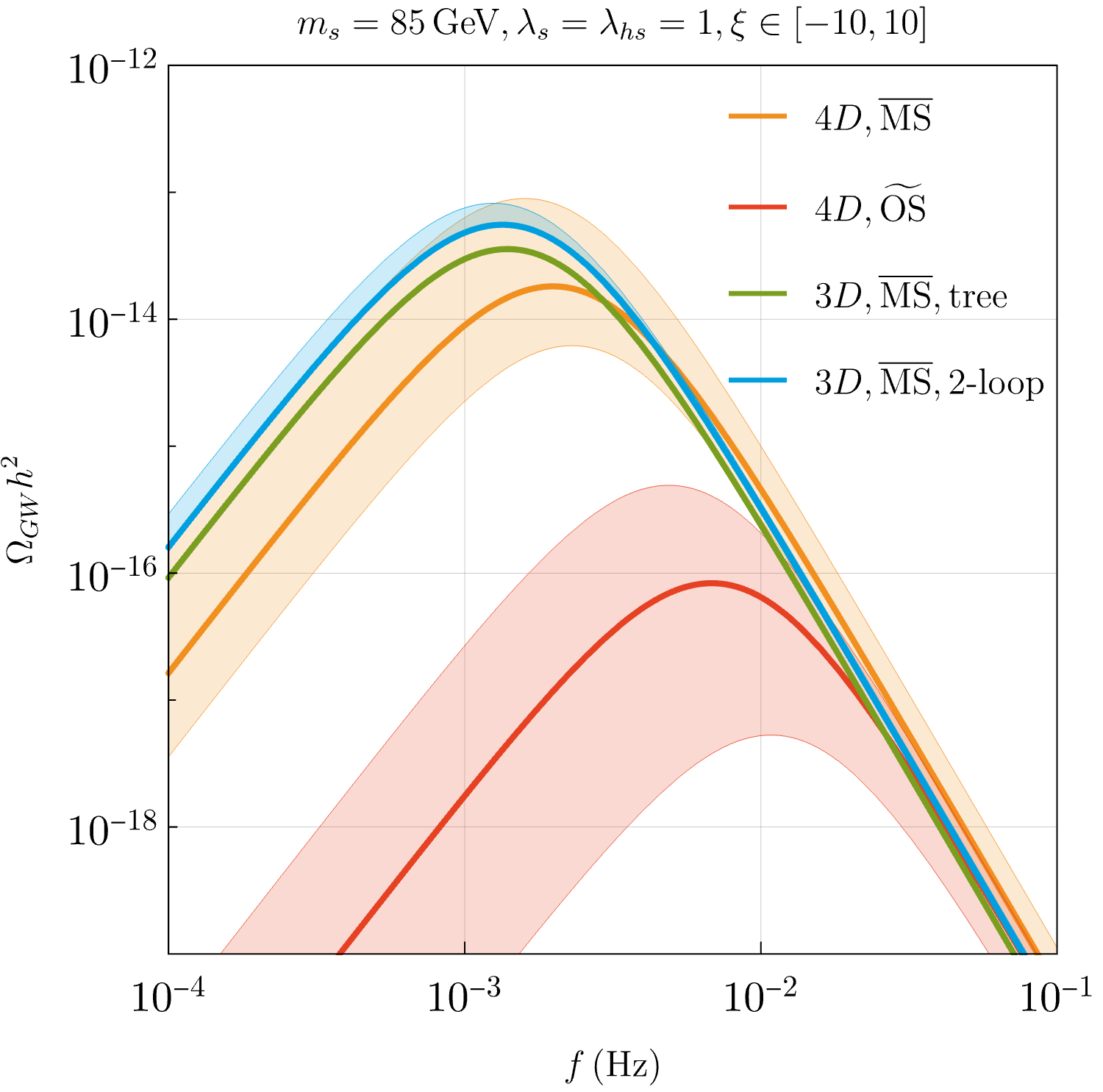}
\includegraphics[width=0.48\textwidth]{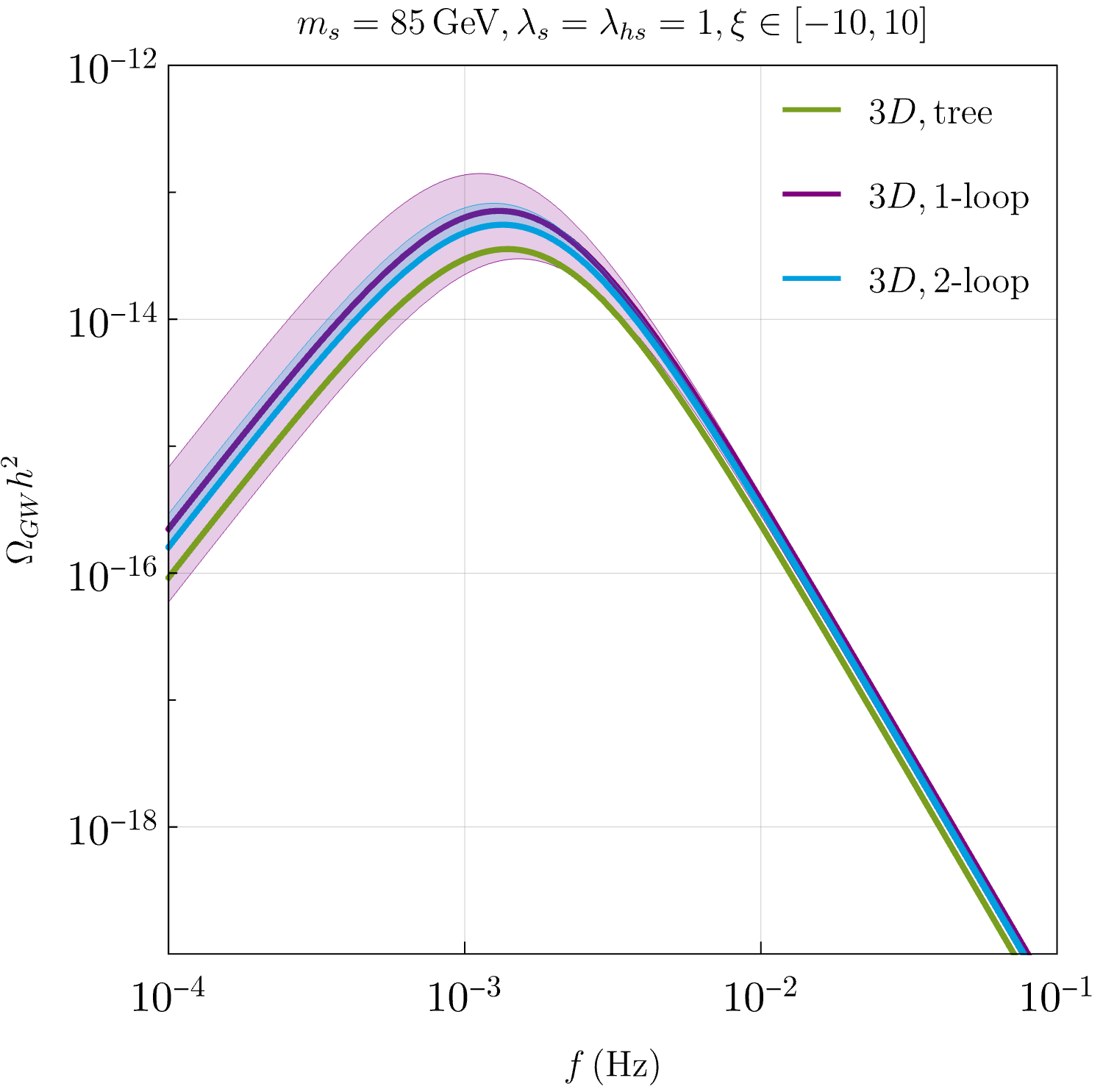}
\caption{
Gauge-parameter variation of $\xi \in [-10,10]$ in the gravitational wave spectrum for $m_s=85\,\text{GeV}$, $\lambda_s=\lambda_{hs}=1$ in $\rxi$-gauge.
In the \textit{left} panel, we compare the $4D$ and $3D$ approaches.
The \textit{right} panel shows a comparison of including different loop orders of the effective potential in the $3D$ approach for the bounce calculation.
}
\label{fig:GWspectrum-xi}
\end{figure}

We now turn to the residual gauge dependence
in the predictions for the GW signal.
Within the $3D$ approach, the bounce can be determined without the spurious insertion of the loop-improved effective potential by simply using the $3D$ tree-level potential, which yields inherently gauge-independent results. 
Nevertheless, we can assess the residual gauge dependence
by studying the impact
of substituting the loop-improved potential within the dimensionally reduced theory to determine the bounce action
and varying the gauge-fixing parameter $\xi$.
In the left plot of \cref{fig:GWspectrum-xi}
we show the predictions for the GW signal using
the different approaches (the colors are defined
as in the left plot of \cref{fig:GWspectrum-mu}).
We here used the background $R_\xi$-gauge in all
cases, and the shown uncertainty bands arise
from a variation of the gauge fixing parameter
in the interval $\xi \in [-10, 10]$, while the
solid lines indicate the predictions in the
Landau gauge with $\xi = 0$.
In a similar fashion to the scale dependence
discussed above, also the residual gauge dependence
is significantly reduced using the $3D$ approach
compared to the $4D$ approach.
The gauge-dependent $3D$ result using the
two-loop level potential for the computation
of the bounce solution is very close to
the inherently gauge-independent result using the
tree-level potential. The largest theory uncertainty
from the residual gauge dependence is again
observed 
for
the $4D$-\os\ approach.

We further scrutinize the residual gauge dependence
in the dimensionally reduced EFT by showing
in the right plot of \cref{fig:GWspectrum-xi} the
predicted GW signals using the tree-level (green),
one-loop level (purple) and two-loop level (blue)
effective potentials for the computation of
the bounce solutions.
It is interesting to note that only the
GW signal predicted using the two-loop level potential
shows a very mild gauge dependence, whereas
the gauge dependence of
the GW signal predicted using the one-loop level
potential is substantially more pronounced.
Moreover, as already visible in the left plot
of \cref{fig:GWspectrum-xi}, the results using
the two-loop level potential are in good agreement
with those obtained using the tree-level potential.
This comparison demonstrates that the numerical impact of using the two-loop-corrected potential is
small and the induced gauge dependence remains very mild.
Therefore, in the following phenomenological
discussion in \cref{sec:parscans}, we use
the gauge-independent
tree-level potential in the $3D$ approach,
where loop corrections are accounted for in the EFT
by integrating out the dominant
thermal fluctuations,
providing a fully consistent computation of the bounce
action.

\subsection{Peak signal strengths in the \texorpdfstring{$(m_s,\lambda_{hs})$-plane}{(mS,lambdahs)-plane} }
\label{sec:parscans}

In this section, we will identify and analyze more broadly the
parameter space regions of the cxSM that can realize
a SFOEWPT, and we quantify the theoretical uncertainty
of the peak amplitude of the predicted GW signals.
As was discussed in \cref{sec:cxsm},
the cxSM has three independent BSM parameters: the
singlet mass $m_s$, the singlet self-coupling $\lambda_s$,
and the portal coupling $\lambda_{hs}$.
While before we fixed $m_s$ and $\lambda_s$ and showed
the predictions for the parameters describing the SFOEWPT
as a function of $\lambda_{hs}$, we now scan parameter planes
with varying $m_s$ and $\lambda_{hs}$,
while $\lambda_s = 1$ is kept fixed.

\begin{figure}
\centering
\includegraphics[width=0.48\textwidth]{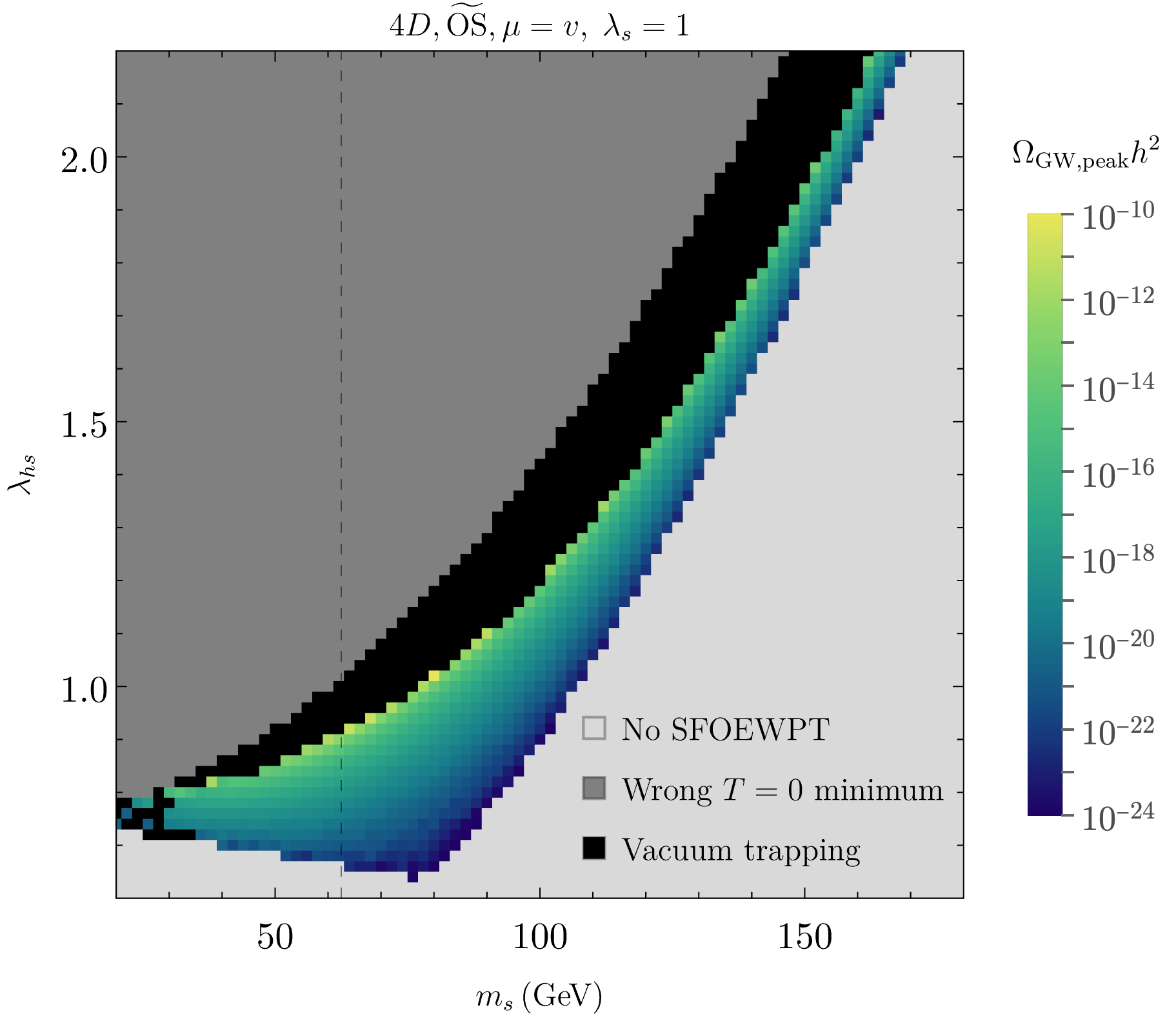}
\includegraphics[width=0.48\textwidth]{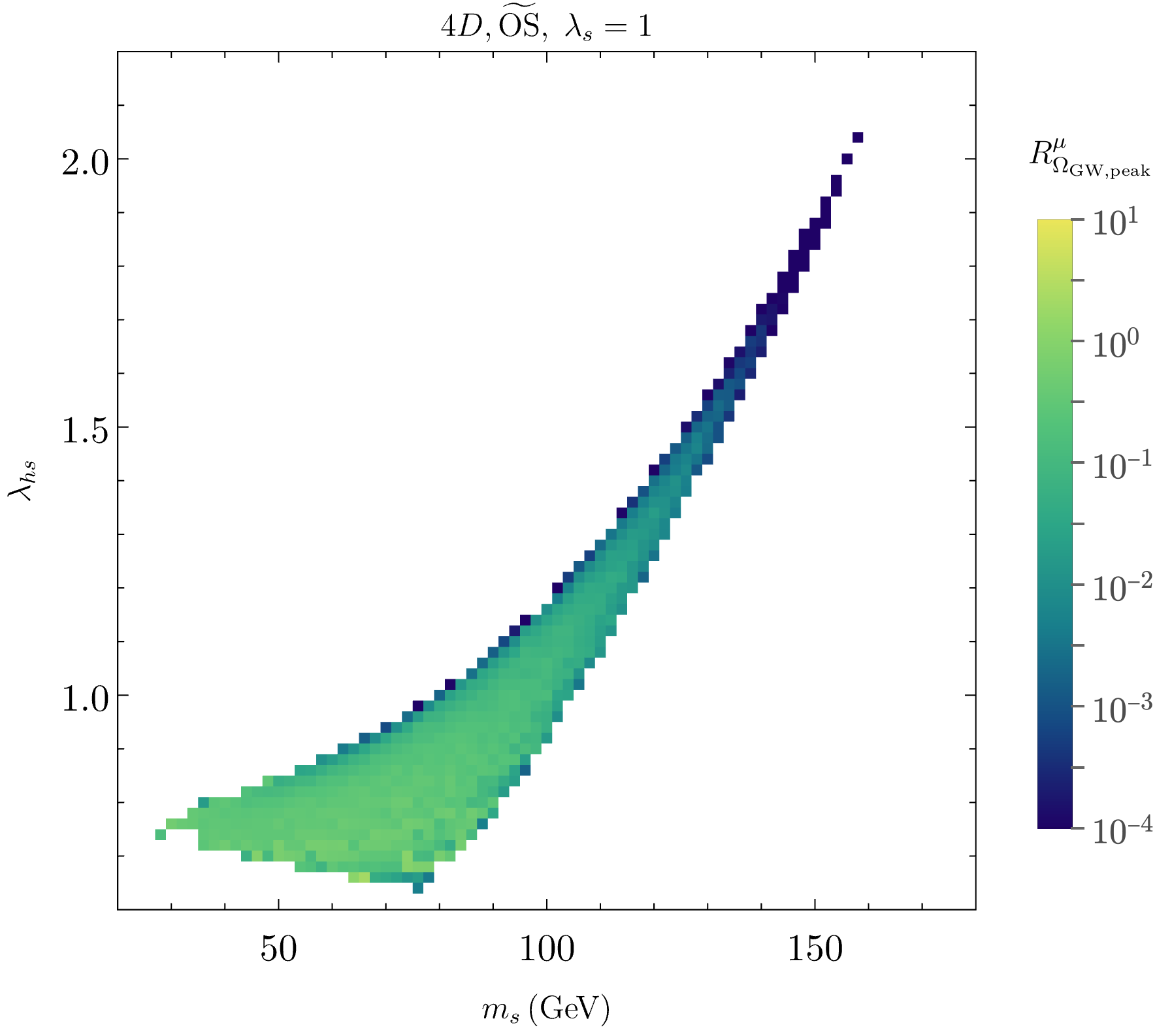}
\includegraphics[width=0.48\textwidth]{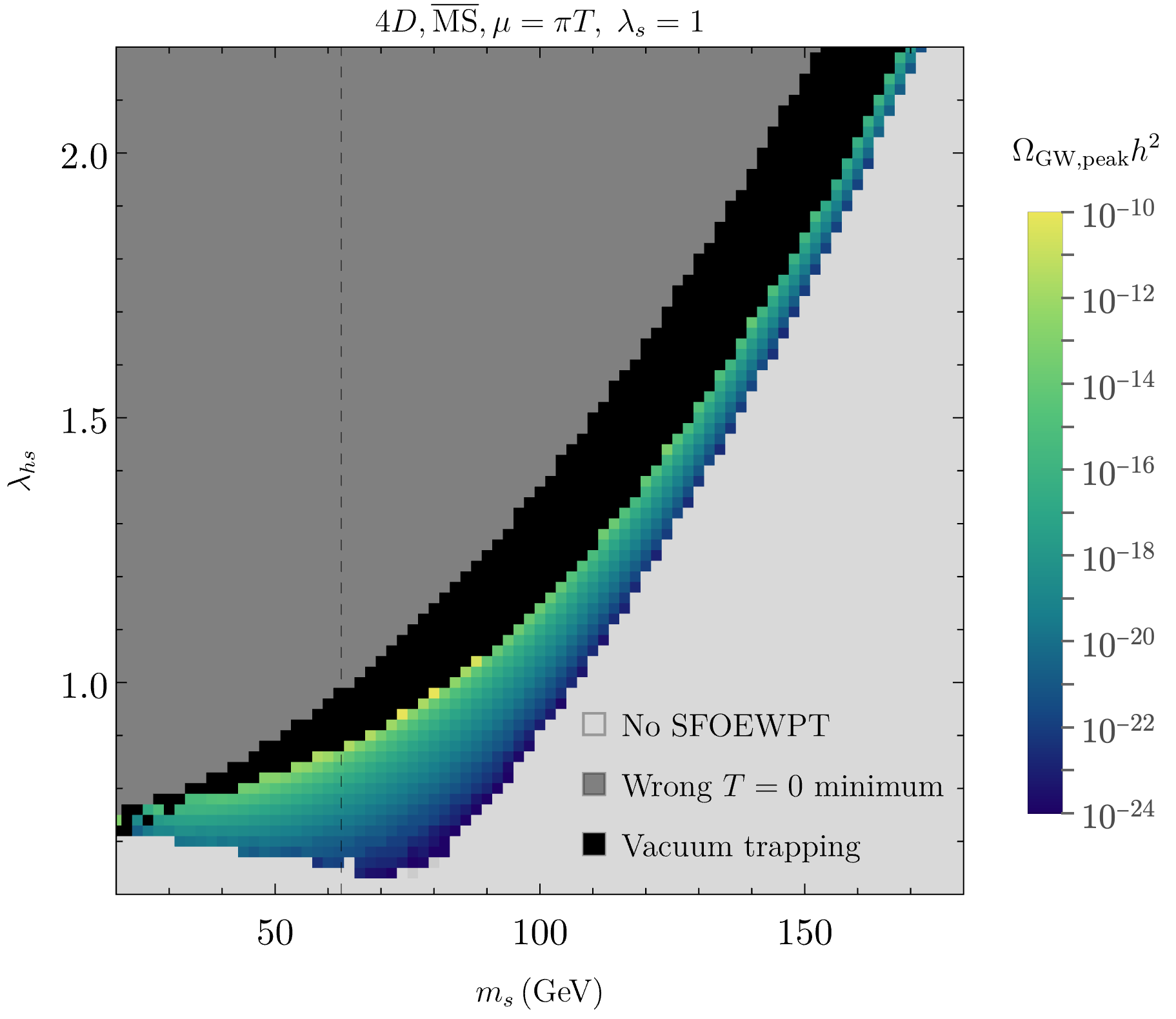}
\includegraphics[width=0.48\textwidth]{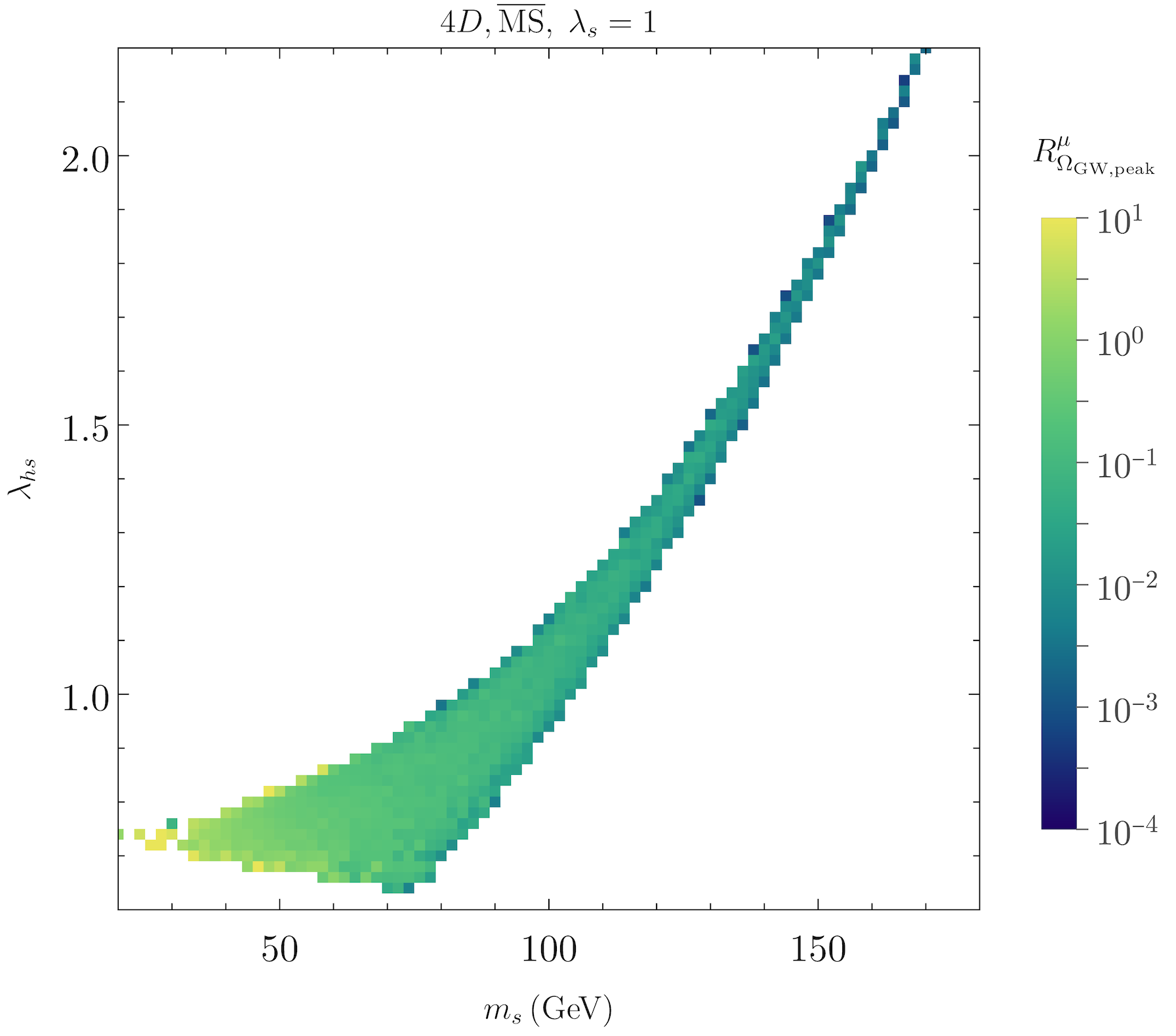}
\includegraphics[width=0.48\textwidth]{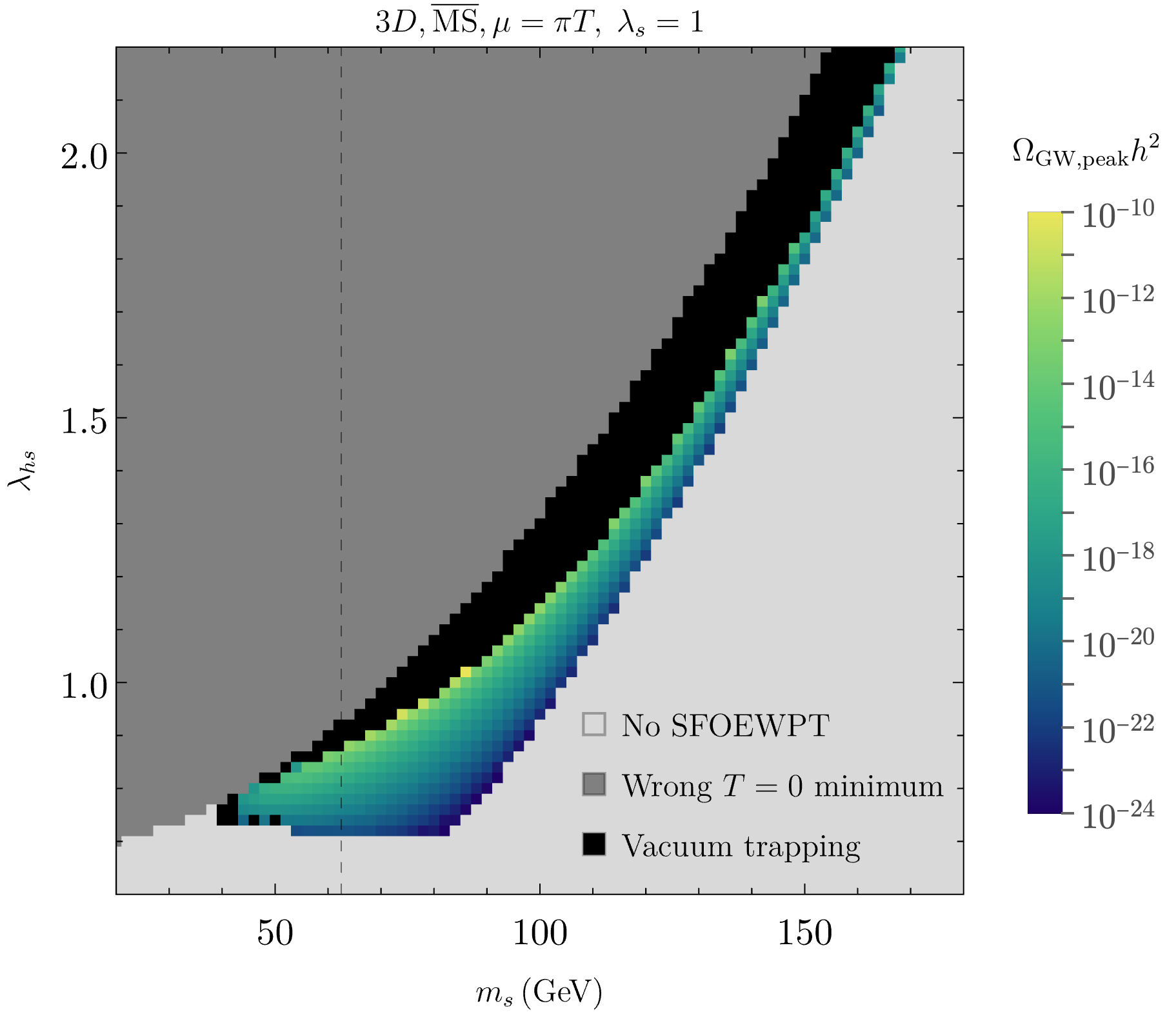}
\includegraphics[width=0.48\textwidth]{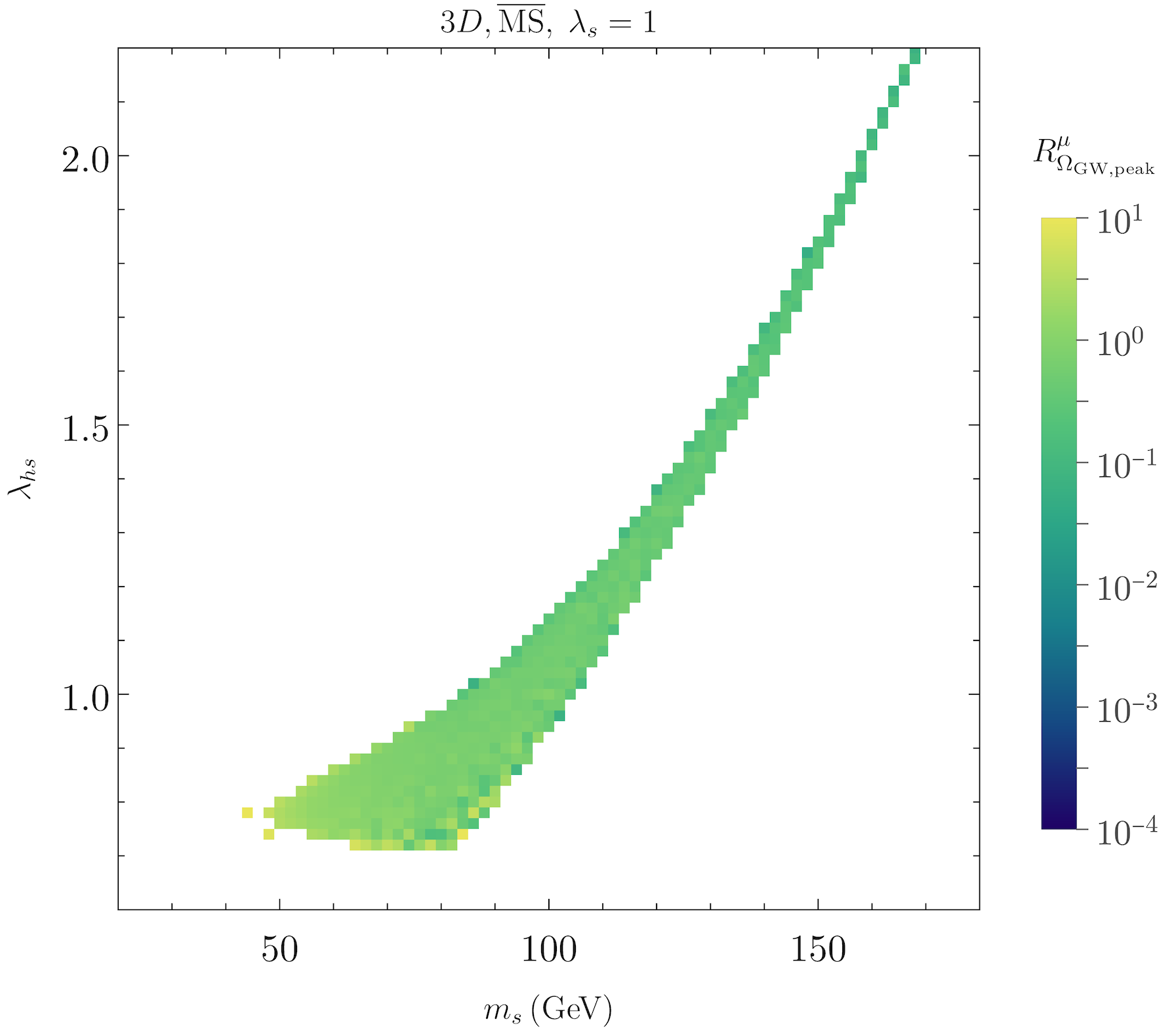}
\caption{
Gravitational wave signal peak strength (\textit{left}) and the corresponding scale variation ratio of \cref{eq:Rmu} (\textit{right}) for the $4D$-\msbar, $4D$-\os and $3D$-\msbar approaches in the Landau gauge.
The dark gray region represents points for which no electroweak phase transition occurs, 
the light gray region represents points without a \emph{strong} first-order electroweak phase transition
(with $\Delta v(T_c) / T_c < 1/2$)
and the black region represents points with vacuum trapping.
The vertical dashed line at $m_s= 62.5 \text{ GeV}$ indicates the region where the invisible Higgs decay channel opens.}
\label{fig:parscanGWmudep}
\end{figure}

In the left column of \cref{fig:parscanGWmudep}, we show
the $\left(m_s , \lambda_{hs} \right)$ parameter plane for
$\lambda_s = 1$, and the three different plots correspond to the
three different methods employed in the previous sections:
$4D$-\os (top), $4D$-\msbar (middle) and $3D$-\msbar (bottom).
The color coding in these plots is as follows:
The light-gray region indicates either weak FOEWPTs with
$\Delta v(T_c) / T_c < 0.5$ or no FOEWPT at all.
The dark-gray region indicates where the minimum giving rise to the singlet vev is the global minimum through the whole thermal evolution of the universe (down to $T=0$),
such that no EW phase transition can
occur. 
The black region indicates a vacuum
trapping~\cite{Biekotter:2021ysx},
where the EW vacuum corresponds to the global minimum at
zero temperature, but the conditions for the onset of
a FOEWPT are never satisfied during the thermal history,
and the universe remains trapped in the false minimum
in which only the singlet field has a vev.\footnote{Such a region
can become viable and feature a FOEWPT once seeded phase
transitions from topological defects, such as from domain
walls or cosmic strings
(e.g.~generated when the singlet field
obtains a vev~\cite{Blasi:2022woz})
are considered.
We do not include
this possibility in our discussion.}
Finally, the colored region indicates FOEWPTs
with $\Delta v(T_c) / T_c \geq 0.5$, where the color coding
shows the value of the peak amplitude of the GW signal
produced during the transition.
Comparing the three plots,
we find that the overall shape and structure of the regions in parameter space where a
FOEWPT is predicted to be very similar
for the different approaches.
Also, the possible range of the peak amplitudes
$\Omega_{\rm GW, peak} h^2$ that we find are similar.
The fact that the regions of parameter space predicting the
strongest FOEWPTs and the corresponding range of peak
amplitudes for the GW signal remain very similar across
the different approaches, despite the large discrepancies observed
at the level of individual parameter points
as shown in \cref{fig:GWspectrum-mu}
and \cref{fig:GWspectrum-xi},
is as mentioned above a consequence of the strong parametric sensitivity
of the GW predictions, which dominates over the
scheme-dependent differences if it is projected to
parameter space regions of the cxSM.

A common feature across all approaches is that parameter points exhibiting
the strongest GW signals lie near the vacuum trapping region (upper left edges of the colored regions).
In this region along the diagonal,
the peak amplitudes are the largest for
lighter singlet masses $m_s$, whereas
the percolation temperature grows with increasing singlet mass, rendering a weaker signal prediction
for larger values of $m_s$.
This fact can be used to define an upper bound on the singlet mass in the sense that the perturbative description of SFOEWPTs would break down for larger values of $m_s$.
By requiring that the scalar couplings remain perturbative at the hard scale, we require the four-point vertex contributions to stay below $4\pi$.
This condition leads to the perturbativity bounds $\lambda_{hs}(\mu = \pi T) < 8\pi$ and $\lambda_s (\mu = \pi T) < 8\pi/3$, which in turn determine the upper limits on the singlet couplings at the input scale $m_Z$. 
Thence, the upper limit on the singlet mass 
beyond
which a SFOEWPT no longer occurs for values of $\lambda_{hs}$ and
$\lambda_s$ that are sufficiently small
is approximately at 500--550\,GeV
(in agreement with the findings of
Ref.~\cite{Curtin:2014jma}).

While for brevity, we show here only parameter planes with
$\lambda_s$ kept fixed, we also investigated the
modifications to the parameter regions predicting
a SFOEWPT for different values of $\lambda_s$.
When varying the quartic singlet coupling $\lambda_s$, we find qualitatively similar features across all approaches considered.
As $\lambda_s$ increases, with the singlet mass held fixed, the parameter region 
giving rise to
a strong
GW signal is shifted
toward higher values of $\lambda_{hs}$.
Conversely, for fixed $\lambda_{hs}$, the mass range yielding a strong signal becomes narrower as $\lambda_s$ decreases.
Moreover, we find that the parameter region with a 
SFOEWPT extends to even larger mass and coupling values
beyond the ranges shown in the plots.\footnote{A parameter
region with large portal couplings and larger
singlet masses was explored
in Ref.~\cite{Niemi:2024vzw}
in a different
version of the singlet extension
of the SM, where the singlet has a vev at zero temperature
and thus mixes with the SM-like Higgs boson.} 

We leave a comprehensive phenomenological study
of the parameter space regions suitable for
GW signals that may be detectable with LISA
considering
the full three-dimensional parameter space
of the cxSM for future work.

After having compared the different approaches for
predicting the GW signals against each other, we now
want to estimate the theoretical uncertainty of the GW predictions
using either of the methods.
In order to quantify
the scale uncertainties in
the predictions for the peak amplitudes of
the GW signals, we introduce the ratio
\begin{equation}
\label{eq:Rmu}
    R_{\Omega_{\text{GW,peak}}}^{\mu} = \frac{\Omega_{\text{GW,peak}}(\mu = 2\mu_0)}{\Omega_{\text{GW,peak}}(\mu = \mu_0/2)} \, ,
\end{equation}
where the central value $\mu_0$ 
is chosen to be
equal to $v$ for the \os- and
to
$\pi T$ for the \msbar-scheme.
In the right plots of \cref{fig:parscanGWmudep}
we show the values of this ratio for the parameter
space regions predicting a SFOEWPT
(as defined in the left plots) across the different
approaches. The regions with colored points
do not coincide exactly between the left and
the right plot in each row, because the ratio
$R_{\Omega_{\rm GW,peak}}^\mu$ can only
be determined if both scale choices $\mu = 2\mu_0$
and $\mu = \mu_0 / 2$ predict a SFOEWPT.
We also note
that in the $3D$-\msbar case, we determine the bounce action using the tree-level potential within the EFT.
We have checked that the renormalization scale uncertainty (in this case, due to variation of the matching scale) 
would remain almost unchanged if the loop improved potential were
taken instead.
The right plots of \cref{fig:parscanGWmudep} show
that, going from top to bottom, \ie from $4D$-\os over $4D$-\msbar to $3D$-\msbar, the
scale dependence of the peak amplitudes of the
GW signals are significantly reduced,
as already apparent from the previous sections.
The scale uncertainties of the
$\Omega_{\rm GW,peak}$
are most
pronounced in the $4D$-\os peak regions,
especially at
the upper left edge of the region shown 
where it can be as large as four orders of magnitude.
This significant scale dependence in the
$4D$-\os\ approach is problematic since it is
most severe in parameter space regions where the GW
signals could be strong enough for detection by LISA. 
On the other hand, exploiting the hierarchies between thermal scales and using a dimensional reduction approach leads to a greatly reduced uncertainty in the
GW
peak amplitude due to the scale variation,
which is at most 
of
one order of magnitude 
as is visible in the
bottom right plot of 
\cref{fig:parscanGWmudep}.

\begin{figure}[t]
	\centering
	\begin{subfigure}[t]{0.48\textwidth}
	\includegraphics[width=\textwidth]{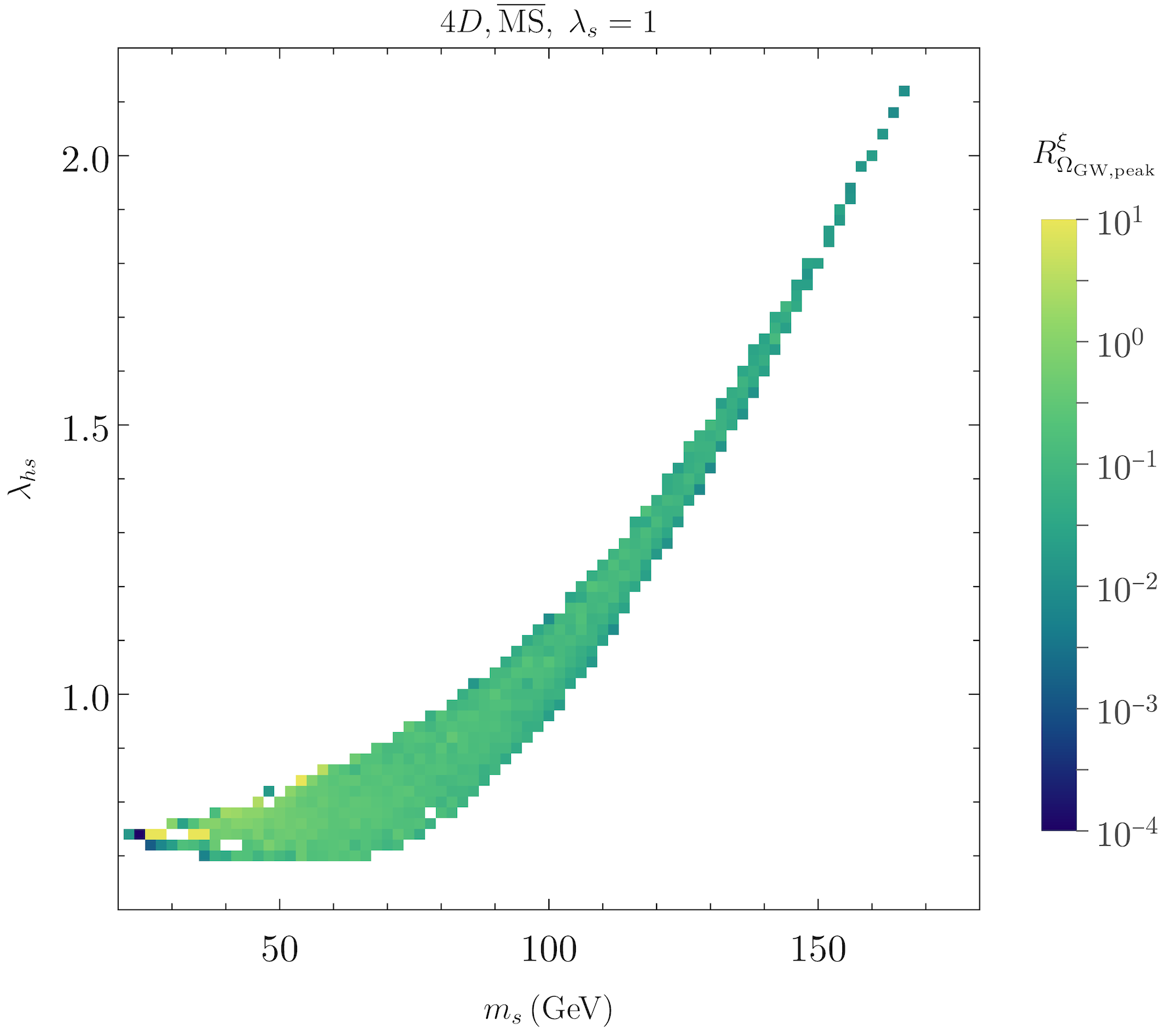}
	\end{subfigure}
	~
	\begin{subfigure}[t]{0.48\textwidth}
	\includegraphics[width=\textwidth]{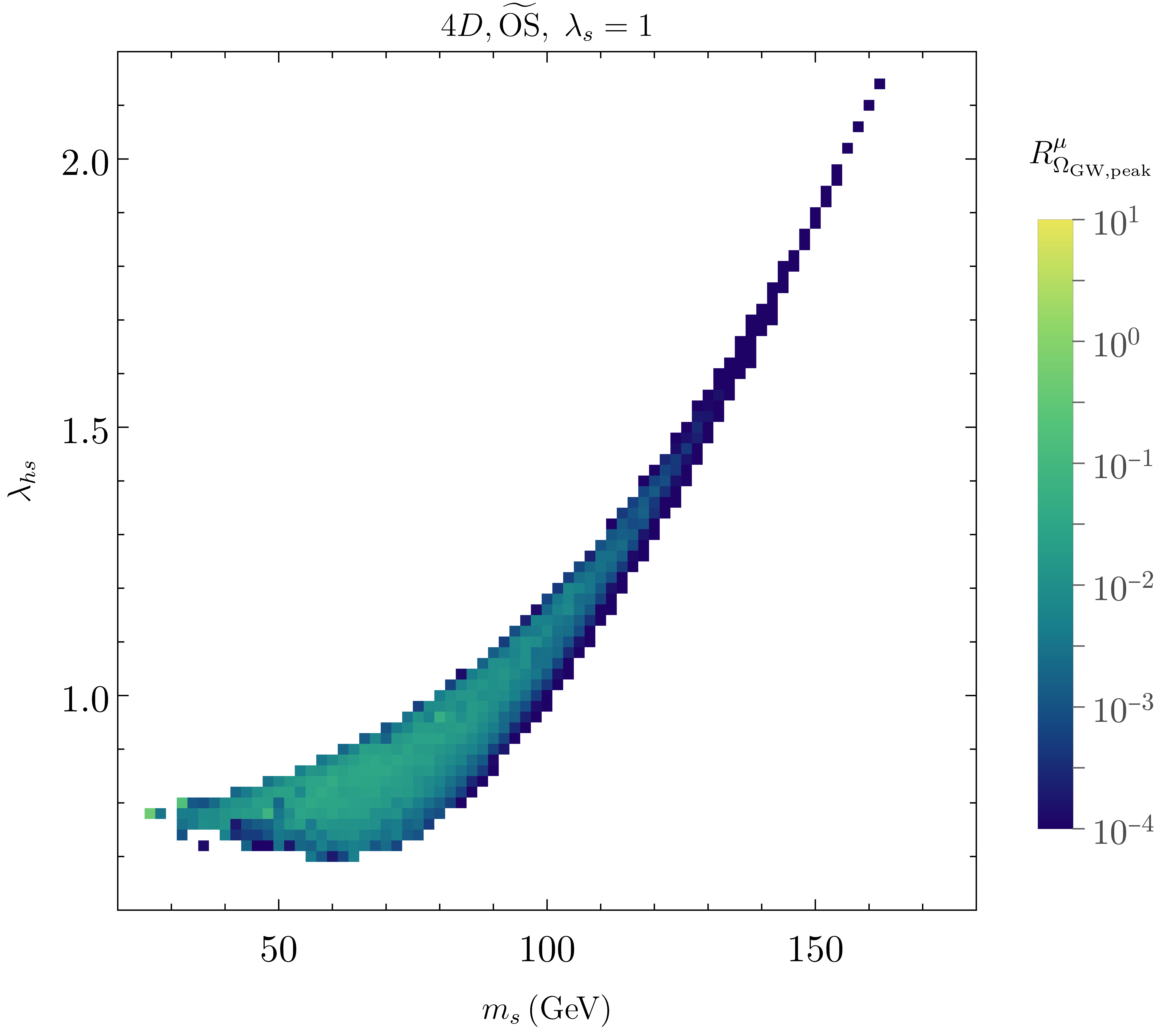}
	\end{subfigure}
	\caption{Uncertainty in the gravitational wave peak strength due to gauge-fixing parameter variation, quantified by \cref{eq:Rxi}, in the $4D$-\msbar (\textit{left}) and $4D$-\os (\textit{right}) schemes. 
	}
	\label{fig:parscanGWxidep}
\end{figure}

Similarly, we assess the shift in the gravitational wave peak signal due to the variation of the gauge-fixing parameter.
We do not show the gauge dependence of the $3D$ approach, because as discussed before, calculating the bounce action using
the tree-level potential
is the power-counting accurate approach here, and this is inherently gauge-independent.
From our checks of individual parameter space points (see \eg \cref{fig:GWspectrum-xi}), it is clear, though, that the gauge dependence is small in the $3D$ approach even when including the loop-improved potential.
To quantify the impact of residual
gauge dependence in the $4D$ approach, we define for
the ratio
\begin{equation}
\label{eq:Rxi}
    R_{\Omega_{\text{GW,peak}}}^{\xi} = \frac{\Omega_{\text{GW,peak}}(\xi = 10)}{\Omega_{\text{GW,peak}}(\xi = -10)}.
\end{equation}
In the left plot of 
\cref{fig:parscanGWxidep}
we show
with the color coding the values of
$R_{\Omega_{\text{GW,peak}}}^{\xi}$ using the
$4D$-\msbar potential, in the same parameter plane that
was already shown in \cref{fig:parscanGWmudep}.
One can observe that
the variation of the gauge-fixing parameter leads to a change in the peak amplitude
by approximately one order of magnitude across the entire parameter space 
giving rise to a SFOEWPT.
This can be compared to the results using the $4D$-\os
approach that is shown in the right plot
of \cref{fig:parscanGWxidep}.
We observe a substantially stronger variation in the
$R_{\Omega_{\text{GW,peak}}}^{\xi}$ values using
the \os-scheme, where
the variation typically reaches at least two orders of magnitude and can rise to four orders of magnitude
in the region
predicting the strongest GW signals.
As a result, the question of whether a parameter
point of the cxSM predicts a GW signal that might be
detectable with LISA can be answered less reliably
using the $4D$-\os approach.
In the following, we present a more detailed
analysis of the detectability
of the GW signals with LISA in terms of the signal-to-noise ratio.

\subsection{LISA SNR}
\label{sec:snr}

If the SFOEWPT
occurs at temperatures around 100~GeV, the resulting GW signal is expected to fall within the sensitivity range of the LISA interferometer.
The detectability of such a signal can be quantified by computing the Signal-to-Noise Ratio (SNR).
It can be computed as~\cite{Caprini:2019egz}:
\begin{equation}
\label{eq:snr}
\mathrm{SNR} = \sqrt{ \mathcal{T}\int_{-\infty}^{\infty} \left(\frac{\Omega_{\text{GW}}(f)h^2}{\Omega_\text{sens}(f)h^2}\right)^2 \, \mathrm{d}f },
\end{equation}
where $\Omega_\text{sens}(f)h^2$ is the nominal sensitivity of LISA to GW signals~\cite{Caprini:2019pxz}, and we assume the duration of the observations to be $\mathcal{T}=4$\,years. 
Roughly speaking, if $\text{SNR}\gtrsim1$, the GW signal is classified as detectable.\footnote{As mentioned
already in \cref{sec:intro}, we do not consider
astrophysical foregrounds in our analysis.
While we adopt a SNR threshold of 1 to assess detectability,
it is worth noting that a threshold of
$\mathcal{O}(10)$ is often used
in a Bayesian statistical interpretation of
potential signals (see also the discussion
in Ref.~\cite{Caprini:2019pxz}).
The precise choice of the SNR threshold
is of minor concern here, as the focus of this work is
not on the absolute detectability of a signal,
but rather on comparing the SNR predictions obtained from
different computational approaches and assessing the
associated theoretical uncertainties.
As we will show, these uncertainties stemming from the
perturbative description of the phase transition
reach two orders of magnitude or more, depending on
the method used, such that using the condition
$\mathrm{SNR} \gtrsim 10$ instead of~1 would not alter our conclusions.}

\begin{figure}[t]
\centering
\includegraphics[width=0.48\textwidth]{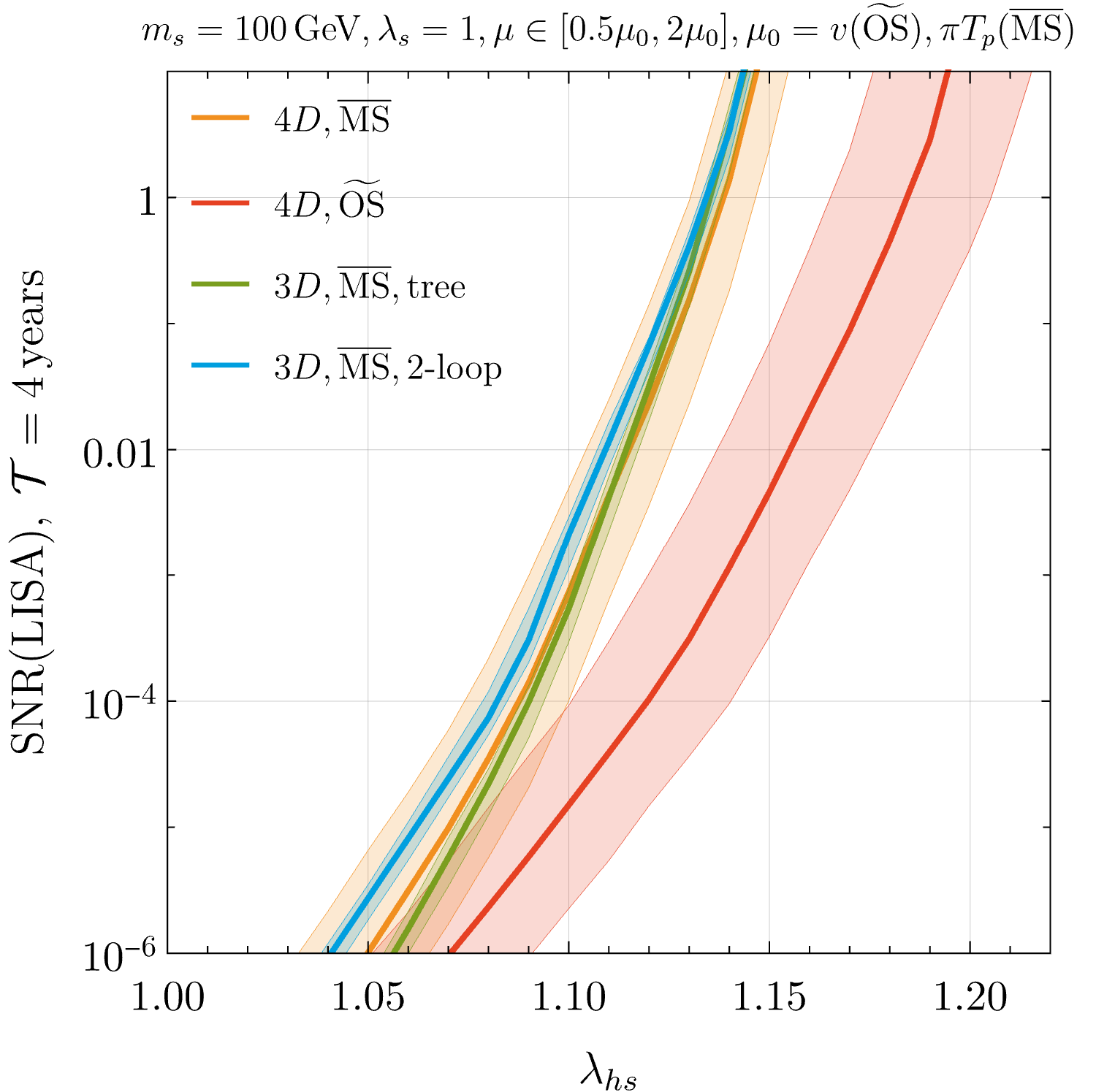}
\includegraphics[width=0.48\textwidth]{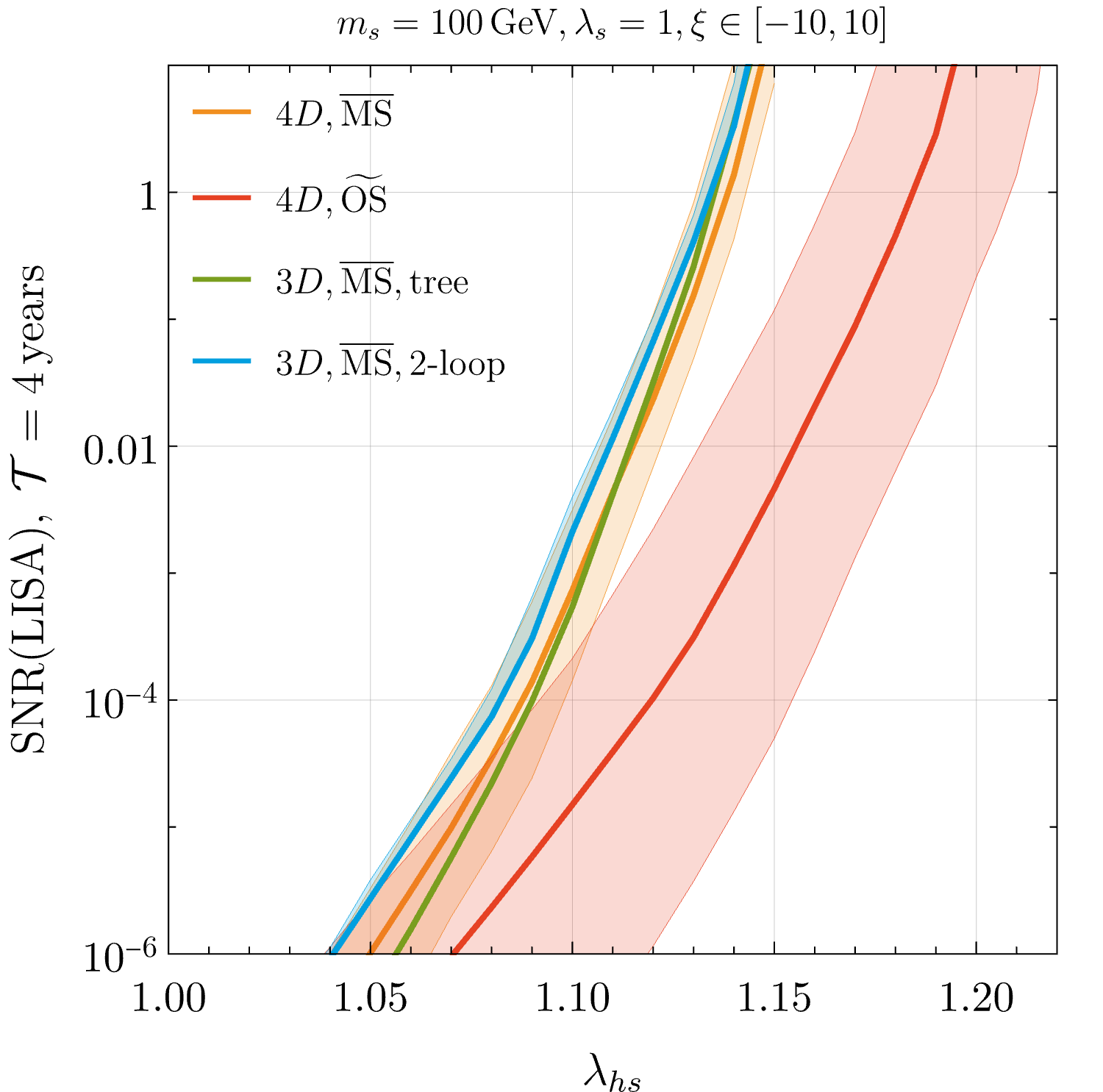}
\caption{
Predicted Signal-to-Noise Ratio (SNR) at LISA for $m_s=100\,\text{GeV}$, $\lambda_s=1$, $\lambda_{hs}\in[1,1.2]$ and its renormalization scale (\textit{left}) and gauge (\textit{right}) dependencies.
}
\label{fig:LisaSNR}
\end{figure}

\begin{figure}[t]
\centering
\includegraphics[width=0.48\textwidth]{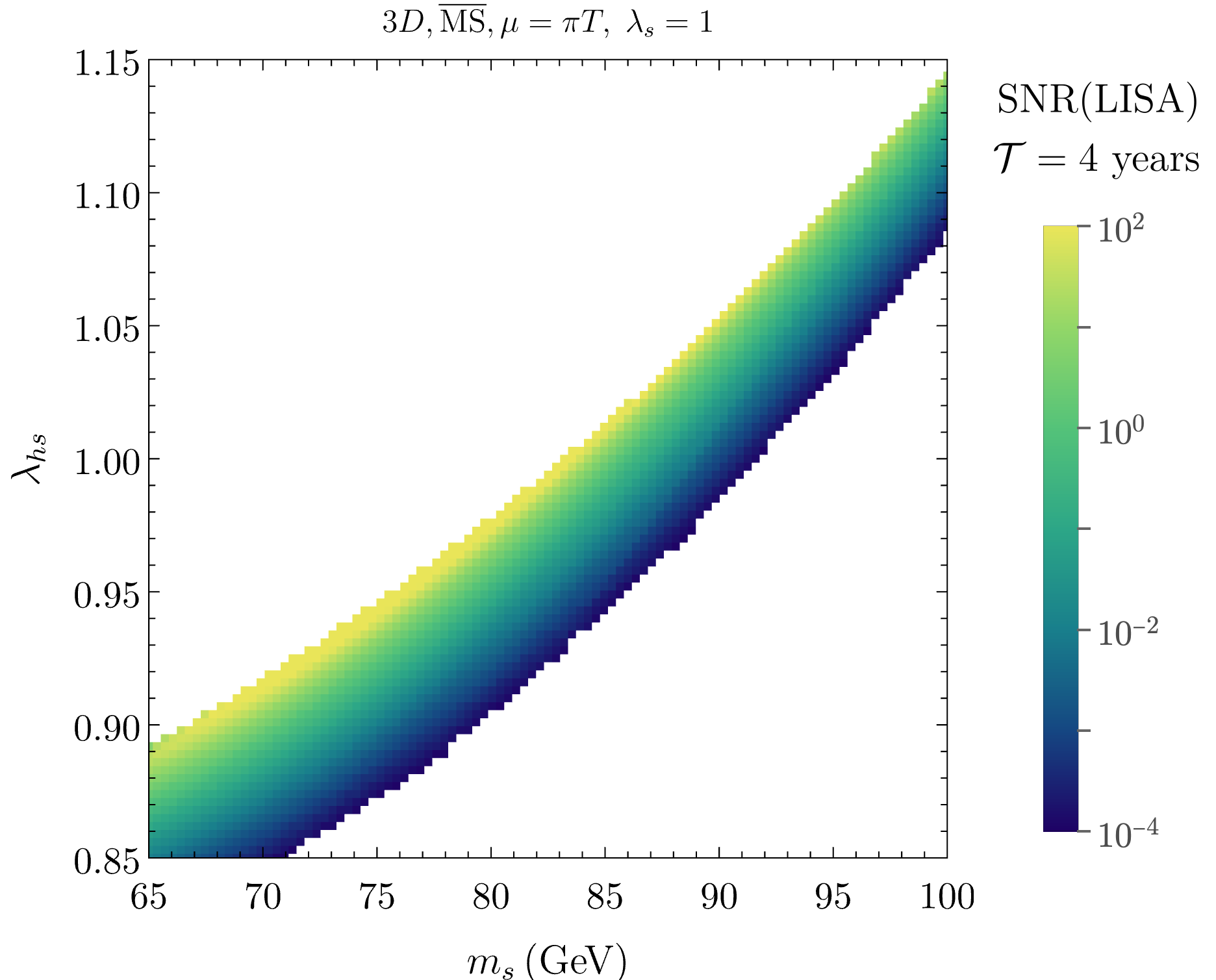}
\caption{
The absolute values of the SNR at LISA for $\lambda_s=1$ obtained in the $3D$-\msbar approach.
We limit the heatmap scale at $\text{SNR}=10^2$
to emphasize variations in lower-SNR regions, and only show points with $\text{SNR}>10^{-4}$.
}
\label{fig:LisaSNRscan}
\end{figure}

\Cref{fig:LisaSNR} presents the predicted SNRs at LISA
determined using the various approaches discussed in \cref{sec:equilibrium} for the parameter points
with $m_s = 100$~GeV and varying coupling
$\lambda_{hs}$ already discussed in
\cref{sec:nucleationthermo}.
The colors represent the different approaches
as in \cref{fig:Tn4d3dmu},
and their theory uncertainties from the residual scale dependence
(left plot) and gauge dependence (right plot)
are indicated with the colored bands (see the
discussion in \cref{sec:nucleationthermo}).
We observe that
the theory uncertainties are greatly reduced
using the $3D$-\msbar approach compared to the $4D$ approaches.
Focusing on the $4D$-approaches, we observe that the
$4D$-\os approach 
shows
a more pronounced
uncertainty from both the scale- and the gauge-dependence
compared to the $4D$-\msbar approach.
Moreover, we observe an
overall shift between the predicted SNRs
from the approaches using the \msbar-scheme
and the $4D$-\os approach.
The shift between the two results can be
attributed to the different choices of central
renormalization/matching scales ($\mu = \pi T$
in the \msbar-scheme and $\mu = v$ in the \os-scheme),
differences in the RGE running of the parameters,
and missing momentum-dependent contributions in
the \os-scheme in the relations to the physical parameters (see \cref{app:renormalization}), 
as was discussed in more detail
in relation to the differences in the prediction
for the critical temperature $T_c$ in
\cref{sec:scale_dependence}.
This shift is 
so
large that the
uncertainty band of the $4D$-\os approach does not overlap
with the results from the $4D$-\msbar approach in most parts
of the displayed range of the coupling $\lambda_{hs}$,
potentially indicating a poor 
behavior
of the
perturbative expansion at the considered one-loop level.

In \cref{fig:LisaSNRscan} 
we show
the SNR values at LISA for $\lambda_s=1$ computed using the $3D$-\msbar approach in the $\left(m_s , \lambda_{hs} \right)$ plane. 
Compared to \cref{fig:parscanGWmudep}
and \cref{fig:parscanGWxidep}, the ranges for the singlet mass and the portal coupling in \cref{fig:LisaSNR} are reduced to
zoom into the region where high SNR values are predicted, and we only show the points predicting a GW signal
with $\text{SNR}>10^{-4}$.
Despite the sizable theoretical uncertainties in
the predicted SNRs across the different approaches,
see the discussion above, the regions of the parameter
space yielding $\mathrm{SNR} > 1$ remain similar
as a result of the extreme sensitivity of the
strength of the GW signals on the model parameters
$m_s$, $\lambda_s$ and $\lambda_{hs}$.
Consequently, we display in this plot only the results
obtained using the $3D$ approach.
As one can expect already from \cref{fig:parscanGWmudep}, the parameter region where $\text{SNR}>1$ occupies only a rather restricted area in parameter space.
Nevertheless,
a detection of a GW signal at LISA consistent with a
SFOEWPT  would not, by itself, allow for a precise
reconstruction of the underlying model parameters
due to the visible approximate flat direction in
the predicted SNRs, where a decrease in the SNR
caused by larger singlet masses $m_s$ can be
compensated to a large extent
by an increase in the portal
coupling $\lambda_{hs}$. In particular,
this degeneracy implies that the singlet mass
cannot be tightly constrained from
an experimental observation of 
a
GW signal alone.

\section{Impact of higher-dimensional operators within the $3D$ approach}
\label{sec:highdim}
An important check for the validity of the dimensional reduction approach is to know 
to what extent
the high-temperature expansion at its core is
justified.
We investigate this question via the effects of higher-dimensional operators in the EFT Lagrangian (see \cref{sec:DR} for details).
Significant effects from these operators on thermodynamic observables and ultimately the GW signal indicate the failure of the truncation of the effective theory, \ie the failure of the high-temperature expansion.

\begin{figure}[t]
	\centering
	\begin{subfigure}[t]{0.45\textwidth}
		\includegraphics[width=\textwidth]{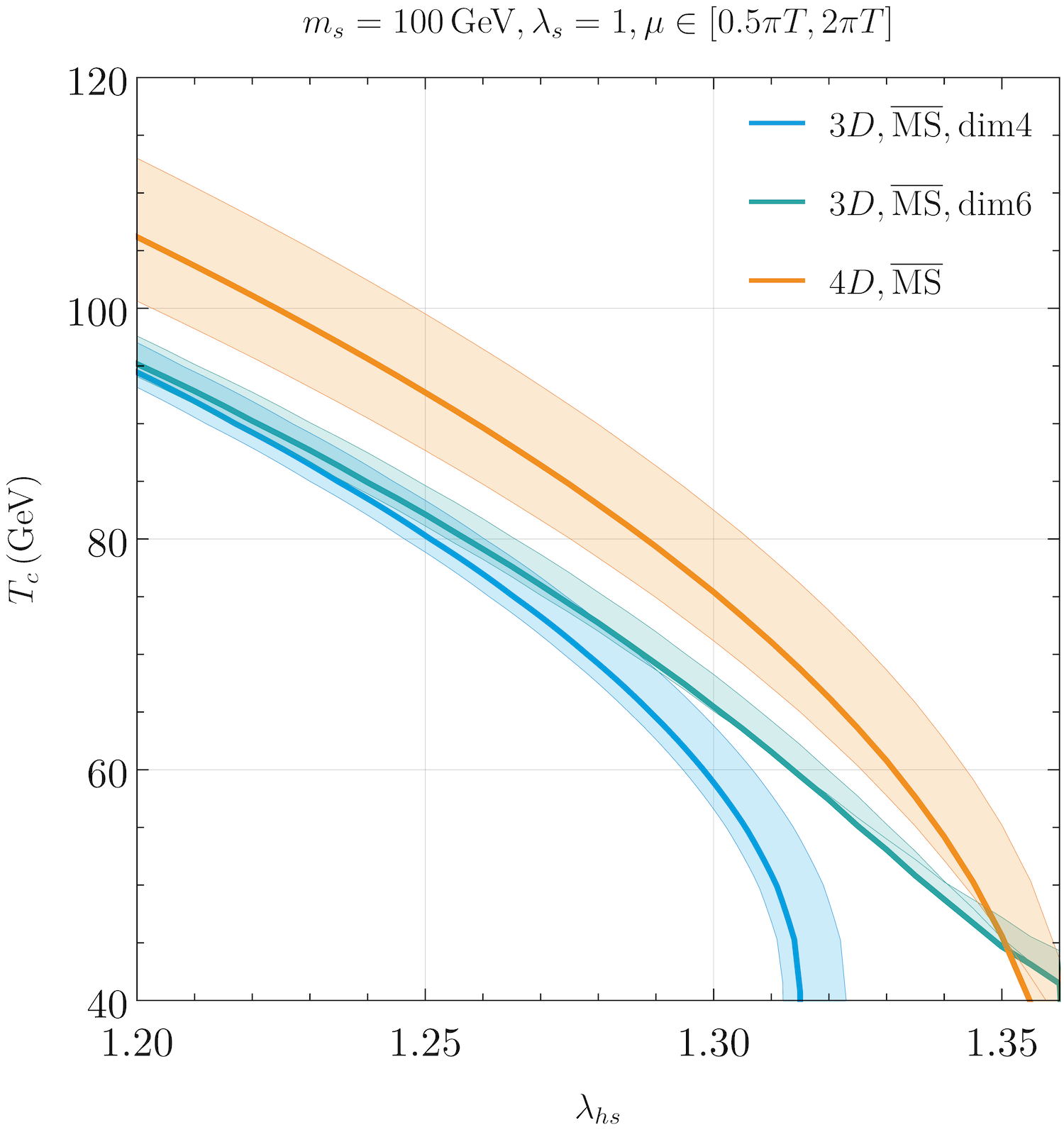}
	\end{subfigure}
	~
	\begin{subfigure}[t]{0.47\textwidth}
		\includegraphics[width=\textwidth]{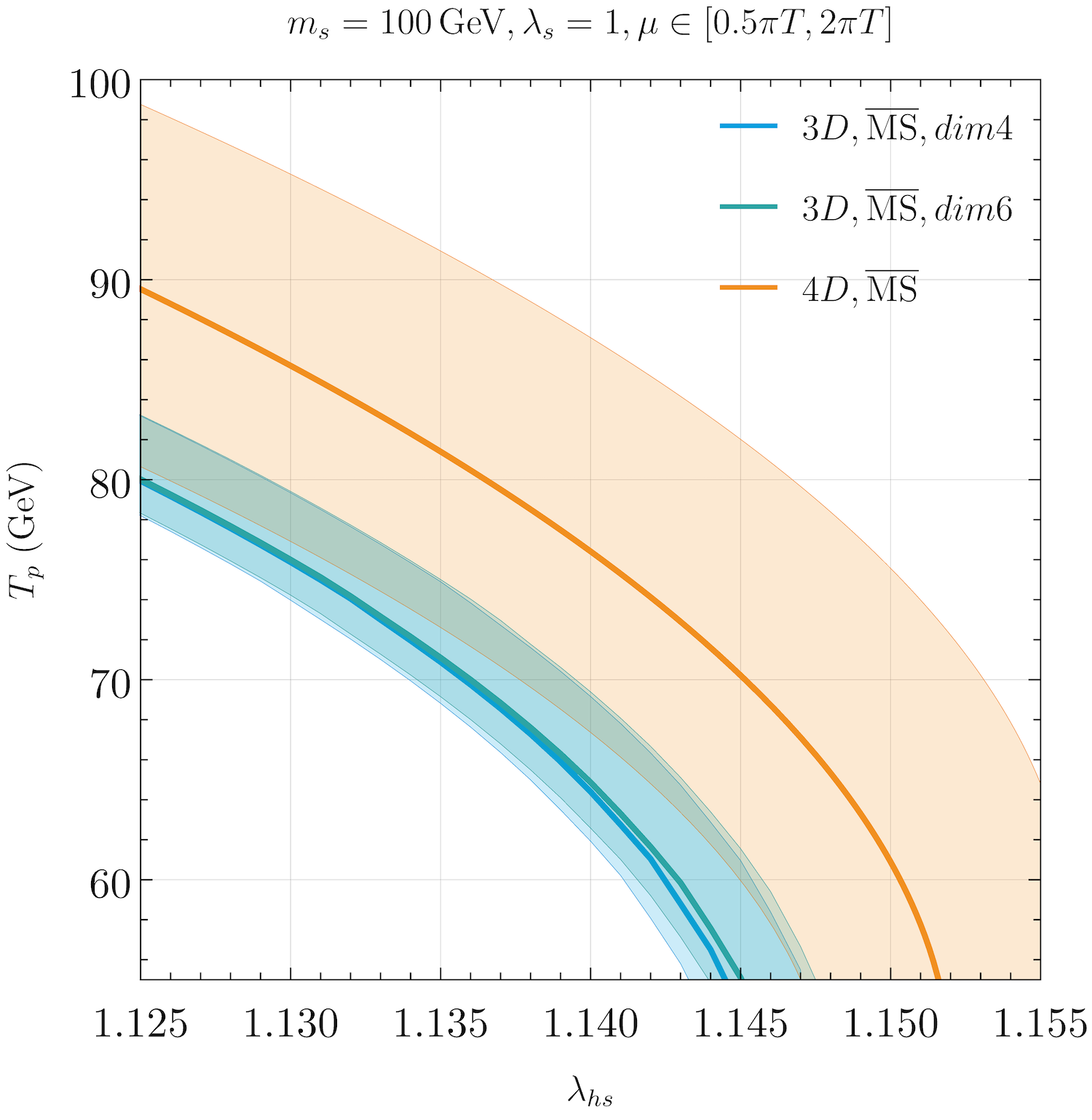}
	\end{subfigure}
	\caption{
		Critical (\textit{left}) and percolation (\textit{right}) temperatures determined 
        within the $4D$ approach in the \msbar-scheme as well as 
        in the $3D$ approach in the \msbar-scheme
        with and without the dimension-6 operators of \cref{eq:dim6oper} 
        in the Landau gauge for $m_s=100\,\text{GeV}$, $\lambda_s=1$.
		The bands represent the renormalization scale uncertainty, as before.
		Note that the parameter ranges differ between the two plots.}
	\label{fig:TDim6}
\end{figure}

In the left
plot of \cref{fig:TDim6}, we show how the inclusion of dimension-six operators in the EFT-Lagrangian
(turquoise) changes the predictions for
the critical temperature
$T_c$ compared to the ones when
truncating the EFT-Lagrangian
at dimension four (blue).
For comparison, we also show the prediction using
the $4D$-\msbar approach (orange), and in all cases
the bands around the solid lines show the theory
uncertainty from the scale dependence.
The comparison between the turquoise and the
blue bands reveals that for higher temperatures adding higher-dimensional operators barely changes the thermodynamic observables, justifying the truncation of the high-temperature EFT Lagrangian in these regions.
However,
the difference between the turquoise and the
blue bands becomes prominent for higher portal couplings, which correspond to lower critical temperatures.
This indicates a breakdown of the high-temperature
expansion in this regime. Notably, this issue arises
precisely where the accuracy 
of the
GW predictions is most critical, since the strength
of the signal increases with decreasing
transition temperature.
We expect the $4D$ approach at low temperatures to
more accurately describe the phase structure, because
(apart from the, in this parameter region, small contribution
from the
AE-type resummation of daisy diagrams)
it does not rely on high-temperature expansion.
In agreement with this expectation, we observe that
adding dimension-six terms to the EFT Lagrangian puts the predicted critical temperature closer to the one obtained within the $4D$ method for low temperatures.

In the right plot of \cref{fig:TDim6} we show
the predictions for the percolation temperature $T_p$,
with the definition of the different bands as in the
left plot discussed above.
In contrast to the critical temperature, the impact of dimension-six operators on the percolation (and nucleation) temperature
is significantly less significant, even for the
parameter points predicting the lowest values of percolation temperature.
A similar behavior was previously observed for phase transitions in the Standard Model Effective Field Theory~(SMEFT)~\cite{Croon:2020cgk,Chala:2025aiz}.

\begin{figure}[t]
	\centering
	\begin{subfigure}[t]{0.46\textwidth}
	\includegraphics[width=\textwidth]{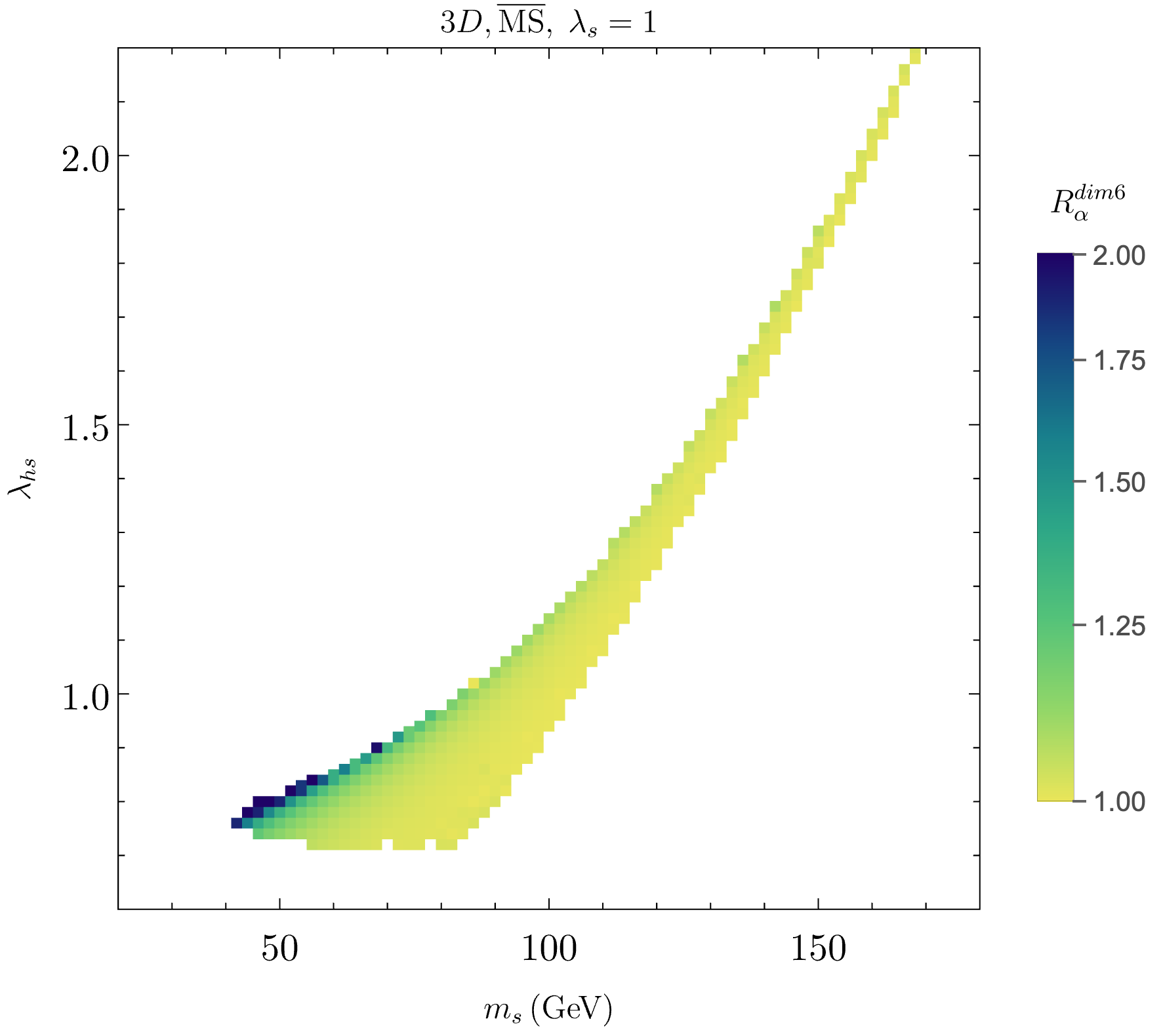}
	\end{subfigure}
	~
	\begin{subfigure}[t]{0.46\textwidth}
	\includegraphics[width=\textwidth]{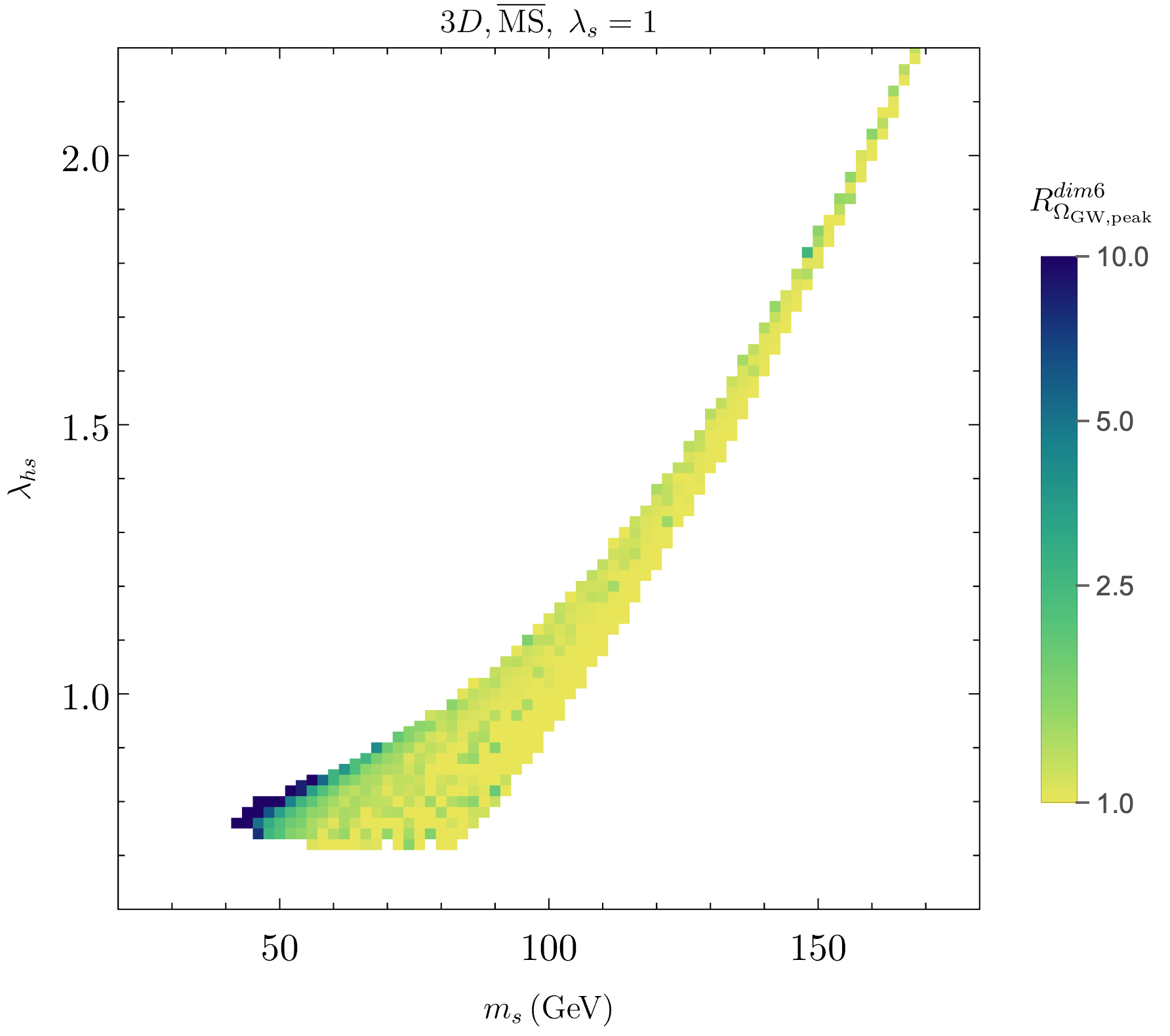}
	\end{subfigure}
	\caption{
		Variation ratio of \cref{eq:Rdim6} for the phase transition strength $\alpha(T_p)$ (\textit{left}) and the gravitational wave peak strength $\Omega _{\text{GW,peak}}$ (\textit{right}) 
        obtained from the inclusion of
        dimension-6 operators within the $3D$ approach. 
        Note that the same colors correspond to
        different ranges in the two plots.
        }
	\label{fig:Dim6}
\end{figure}

The thermodynamic observable that changes most significantly
under the inclusion of the dimension-six operators
in the high-$T$ EFT
is the phase transition strength $\alpha$
defined in \cref{eq:alpha}.
\Cref{fig:Dim6} illustrates the
modifications 
of
the phase transition strength (\textit{left} plot)
and the peak amplitude of the GW signal (\textit{right} plot),
which we respectively quantify in terms of the ratios
\begin{equation}
\label{eq:Rdim6}
    R_{\alpha}^{\mathrm{dim}6} = \frac{\alpha [3D, \mathrm{dim}6]}{\alpha[3D, \mathrm{dim}4]} \quad \textrm{and} \quad R_{\Omega_{\text{GW,peak}}}^{\mathrm{dim}6} = \frac{\Omega_{\text{GW,peak}}[3D, \mathrm{dim}6]}{\Omega_{\text{GW,peak}}[3D, \mathrm{dim}4]} \, .
\end{equation}
The parameter space region predicting the strongest
GW signals corresponds to the region of the lowest percolation temperatures, which in the $(m_s,\lambda_{hs})$ plane
is located at the lower left corner of the banana-shaped
band of points predicting a SFOEWPT
(see \cref{fig:parscanGWmudep}).

The 
plots of \cref{fig:Dim6} demonstrate that
this is precisely the region
where the truncation of the EFT Lagrangian becomes uncertain,
as in this region the ratio $R_{\alpha}^{\mathrm{dim}6}$
reaches values of about two (yellow points in the left plot).
Accordingly, the peak amplitude of the GW signal
given in \cref{eq:peakampli},
which for $\alpha \ll 1$ depends approximately quadratically
on $\alpha$, is also significantly altered.
For these points, the inclusion of dimension-six operators
give rise to an increase in the peak amplitudes
by up to an order of magnitude, as is visible in the
right plot of \cref{fig:Dim6}.

We checked that for both the strength $\alpha$ and
the peak amplitude $\Omega_{\rm GW,peak}$ the inclusion of dimension-six
operators
brings the
predictions for these observables closer to the ones calculated within the $4D$-\msbar approach.
This observation could be interpreted as a hint
that an accurate description of the phase transition
just requires the inclusion of higher-dimensional
operators beyond the renormalizable ones, instead
of an indication
of a complete breakdown of the high-$T$ EFT framework.
In this view, dimension-six operators act as leading
corrections that effectively capture relevant physics
missing from the EFT truncated at dimension four.
However, without systematically including and assessing
the impact of operators of even higher dimension
(such as dimension-eight terms), whose effects are expected
to be smaller but not necessarily negligible,
it is not possible to definitively conclude whether
the observed improvement in agreement with the
$4D$-\msbar approach genuinely reflects increased
predictive power,
or instead indicates the onset
of the breakdown of the EFT expansion.

\section{Summary and conclusions}
\label{sec:conclus}

In this work,
we have presented a
comprehensive study of the dynamics of the
EW phase transition
and the associated theory uncertainties of the predictions for physical observables
characterizing the transitions in the cxSM, the
SM extended by a complex gauge singlet scalar field.
We focused on a scenario in which SFOEWPTs are realized
in such a way that the singlet field has a non-vanishing vev
before the transition and a vanishing vev after the transition.
This is the simplest realization of a SFOEWPT that,
if the singlet mass is larger than twice the Higgs boson mass,
is expected to leave
no detectable traces at the LHC.
In order to investigate the robustness and precision
of the perturbative description of the SFOEWPT, we compared
a set of commonly used approaches.
This set consists of two different input 
schemes
in the $4D$ approach, the \msbar- and the \os-scheme, 
and of
the $3D$ or high-temperature EFT approach, in which heavy modes are
systematically integrated out, leading to an EFT in three
spatial dimensions that captures the infrared dynamics
of the transition.

We presented an analysis of uncertainty bands arising from both residual scale and gauge dependence,
affecting the predictions for thermodynamic observables, such as the critical
temperature and latent heat, as well as the predicted GW
signals and the corresponding SNRs at LISA.
We showed that the uncertainties from the residual scale and gauge dependence
are significantly reduced in the $3D$ approach.
Moreover, focusing on the $4D$ approaches,
we found
that the gauge-parameter dependence in the \os-scheme is enhanced compared to the one in the \msbar-scheme.
We showed how
with increasing loop order of the effective potential
in the $3D$ approach
(up to the full two-loop effective potential in $\rxi$ gauge)
the gauge dependence and the associated theory uncertainty
is reduced order by order, and
we discussed in detail how this follows from
the gauge-parameter dependence
of the effective potential governed by
the Nielsen identity.

In both the $4D$ and the dimensionally reduced
$3D$ frameworks, we obtained gauge-independent results for
the quantities characterizing the phase transition
and the resulting GW signals by applying the $\hbar$-expansion.
In the $4D$ approach, we observed that the gauge-independent
results at the considered one-loop level exhibit significant
numerical deviations from the gauge-dependent result
where one directly minimizes the loop-corrected effective
potential, indicating the need for higher-loop corrections
to achieve a gauge-independent result with sufficient
numerical precision.
On the other hand, in the $3D$ EFT, the gauge-independent results
obtained by making use of the $\hbar$-expansion are in excellent
numerical agreement with those derived from a direct
minimization of the loop-corrected potential in a
gauge-dependent way.
As a result, using the $\hbar$-expansion,
one-loop \msbar-relations between the model parameters,
dimensional reduction with NLO matching, integrating
out hard and soft scales,
and the effective potential at the two-loop level,
we achieved a gauge-independent
and numerically precise
description of SFOEWPTs in the cxSM.
As the main remaining theoretical uncertainty in
the parameter space regions predicting detectable
GW signals at LISA, we identified the impact of
higher-dimensional
operators in the dimensionally
reduced EFT, see the discussion below.

Despite the sizable uncertainties
associated to the different approaches
used to describe the phase transition
dynamics,
the parameter regions where
SFOEWPTs occur
remain similar.
This is a result of the large parametric
dependence of the transition strengths
on the model parameters of the cxSM.
While a SFOEWPT can be accommodated over wide parameter regions 
of the singlet mass and couplings, generating a GW signal that is potentially detectable with LISA, requires a significant amount of fine-tuning among 
the cxSM parameters $m_s$, $\lambda_{hs}$
and $\lambda_s$.
Regarding the occurrence of a SFOEWPT,
using the perturbativity limit on scalar couplings, we obtained an upper bound on the singlet mass
of $m_s \lesssim 500$--550~GeV.
For larger masses of the singlet state,
the coupling values that would be required
to realize a SFOEWPT fall outside of the
perturbative regime, such that non-perturbative (lattice) computations would
be required to study whether the cxSM
allows for a SFOEWPT and an associated
GW signal in reach of LISA.

Although the parameter space regions
predicting potentially detectable GW
signals are largely consistent
across the different approaches,
within these parameter regions
the magnitude of the predicted
GW signals can vary by several orders of
magnitude, depending on the specific perturbative framework employed.
In particular, the $4D$ approach using the \os-scheme
predicts significantly smaller SNRs, typically by about two
orders of magnitude, compared to the \msbar-scheme.
Moreover, the $4D$-\os approach is subject to considerably larger theoretical uncertainties. A variation of
the renormalization scale by a factor of two
can induce changes in the predicted SNRs of up to four
orders of magnitude.
On the other hand, the $4D$ approach using the \msbar-scheme
yields SNRs that are in fairly good agreement with those
obtained in the dimensionally reduced $3D$ approach.
The differences between these two approaches
are smaller than the intrinsic theoretical uncertainties
from the residual scale and gauge dependence
associated with the $4D$-\msbar approach,
which are at the level of about one order of magnitude.
Ultimately, the $3D$ EFT approach offers the most stable
predictions in this regard. When employing NLO matching at
$\mathcal{O}(g^4)$, the residual scale dependence in the SNRs is reduced by about one order of magnitude.
In combination with the fact that the $3D$ approach
can be combined with the $\hbar$-expansion to remove
residual gauge dependence without discarding numerically
important contributions, the application of the high-$T$
EFT is therefore in most parts of the parameter space
the most reliable and precise framework to study
SFOEWPTs in the cxSM.

However, the $3D$ EFT approach is subject to an additional
intrinsic theory uncertainty from truncating the EFT at finite
operator dimension. To assess this uncertainty, we compared results
including only renormalizable/dimension-four operators with those
also incorporating dimension-six operators. At high temperatures, both types of results agree well, indicating 
a reliable 
behavior
of the EFT.
However, for lower temperatures, particularly below
$T_c \lesssim 80$~GeV, the dimension-six operators noticeably
affect the critical temperature and enhance the predicted strength
of the phase transition and GW signals.
We found that the impact on the critical temperature is more
pronounced than on the percolation temperature.
Nevertheless, the effects on the GW signals (which are
computed at percolation) remain sizable.
Specifically, we observed changes in the transition strength
$\alpha$ of up to 100\% and shifts in the GW peak amplitude by
up to an order of magnitude. Since stronger GW signals correlate
with lower transition temperatures, these effects become relevant
precisely where the prospects for a GW detection with LISA are best.
Interestingly, the inclusion of dimension-six operators brings the
predictions of the $3D$ approach closer to the results obtained
using the $4D$-\msbar approach, which may be more reliable in
this regime since it relies to a much lesser extent on the high-$T$ expansion.
Still, since we included only a subset of all possible dimension-six
operators, and since higher-dimensional terms are not considered
with which the 
behavior
of the EFT could be further checked,
it remains unclear whether this signals a genuine improvement
or merely a breakdown of the EFT.
We leave a dedicated study in this direction for future work.

In addition to this 
study of the 
thermodynamic observables of the
cxSM, we also discussed many of the technical details. 
This includes a demonstration of
the cancellation of the scale dependence by including
the RGE evolution of the couplings in the $4D$-\msbar approach;
the reduction of the scale dependence
with increasing matching order in the $3D$ EFT;
a comparison of the results from a direct minimization of
the effective potential compared to the 
gauge-independent results using the $\hbar$-expansion,
showing 
good agreement between the two types of results
only in the $3D$ approach;
a comparison of different gauge-fixing choices
($R_\xi$ and Fermi gauges); and
the reduction of explicit scale and gauge dependence when including
higher loop orders of the effective potential.

\acknowledgments
We thank Andreas Ekstedt, Guilherme Guedes, Panagiotis Stylianou, Johannes Braathen, Philipp Schicho and Peter Athron for useful discussions and comments.
A.D., M.L.~and G.W.~acknowledge the support of the Deutsche
Forschungsgemeinschaft (DFG, German Research Association) under
Germany's Excellence Strategy-EXC 2121 ``Quantum Universe''-390833306.
This work
has also been funded by the Deutsche Forschungsgemeinschaft (DFG,
German Research Foundation) -- 491245950.
T.B.~acknowledges the support of the Spanish Agencia
Estatal de Investigaci\'on through the grant
``IFT Centro de Excelencia Severo Ochoa CEX2020-001007-S''.
The project that gave rise to these
results received the support of a
fellowship from the ``la Caixa''
Foundation (ID 100010434). The
fellowship code is LCF/BQ/PI24/12040018.
A.D.~acknowledges the support by Consejería de Universidad, Investigación e Innovación, Gobierno de España and Unión Europea – NextGenerationEU under grant AST2\_6.5.

\begin{appendices}

\appendix
\section{Lagrange parameter input scheme and renormalization}
\label{app:renormalization}

In this section, we provide details about the relations between the physical parameters and those
parameters entering the Lagrangian and calculations.
Explicit relations are given
at the one-loop level.
These relations depend on the renormalization scheme used, 
for which 
we 
discuss two options here: the \msbar- and \os- prescription.

\subsection{Input parameters}
\label{app:parameters}

We study the dynamics of electroweak phase transitions using the effective potential, which is defined in terms of the Lagrangian parameters
\begin{equation}
\label{eq:paras-L}
    P_\mathcal{L} = \{\mu_h^2, \lambda_h, \mu_s^2, g, g', y_t, \lambda_s, \lambda_{hs} \} \, .
\end{equation}
In order to define a scheme yielding
sensible input values for those 
parameters
we use their relation to the following set of 
input parameters
\begin{equation}
    P_\mathcal{I} = \{G_F, \mh, \ms, m_Z, m_W, m_t, \lambda_s, \lambda_{hs}\} \, ,
\end{equation}
where $G_F$ and the physical masses $\mh, \ms, m_Z, m_W, m_t$ 
are in principle (i.e., if an appropriate singlet state is detected)
directly accessible in measurements.
Since for the cxSM there are too few physical masses to 
determine
all parameters of the theory 
from them, we use
the portal coupling $\lambda_{hs}$ and the 
quartic coupling $\lambda_s$ 
of the singlet
as free input parameters.
The tree-level relations for the two parameter sets are
\begin{subequations}
 \begin{align}
 	g^2 &= 4 \frac{m_W^2}{v_h^2},\\
 	g'^2 &= 4 \frac{m_Z^2-m_W^2}{v_h^2},\\
 	y_t^2 &= 2\frac{m_t^2}{v_h^2}, \\
     \mu_h^2 &= - \frac{\mh^2}{2}, \\
    \lambda_h &= \frac{\mh^2 }{2 v_h^2},\\
    \mu_s^2 &= 2 \ms^2 - \frac{v_h^2 \, \lambda_{hs}}{2},
\end{align}
\label{eq:tree-level-rels}
\end{subequations}
where 
\begin{equation}
	v_h^2 =  \frac{1 }{\sqrt{2}G_F} \ .
\end{equation}
We note that in this section $v_h$ denotes the
Higgs vev at zero temperature, \ie $v_h = v_{\rm EW}$
as defined in \cref{eq:vevdef}.
The numerical input values we use in our analysis are
\begin{equation}
	\{m_W, m_Z, m_h, m_t\} = \{80.379,\, 91.1876,\, 125.1,\, 172.5\}\,\text{GeV}, \; G_F = 1.1663787\times 10^{-5}\,\text{GeV}^{-2}.
\end{equation}
The above
relations undergo modifications when working beyond the tree-level approximation and will depend on the renormalization scheme used. 
We identify the physical input masses as the pole masses of the theory, \ie the propagator poles of the 
two-point functions of the
corresponding particles.
According to the renormalization prescription (RP) used to define the counterterms of the theory, the poles will appear at different values of masses $m_i^{(\text{RP})}$, \ie from
\begin{equation}
\Delta_i (p^2) = \frac{i}{p^2 - \left(m_i^{(\text{RP})}\right)^2 + \Sigma_i^{(\text{RP})} (p^2)}
\end{equation}
we get
\begin{equation}
\label{eq:mRP}
     \left(m_i^{(\text{RP})}\right)^2 =  m_{i,\text{pole}}^2 + \Sigma_i^{(\text{RP}) }\left(p^2 = m_{i,\text{pole}}^2\right)
\end{equation}
and identify 
$m_{i,\text{pole}}$ with the physical input masses $m_i$.
The self-energies consist of two parts
\begin{equation}
    \Sigma_i^{(\text{RP})} (p^2) = \Sigma_{i,\text{proper}}^{(\text{RP})}(p^2) + \Sigma_{i,\text{tad}}^{(\text{RP})}(p^2), 
\end{equation}
where $\Sigma_{i,\text{proper}}^{(\text{RP})}$ 
are the 
(real parts of the)
``proper'' 1PI-irreducible 
self-energy contributions,
and $\Sigma_{i,\text{tad}}^{(\text{RP})}$ 
are the tadpole contributions.
The (non)-appearance of the latter is dictated by the choice of the tadpole scheme.
Now, in order to define the correct values of Lagrangian parameters at a given loop order according to some renormalization prescription RP, we simply replace the masses in \eqref{eq:tree-level-rels} by the values which follow from \eqref{eq:mRP} and adjust the vev according to the scheme of choice.

We study two different RP-options, namely the \msbar-scheme and the ``on-shell like'' \os-scheme
as originally mentioned in~\cite{Weinberg:1973am} and as used \eg in the phase transition tool \texttt{BSMPT}~\cite{Basler:2024aaf} and further explained in~\cite{Basler:2016obg}. 
Other appearances of this scheme are \eg in~\cite{Lewicki:2024xan}
(called, somewhat confusingly, 
\msbar-scheme there), or in a varied form in~\cite{Delaunay_2008}.

\subsubsection*{\texorpdfstring{\msbar}{MSbar} scheme}
In the \msbar-scheme, we have
\begin{equation}
\label{eq:m-msbar}
     \left(m_i^{(\msbar)}\right)^2(\mu) =  m_i^2 +  \Sigma_{i,\text{proper}}^{(\msbar)}(m_i^2,\mu) +  \Sigma_{i,\text{tad}}^{(\msbar)}(m_i^2,\mu),
\end{equation}
where $\mu$ is the renormalization scale.
Note that the tadpole contributions are also renormalized in 
the \msbar scheme here, so their finite remainders enter in our calculation.
Explicitly, the \msbar one-loop relations for \eqref{eq:tree-level-rels} read
\begin{subequations}
	\label{eq:one-loop-rels}
	\begin{align}
		g^2 &= \frac{4}{v_h^2} \left[ m_W^2\left(1-\frac{\delta v_h^2}{v_h^2}\right) + \Sigma_W^{(\msbar)}(m_W^2)\right],\\
		g'^2 &=\frac{4}{v_h^2} \left[(m_Z^2-m_W^2)\left(1-\frac{\delta v_h^2}{v_h^2}\right) + \Sigma_Z^{(\msbar)}(m_Z^2)-\Sigma_W^{(\msbar)}(m_W^2) \right],\\
		y_t^2 &= \frac{2m_t}{v_h^2} \left[m_t\left(1- \frac{\delta v_h^2}{v_h^2}\right) + 2 \Sigma_t^{(\msbar)}(m_t)\right], \\
		\mu_h^2 &= - \frac{1}{2} \left[\mh^2 + \Sigma^{(\msbar)}_\phi (m_h^2)\right], \\
		\lambda_h &= \frac{1 }{2 v_h^2}\left[m_h^2\left(1-\frac{\delta v_h^2}{v_h^2}\right) + \Sigma^{(\msbar)}_\phi (m_h^2)\right],\\
		\mu_s^2 &= 2 \left(\ms^2 + \Sigma_s^{(\msbar)}(m_s^2)\right) - \frac{v_h^2 \lambda_{hs}}{2}\left(1 + \frac{\delta v_h^2}{v_h^2}\right),
		\label{eq:one-loop-rels-muSsq}
	\end{align}
\end{subequations}
in which $\Sigma_W$ and $\Sigma_Z$ are the transverse parts of the vector boson self-energies.
For the top self-energy, we use 
\begin{equation}
	\Sigma_t(m_t) = \Sigma_s (m_t) + \Sigma_v(m_t),
\end{equation}
following the decomposition
\begin{equation}
	\Sigma_f (p) = m_f \Sigma_s (p^2) + \slashed{p} \Sigma_v(p^2) + \slashed{p} \gamma_5  \Sigma_a(p^2).
\end{equation}
Concerning the vev, we choose to work in the so-called $(G_F,m_W,m_Z)$-scheme.
We write the radiative vev shift as\footnote{Inserting this into \eqref{eq:one-loop-rels} yields the results presented in~\cite{Niemi:2021qvp}.}
\begin{equation}
	\frac{\delta v_h^2}{v_h^2} =  \frac{\Sigma_W^{(\msbar)} (m_W^2)}{m_W^2} - \frac{\delta g^2}{g^2},
\end{equation}
which follows from the relation $v_h = 2 m_W/g$.
The gauge coupling can, in turn, be expressed as 
\begin{equation}
	g^2 = \frac{8 m_W^2 G_F}{ \sqrt{2}}
\end{equation}
and be experimentally accessed by measuring the Fermi coupling $G_F$ 
in muon decay.
One-loop corrections to muon decay
in the \msbar-scheme 
yield
\begin{equation}
	\frac{\delta g^2}{g^2} = \frac{\Sigma_W^{(\msbar)}(m_W^2)-\Sigma_W^{(\msbar)}(0)}{m_W^2} -\frac{g^2}{16 \pi^2} \left[4 \ln \frac{\mu^2}{m_W^2} + \left(\frac{7}{2} \frac{m_Z^2}{m_Z^2 - m_W^2} -2\right)\ln\frac{m_W^2}{m_Z^2} + 6\right].
\end{equation}

Note that the expressions in~\cref{eq:one-loop-rels} are renormalization scale dependent.
This dependence is fully governed by the $\beta$-functions of the respective couplings,
except for~\cref{eq:one-loop-rels-muSsq}.
The reason is that we do not 
make use of
a relation of $\lambda_{hs}$ to a physical observable, as we do for the other parameters.
Thence, $\lambda_{hs}$ is directly used as an input parameter and does not have a one-loop relation to the other parameters of the theory, so its running is not included in~\cref{eq:one-loop-rels-muSsq}.
Apart from that, we explicitly checked that the scale dependence in~\cref{eq:one-loop-rels} exactly reproduces the running as expected from the known $\beta$-functions of the theory.
We 
list
those in~\cref{sec:beta-functions}.

\subsubsection*{OS-like scheme}
In the OS-like scheme (denoted \os in what follows)---after the \msbar-renormalization of $V_1^{T=0}$---one uses renormalization conditions on the effective potential in order to fix a \emph{finite} counterterm contribution to the potential, $V_{\rm CT}$, such that the minima and scalar masses of the potential do not get shifted away from their tree-level values, \viz 
\begin{subequations}
	\label{eq:OS-conditions}
    \begin{align}
        \frac{\partial }{\partial \phi_i} \left(V_1^{T=0} + V_{\rm CT} \right)\,\bigg|_{\phi_i=\langle \phi_i\rangle} &= 0, \label{eq:OS-condition-tad}\\
        \frac{\partial^2 }{\partial \phi_i \partial \phi_j} \left(V_1^{T=0} + V_{\rm CT} \right)\,\bigg|_{\phi_i=\langle \phi_i\rangle} &= 0.
    \end{align}  
\end{subequations}
This can be useful for simplifying the scanning procedure over large ranges of input values for masses~\cite{Basler:2024aaf}.
Note, though, that not all field combinations in \eqref{eq:OS-conditions} yield independent or non-trivial equations, so only a subset of fields will contribute here.
The relevant ones are the tadpole equation for the field $\phi$ (because the vev for this component is non-zero) and the mass equations for $\phi$ and $s$.
All mixed derivatives of $\phi$ and $s$ vanish, and the equations involving derivatives with respect to the Goldstone fields are not independent from the rest, which can be seen by virtue of the corresponding Ward identities, \eg in \cite{Denner:1994xt}.
Lastly, the tadpole equation for the scalar field $s$ is fulfilled trivially in the parameter setup we study, where $v_s = 0$ and does not receive radiative corrections.

As an aside, we note that derivatives of the effective potential in the Landau gauge must be handled with care, as the presence of massless Goldstone boson degrees of freedom can lead to infrared divergences that require proper treatment~\cite{Cline:1996mga, Cline:2011mm}.

The counterterm potential for the cxSM reads
\begin{equation}
	\label{eq:VCT}
	V_{\rm CT} =  \delta\mu_h^2 \Phi^\dagger \Phi + \delta\lambda_h (\Phi^\dagger \Phi)^2 +  \frac{1}{2} \delta \mu_s^2 |S|^2.
\end{equation}
We only need to adjust the three counterterms in \eqref{eq:VCT} in order to keep $m_h$, $m_S$ and $v_h$ at their tree-level values.
This is why no other counterterms are taken into account explicitly.
From \eqref{eq:OS-conditions} we then get
\begin{subequations}
	\label{eq:OS-CTs}
	\begin{align}
		\delta \mu_s^2 &= -2 \frac{\partial^2 V_1^{T=0}}{\partial \phi_s^2}\,\bigg|_{\phi_h = v_h,  \phi_s = 0},\\
		\label{eq:CT-muhsq}
		\delta \mu_h^2 &= \left(\frac{\partial^2 V_1^{T=0}}{\partial \phi_h^2} - \frac{3}{v_h}\frac{\partial V_1^{T=0}}{\partial \phi_h}\right)\,\bigg|_{\phi_h = v_h,  \phi_s = 0},\\
		\delta \lambda_h  &= -\frac{2}{v_h^3} \left( v_h \, \frac{\partial^2 V_1^{T=0}}{\partial \phi_h^2} - \frac{\partial V_1^{T=0}}{\partial \phi_h}\right)\,\bigg|_{\phi_h = v_h,  \phi_s = 0}.
	\end{align}
\end{subequations}
Knowing that field derivatives of the effective potential correspond to 1PI $n$-point functions at zero momentum, we can identify
\begin{subequations}
	\begin{align}
		\frac{\partial^2 V_1^{T=0}}{\partial \phi_s^2}\,\bigg|_{\phi_h = v_h, \phi_s = 0} &= \Sigma_{\phi_s,\text{proper}}^{(\msbar)} (0), \\
		\frac{\partial^2 V_1^{T=0}}{\partial \phi_h^2}\,\bigg|_{\phi_h = v_h, \phi_s=0} &= \Sigma_{\phi_h,\text{proper}}^{(\msbar)} (0), \\
		\frac{\partial V_1^{T=0}}{\partial \phi_h}\,\bigg|_{\phi_h = v_h, \phi_s=0} &= T_{\phi_h}^{(\msbar)},
	\end{align}
\end{subequations}
where $T_{\phi_h}$ are the tadpole diagrams of $\phi_h$.
Moreover, the trilinear Higgs coupling in this model times the zero-momentum Higgs propagator is 
\begin{equation}
	g_{\phi_h^3} \times \Delta_{\phi_h}(p^2=0) =
	-\frac{3}{v_h},
\end{equation}
and we can write
\begin{equation}
	- \frac{3}{v_h}\frac{\partial V_1^{T=0}}{\partial \phi_h} = \Sigma_{\phi_h,\text{tad}}^{(\msbar)}(0).
\end{equation}
Using this, \eqref{eq:OS-CTs} can be written as
\begin{subequations}
	\label{eq:OS-one-loop-rels}
\begin{align}
	\delta \mu_s^2 &= -2 \Sigma^{(\msbar)}_S(0) \,, \\
	\label{eq:delta-muh-Pi}
	\delta \mu_h^2 &= \Sigma_{\phi,\text{proper}}^{(\msbar)} (0) + \Sigma_{\phi,\text{tad}}^{(\msbar)} (0)\, , \\
	\delta \lambda_h  &= -\frac{2}{v^2} \left( \,  \Sigma_{\phi,\text{proper}}^{(\msbar)} (0) +\frac{1}{3}  \Sigma_{\phi,\text{tad}}^{(\msbar)} (0) \right)\, .
\end{align}
\end{subequations}
This might seem counterintuitive because, despite canceling the one-loop tadpoles in \eqref{eq:OS-condition-tad} via $V_{\rm CT}$, tadpole contributions still appear explicitly in \eqref{eq:delta-muh-Pi}.

At this point, we are able to compare the two different renormalization prescriptions, \ie we compare \eqref{eq:one-loop-rels} with \eqref{eq:OS-one-loop-rels}.
For $\mu_h^2$ we find
\begin{equation}
\label{eq:osms}
	(\mu_h^2)^{(\msbar)} - (\mu_h^2)^{(\os)} = \Sigma^{(\msbar)}_\phi (m_h^2) - \Sigma^{(\msbar)}_\phi (0),
\end{equation}
so the two prescriptions merely differ in the momentum of the self-energies. 
Notice, though, that the differences for $\lambda_h$ can not be expressed in such a simple manner.
Notably, the corrections in the \os-scheme are gauge-dependent\footnote{
	The gauge dependence of Lagrange parameters in the \os-scheme propagates to the gauge dependence of 
    the predictions for
    physical observables, 
    even 
    for the case where
    the $\hbar$-expansion of the effective potential
    is applied.
	}
despite the inclusion of tadpole contributions, because the self-energies can only be gauge-independent at the poles.
Another major difference is that in the \msbar-scheme
also the gauge and Yukawa couplings
receive radiative shifts 
from their relations to the physical input quantities,
while in the \os-scheme only $\mu_s^2$, $\mu_h^2$ and $\lambda_h$ do.
Lastly,  both schemes introduce a renormalization scale dependence.
However, in the analysis where the \os-scheme is implemented, the renormalization scale is fixed to $v_h=v_{\text{EW}}$ and is not run to the thermal scale.

\subsection{\texorpdfstring{$\beta$}{Beta}-functions for the cxSM}
\label{sec:beta-functions}
Here, we list the one-loop $\beta$-functions used in our analysis. 
We used the internal \texttt{DRalgo}~\cite{Ekstedt:2022bff} functions to derive them and to cross-check against our one-loop results.
In our analysis, we use $n_g=3$
for the number of fermion generations and keep all fermions massless, except for the top-quark.
Moreover, we do not include the 
anomalous dimensions of the fields here.\footnote{
	The $\beta$-functions listed here correspond at one-loop order to $(-2 \varepsilon)$ times the UV-pole, \eg for the bilinear Higgs term, where the renormalization transformation reads $\mu_h^2 \Phi^\dagger \Phi \to Z_\Phi Z_{\mu^2}\mu_h^2 \Phi^\dagger \Phi$, with $Z_\Phi$ being the field renormalization constant.
	}
The $\beta$-functions read
\begin{subequations}
	\begin{align}
		\beta_{g^2} &= \frac{1}{16\pi^2} \left[ -\frac{19 g^4}{3} \right],\\
		\beta_{g'^2} &=\frac{1}{16\pi^2} \left[  \frac{41 g'^4}{3}\right],\\
		\beta_{y_t^2} &= \frac{1}{16\pi^2} \left[ \frac{9 y_t^4}{2} - y_t^2 \left(\frac{9 g^2}{4} + \frac{17 g'^2}{12} + 8 g_S^2\right) \right], \\
		\beta_{\mu_h^2} &= \frac{1}{16\pi^2} \left[\frac{3}{2}\mu_h^2\left(8 \lambda_h + 4y_t^2-3g^2-g'^2\right) + \frac{1}{2}\lambda_{hs} \mu_s^2\right], \\
		\beta_{\lambda_h} &=  \frac{1}{16\pi^2} \left[  \frac{9g^4}{8} + \frac{3 g^2 g'^2}{4} + \frac{3g'^4}{4} - 6 y_t^4 + \lambda_h \left(12 y_t^2 -9 g^2-3g'^2\right) +24 \lambda_h^2 +\frac{\lambda_{hs}^2}{4} \right],\\
		\beta_{\mu_s^2} &= \frac{1}{16\pi^2} \left[ 2 \lambda_{hs} \mu_h^2+4 \lambda_s \mu_s^2 \right],\\
        \beta_{\lambda_{s}} &=  \frac{1}{16\pi^2} \left[ 5 \lambda_{s} +2\lambda_{hs}\right],\\
        \beta_{\lambda_{hs}} &=  \frac{1}{16\pi^2} \left[ \lambda_{hs} \left(2\lambda_s-\frac{9}{2} g^2-\frac{3}{2}g'^2+6y_t^2+2\lambda_{hs}+12\lambda_h\right) \right].
	\end{align}
	\label{eq:betas}
\end{subequations}
The one-loop beta functions are gauge-independent; however, they become gauge-dependent at two loops. The gauge fixing parameters also run, starting from one loop. For instance, the two-loop $\beta$-functions for the SM in a generalized gauge fixing are available in~\cite{Martin:2018emo}, where they have been derived from the effective potential.

\section{Mass eigenstates in $\rxi$ and Fermi gauges}
\label{app:masseigenstates}

In this appendix, we 
list the tree-level mass eigenvalues, depending on the background fields ($\phi_h$ and $\phi_s$),
of the scalars and vectors in the cxSM in various gauges, defined in \cref{sec:gauge}. 

The singlet field is not directly coupled to the gauge sector, hence its mass eigenstates, together with the physical Higgs degree of freedom, are not affected by the gauge-fixing conditions and stay as in \cref{eq:scalarmixmatrix,eq:Amassvalue}.
Using \cref{eq:gaugefixingCXSM} as the gauge-fixing condition, we obtain the gauge-dependent mass eigenvalues of vectors, Goldstone bosons, and ghosts.
The physical vector degrees of freedom have mass eigenvalues:
\begin{align}
    M_{W}^2 &= \frac{g^2\phi_h^2}{4},\\
    M_{Z}^2 &= \frac{(g^2+g'^2)\phi_h^2}{4}.
\end{align}
On the other hand, the 
unphysical degrees of freedom of the vector bosons mix with Goldstone modes and read (for compactness, we introduce the subscript $V = \{W,Z\}$):
\begin{align}
    M_{V,\pm}^2 & =\frac{1}{2} \left( 
M_G^2 + 2 M_V^2 \frac{\phi^{GF}}{\phi_h} \sqrt{ \xi^V_1 \xi^V_2} \pm M_G\sqrt{ M_G^2 - 4 M_V^2 \left( \xi^V_1 - \frac{\phi^{GF}}{\phi_h} \sqrt{ \xi^V_1 \xi^V_2} \right) }\right),\\
    M_{c_V}^2 &= M_V^2 \frac{\phi^{GF}}{\phi_h} \sqrt{ \xi^V_1 \xi^V_2},
\end{align}
where $M_G^2 = \mu_h^2+\lambda_h \phi_h^2 +\frac{\lambda_{hs} s^2}{4}$, and $M_{c_V}^2$ is the ghost mass eigenvalue.
Then, for the gauge-fixing choices discussed in \cref{sec:gauge}, we have 
\begin{enumerate}
    \item \textbf{Fermi gauge} ($\xi^V_1 = \xi^V$, $\xi^V_2 = 0$):
    \begin{equation} 
        M_{V,\pm}^2 =\frac{1}{2} \left(M_G^2 \pm M_G \sqrt{ M_G^2 - 4 M_V^2 \xi^V_1}\right), \quad M_{c_V}^2 = 0
    \end{equation}
	
    \item \textbf{Background $\rxi$-gauge} ($\xi^V_1 = \xi^V_2 =\xi^V,\phi^{GF} = \phi_h$):
    \begin{equation}
        M_{V,+}^2=M_G^2+\xi^V M_V^2, \quad M_{V,-}^2 =\xi^V M_V^2, \quad M_{c_V}^2 = \xi^V M_V^2
    \end{equation}
    
    \item \textbf{Standard $\rxi$-gauge} ($\xi^V_1 = \xi^V_2 =\xi^V,\phi^{GF} = v_h$):
    \begin{equation}
        M_{V,\pm}^2= \frac{1}{2}\left(M_G^2 + 2 M_V^2 \xi^V \frac{v_h}{\phi_h} \pm M_G\sqrt{ M_G^2 - 4 M_V^2 \xi^V\frac{\phi_h-v_h}{\phi_h} }\right) , \quad M_{c_V}^2 = \xi^V M_V^2 \frac{v_h}{\phi_h}
    \end{equation}
	
    \item \textbf{Landau gauge} ($\xi^V_1 = \xi^V_2 = 0$): 
    \begin{equation}
        M_{V,+}^2=M_G^2, \quad M_{V,-}^2 =0 , \quad M_{c_V}^2 = 0 .
    \end{equation}
\end{enumerate}
In general, 
the relations between
gauge-fixing parameters are not preserved by the RGE evolution, except for the fixed points \cite{DiLuzio:2014bua, Martin:2014bca}.

\end{appendices}

\bibliographystyle{JHEP}
\bibliography{bibl}

\end{document}